\documentclass[nofootinbib,preprintnumbers,superscriptaddress]{revtex4-1}
\usepackage{amsmath}
\usepackage{mathrsfs} 
\usepackage{amssymb}
\usepackage{graphicx}
\usepackage{hyperref}
\usepackage{xspace}
\usepackage{bm}
\usepackage{slashed}
\usepackage{dsfont}
\usepackage{float} 
\usepackage{caption}
\usepackage{subcaption}
\usepackage{xcolor}
\usepackage{natbib}
\usepackage{url}
\usepackage{mathtools}
\usepackage{multirow}
\graphicspath{{./figs/}}

\definecolor{light-gray}{gray}{0.8}

\def\ie{{\it i.e.}\ }
\def\eg{{\it e.g.}\ }

\newcommand{\vl}{\mathbf{v}}
\newcommand{\hl}{\mathbf{h}}
\newcommand{\Evh}{E(\vl, \hl)}
\newcommand{\zob}{$\{ 0, 1 \}$ }
\newcommand{\plusmb}{$\{ -1, 1 \}$ }

\newcommand{\EDaff}{Higgs Centre for Theoretical Physics, School of Physics \& Astronomy,
  University of Edinburgh, Edinburgh EH9 3FD, United Kingdom.}
\newcommand{\IGMMaff}{MRC Human Genetics Unit, Institute of Genetics \& Molecular Medicine, University of Edinburgh, Edinburgh EH4 2XU, United Kingdom.}
\newcommand{\Ascentaff}{Ascent Robotics, Shibuya-ku, Tokyo 150-0012, Japan.}
\newcommand{\ATIaff}{Alan Turing Institute, London NW1 2DB, United Kingdom.}

\begin{document}

\title{Machine learning determination of dynamical parameters: The Ising model case}

\preprint{xxx}

\author{Guido Cossu}
\affiliation{\EDaff}
\affiliation{\ATIaff}
\affiliation{\Ascentaff}

\author{Luigi Del Debbio}
\affiliation{\EDaff}

\author{Tommaso Giani}
\affiliation{\EDaff}

\author{Ava Khamseh}
\affiliation{\IGMMaff}

\author{Michael Wilson}
\affiliation{\EDaff}

\begin{abstract}
\vspace{0.5cm}
	We train a set of Restricted Boltzmann Machines (RBMs) on one- and two-dimensional Ising spin configurations at various values of temperature, generated using Monte Carlo simulations. 
	We validate the training procedure by monitoring several estimators, including measurements of the log-likelihood, with the corresponding partition functions estimated using annealed importance sampling. 
	The effects of various choices of hyper-parameters on training the RBM are discussed in detail, with a generic prescription provided. 
	Finally, we present a closed form expression for extracting the values of couplings, for every $n$-point interaction between the visible nodes of an RBM, in a binary system such as the Ising model. We aim at using this study as the foundation for further investigations of less well-known systems.  
\end{abstract}


\maketitle

\section{Introduction}
A structured probabilistic model is a formalism used in machine learning, in particular deep learning, to describe the joint probability distribution of a set of random variables of interest and their interaction, via mathematical graphs. 
These models are often referred to as graphical models and consist of a set of vertices, connected by a set of edges. Over the years, various graphical models, together with their training and inference algorithms, have been developed by the machine learning community \cite{Goodfellow-et-al-2016}. 
{\it Undirected} models are a popular subset of graphical models, often referred to as Markov Random Fields (MRFs), for which there is no directionality to the edges of the graph connecting the nodes \cite{kindermann1980markov}. 
A class of undirected models, with probability distribution of the form
\begin{align}
p({\bf v})=\frac{e^{-E({\bf v})}}{Z} \ ,
\end{align}
are referred to as energy-based models (EBMs), or Boltzmann machines (BMs) if the models contains latent (hidden) variables \cite{Fahlman:1983:MPA:2886844.2886868,ACKLEY1985147,hinton:boltzmann}. 
In the above equation, $E({\bf v})$ is the positive energy function and $Z$ denotes the partition function. In this work, we focus on Restricted Boltzmann Machines (RBMs), a subset of BMs where the variables {\it within} the visible and hidden nodes are taken to be independent of each other \cite{Smolensky:1986:IPD:104279.104290}. 
RBMs are amongst the most common building blocks of deep probabilistic models. Our aim here is two-fold: the first is to be able to describe criteria that guarantee the machine is being adequately trained, as well as testing its limitations; the second is to derive predictions from the RBM that cannot be directly obtained from the data. \\

The explicit form of the RBMs, together with their properties and training algorithms, are presented in Sec.~\ref{sec:RBMtrain}. The RBM is trained for the 1- and 2-dimensional Ising models with volumes $L=6$, $L^2=8\times8$ and $L^2=16\times16$, at various values of temperature. 
The Ising model and its properties are a paradigm in statistical physics literature, making it a suitable system 
on which to examine the training procedure and limitations of the RBMs \cite{DBLP:journals/corr/MehtaS14,PhysRevB.94.165134,Morningstar2017DeepLT,PhysRevE.97.053304}. 
Moreover, it is possible to generate a large number of Ising configurations using simple Monte Carlo simulations, avoiding the problem of small training sets. 

We dedicate Sec.~\ref{sec:valid-1d} and Sec.~\ref{sec:2d-ising} to the details of our training procedure and the extraction of observables from the states generated by the trained RBM, such as magnetisation, energy, susceptibility and heat capacity. 
In particular, in Sec.~\ref{sec:2d-ising}, we recommend a prescription to ensure the machine is being trained correctly, by monitoring certain quantities during training. 
These quantities include measurements of the log-likelihood and the loss function, as well as the first and second moments of the distribution generated by the RBM. 
Following our training procedure, these observables are then shown to agree well with the expected results, obtained directly from the training data. 
Finally, we present a method for extracting the couplings between the spins of the Ising system, based on an observation in Ref.~\cite{Mehta2018AHL}. 
Using this method, RBMs not only give us a model for describing the distribution of the data, but also provide us with the predictive power of estimating the relative strength of the couplings
between the visible nodes. Notice that due to the indirect all-to-all connections between visible nodes of an RBM, there are no pair-wise coupling assumptions in measuring the strength of connection between the nodes.


\section{Training the RBM \& making predictions}
\label{sec:RBMtrain}

\subsection{Restricted Boltzmann Machines}\label{sec:rbm-formalism}
A Restricted Boltzmann Machine (RBM) is a type of undirected Markov Random Field
(MRF) with a two layer architecture. The detailed derivation of some of the main features of the RBMs is clearly presented in Ref.~\cite{Fischer2012AnIT}, and we follow their notation. An RBM consists of $m$ visible nodes $v_j$,
$j\in\{1,\cdots,m\}$, collectively denoted by ${\bf v}$ and representing the
observed input data, and $n$ hidden nodes $h_i$, $i\in\{1,\cdots,n\}$,
collectively denoted by ${\bf h}$. In this work, we will focus on RBMs with
binary variables, \ie $v_j,h_i \in\{0,1\}$. The energy of the joint state
$\{{\bf v},{\bf h}\}$ of the machine is as follows:
\begin{align}
    E({\bf v},{\bf h};\theta)
    =-\sum_{i=1}^n\sum_{j=1}^m h_iw_{ij}v_j -
    \sum_{j=1}^m b_jv_j - \sum_{i=1}^n c_ih_i \, ,
\end{align}
and we collectively call $\theta = \{\bf w, \bf b, \bf c\}$ the model
parameters. The matrix $\bf w$ represents the undirected interaction between the
visible and the hidden layers; with a slight abuse of language, in what follows
a non-vanishing interaction between nodes will often be called a {\em
connection}. The vectors $\bf b, \bf c$ are the biases of the visible and hidden
layers respectively. Nodes on the same layer are not connected (making the
machine \emph{restricted}), leaving the interaction between them to be mediated
by the connections to the other layer.

The RBM is used to encode the joint conditional probability distribution of a
state $\{{\bf v},{\bf h}\}$, given a set of parameters $\theta$:
\begin{align}
    \label{eq:p joint rbm}
    P({\bf v},{\bf h}|\theta)=
    \frac{e^{-E({\bf v},{\bf h}; \theta)}}{\mathcal{Z}(\theta)} \, ,
\end{align}
where the partition function $\mathcal{Z}(\theta)$ normalises the probability
distribution and contains the sums over all possible states,
\begin{align}
    \mathcal{Z}(\theta) = \sum_{{\bf v},{\bf h}} 
    e^{-E({\bf v},{\bf h};\theta)} \, .
\end{align}
The probability distribution of the variables in the visible layer is obtained
by marginalising over the binary hidden variables $h_i$ (see e.g.
\cite{Fischer2012AnIT}) 
\begin{align}
\label{eq:prbm}
    p({\bf v}|\theta)=\sum_{\bf h}P({\bf v},{\bf h}|\theta)=
    \frac{1}{\mathcal{Z}(\theta)}\sum_{\bf h}e^{-E({\bf v},{\bf h};\theta)} 
    =\frac{1}{\mathcal{Z}(\theta)}\,
    \prod_{j=1}^m \left(e^{b_jv_j}\right)\,
    \prod_{i=1}^n 
    \left(1+e^{c_i+\sum_{j=1}^m w_{ij}v_j}\right) \, .
\end{align}
The conditional probability of $\vl$, given $\hl$ and $\theta$, and the one of
$\hl$, given $\vl$ and $\theta$, will be needed for training the RBM, and for
generating configurations. Because the nodes of a given layer of an RBM are not
connected these conditional probabilities factorise
\begin{align}
    p({\bf h}|{\bf v},\theta)=\prod_{i=1}^n p(h_i|{\bf v},\theta)
    \ \ \ \text{and} \ \ \ 
    p({\bf v}|{\bf h},\theta)=\prod_{j=1}^m p(v_j|{\bf h},\theta) \ ,
\end{align}
and can be readily computed:
\begin{align}
    \label{eq:h-from-v}
    p(h_i=1|{\bf v};\theta)=\sigma\left(\sum_{j=1}^m w_{ij}v_j + c_i\right) \ ,
\end{align}
and 
\begin{align}
    \label{eq:v-from-h}
    p(v_j=1|{\bf h},\theta)=\sigma\left(\sum_{i=1}^n w_{ij}h_i + b_j\right) \ ,
\end{align}
where $\sigma(x) = 1/(1 + e^{-x})$ is the logistic function. 

The training dataset is partitioned into batches, \ie each batch $S$ contains a
subset of the training data, $S=\{{\bf v}_1,...,{\bf v}_l\}$. The RBM is trained by maximising the
log-likelihood function on a given batch $S$ with respect to the parameters
${\bf \theta}=\{w_{ij},b_j,c_i\}$
\begin{align}\label{eq:LL-batch}
    \log \mathcal{L}(\theta| S ) = \log \prod_{v \in S} p({\bf v}|\theta) =  
    \sum_{v \in S} \log p({\bf v}|\theta) \, .
\end{align}
Given an input vector $\bf v$, 
\begin{align}
    \frac{\partial\log \mathcal{L}(\theta|{\bf v})}{\partial \theta}=
    \frac{\partial\log p({\bf v}|\theta)}{\partial \theta}=
    - \sum_{\bf h} p({\bf h}|{\bf v};\theta)
    \frac{\partial E({\bf v},{\bf h};\theta)}{\partial\theta}
    + \sum_{\bf v^\prime,h} P({\bf v^\prime,h}|\theta)
    \frac{\partial E({\bf v^\prime,h;\theta})}{\partial \theta} \, .
\end{align}
More explicitly, the gradient with respect to the weights $w_{ij}$ takes the
form:
\begin{align}
    \frac{\partial\log \mathcal{L}(\theta|{\bf v})}{\partial w_{ij}}
    &= p(h_i=1|{\bf v};\theta) v_j - \sum_{\bf v^\prime} 
    p({\bf v^\prime}|\theta) p(h_i=1|{\bf v^\prime};\theta) v^\prime_j \\
    &= p(h_i=1|{\bf v};\theta) v_j - 
    \left\langle p(h_i=1|{\bf v}^\prime;\theta) v_j\right\rangle_{p({\bf v}^\prime)}\, .
\end{align}
We can then average the gradients over a given batch, so that
\begin{align}\label{eq:L-ave-wij}
    \frac{1}{\left|S\right|} \sum_{{\bf v}\in S}
    \frac{\partial\log p({\bf v}|\theta)}{\partial w_{ij}}=
    \mathbb{E}_\text{data} \left[p(h_i=1|{\bf v};\theta) v_j\right]-
    \mathbb{E}_\text{model}
    \left[p(h_i=1|{\bf v^\prime};\theta) v^\prime_j\right]\, , 
\end{align}
where we have introduced the average over the dataset, 
\begin{equation}
    \label{eq:DataAvrg}
    \mathbb{E}_\text{data} \left[F({\bf v})\right] = 
    \frac{1}{\left|S\right|} \sum_{{\bf v}\in S} F({\bf v})\, ,
\end{equation}
and the average over the probability distribution $p({\bf v}|\theta)$, 
\begin{equation}
    \label{eq:ModelAvrg}
    \mathbb{E}_\text{model}
    \left[F({\bf v})\right] = \sum_{{\bf v}} p(\vl|\theta) F(\vl)\, .
\end{equation}
Similarly, for the bias parameters,
\begin{align}
    \label{eq:L-ave-ci}
    \frac{1}{\left|S\right|} \sum_{{\bf v}\in S} 
    \frac{\partial\log p({\bf v}|\theta)}{\partial b_j}
    &= \frac{1}{\left|S\right|} \sum_{{\bf v}\in S} v_j - \sum_{\bf v^\prime}
    p({\bf v^\prime}|\theta){\bf v^\prime} \\
    &= \mathbb{E}_\text{data}\left[v_j\right]
    -\mathbb{E}_\text{model}\left[v^\prime_j\right] \ , \\
    \label{eq:v-from-j}
    \frac{1}{\left|S\right|}\sum_{{\bf v}\in S}
    \frac{\partial\log p({\bf v};\theta)}{\partial c_i}
    &= \frac{1}{\left|S\right|} \sum_{{\bf v}\in S} p(h_i=1|{\bf v};\theta) -
    \sum_{\bf v^\prime} p({\bf v^\prime}) p(h_i=1|{\bf v}^\prime;\theta) \\
    &= \mathbb{E}_\text{data}\left[p(h_i=1|{\bf v};\theta)\right] -
    \mathbb{E}_\text{model}\left[p(h_i=1|{\bf v^\prime};\theta)\right]\, .
\end{align}
As shown above, the second terms in Eqs.~\ref{eq:L-ave-wij}, \ref{eq:L-ave-ci}
and \ref{eq:v-from-j}, \ie terms that involve a model average
$\mathbb{E}_\text{model}[.]$, contain sums over all possible states of the
RBM; they are independent of the choice of the batch, and are computationally
challenging. In practice, an approximation to the gradients is necessary, here
we use the Contrastive Divergence (CD) algorithm
\cite{Hinton:2002:TPE:639729.639730}. The algorithm starts by choosing a
training vector as the initial state of the visible units. Then a new state of
the hidden units is generated using Eq.~\ref{eq:h-from-v}. 
This is done by setting a hidden unit $h_i$ to be equal to 1 if the
probability in Eq.~\ref{eq:h-from-v} is greater than a random number uniformly
distributed between 0 and 1; otherwise $h_i$ is set to zero. 
Once the hidden units are chosen, a state of visible units, ${\vl}^{(1)}$, is
sampled according to the probability distribution Eq.~\ref{eq:v-from-h}. These
steps can be iterated several times, the number of iterations is controlled by a
parameter, which we denote by $k$ in what follows. Finally, the gradient is
computed using Eq.~\ref{eq:L-ave-wij}, where ${\bf v}^\prime={\mathbf v}^{(k)}$
is the state sampled at the end of the CD loop~\cite{Hinton10apractical}. Given the
gradient the parameters can be updated by any form of stochastic gradient
ascent. In this work the parameters $\theta$ are updated by just taking a step
of size $\alpha$ in the direction of the gradient -- we refer to $\alpha$ as the
{\em learning rate} of the training.

A cycle over all training batches is called a {\em training epoch}, resulting in
a number of parameter updates equal to the number of batches. It is important
for the purposes of our study to be able to quantify the quality of the training
procedure. Therefore, at each epoch during the training procedure, we monitor
several indicators: 

\begin{enumerate}
    \item The maximisation of the average log-likelihood. The average is
    computed first within each batch and then averaged over all batches. The
    computation of the  log-likelihood requires the intractable partition
    function to be estimated. This estimate has been done using the annealed
    importance sampling procedure, discussed in detail in
    Sec.~\ref{sec:annealed-imp-sample}. 
    
    \item For a given state $\vl$, $F\left(\bf v;\theta\right)$ is defined as
    \begin{equation}
        F\left(\bf v;\theta\right) = \log \sum_{\bf h} e^{-E({\bf v},{\bf h}; \theta)}\, .
    \end{equation} 
    This quantity is averaged first across the states in each batch and then
    across all the batches, and it is expected to reach a constant once the machine is trained. To see this more clearly, notice that we can write the log-likelihood in Eq.~\ref{eq:LL-batch} as: 
\begin{align} 
    \sum_{v \in S} \log p({\bf v}|\theta) = \sum_{v \in S} \log \sum_{\bf h} e^{-E({\bf v},{\bf h}; \theta)} -\log(Z)=   \sum_{v \in S} F \left(\bf v;\theta\right)-\log(Z)\, .
\end{align}
Since the log-likelihood is a quantity we are trying to maximise, once the
machine is trained, it converges to a constant. Similarly for the estimation of
the partition function $\log(Z)$. Therefore,
during the training procedure, we can monitor how $F\left(\bf v;\theta\right)$
is also progressing to a constant.     
    \item The loss function is defined as the average over the batches of the
    quantity $F\left({\bf v}^{(0)};\theta \right) - F\left({\bf
    v}^{(k)};\theta\right)$,
    \begin{equation}
        \label{eq:LossTrainig}
        \text{Loss}= \frac{1}{N_{\text{batch}}}\sum_{\text{batches } b} \frac{1}{|S^{(b)}|}\sum_{v \in S^{(b)}}\left(F\left({\bf v}^{(0)};\theta \right) - F\left({\bf v}^{(k)};\theta \right)\right)
    \end{equation}
    where ${\bf v}^{(0)}$ is the input data from the training set, $\vl^{(k)}$
    is the state reconstructed by the machine using $k$ Gibbs sampling steps
    ($k$ is referred to as the CD parameter used for training the machine), and
    $N_{\text{batch}}$ is the total number of batches. 
    
    \item The reconstruction error $\epsilon $, is defined as
    \begin{equation}
        \label{eq:ReconstructionErrorTrainig}
        \epsilon= \frac{1}{N_{\text{batch}}}\sum_{\text{batches } b}\frac{1}{|S^{(b)}|}\sum_{v \in S^{(b)}}|{\bf v}^{(0)} - {\bf v}^{(1)} |^2
    \end{equation}
    where ${\bf v}^{(0)}$ is as stated above and ${\bf v}^{(1)}$ is the state
    reconstructed by the machine using a single step of Gibbs sampling.        
\end{enumerate}
     
In summary, if the training procedure is correct, the log-likelihood should be
an increasing function of the training steps, approaching a plateau where it
stabilises at a fixed value. The quantity $F\left(\bf v;\theta\right)$ is
expected to reach a constant value as described above. The loss function and the
reconstruction error are expected to decrease along the training, as the machine
is required to reduce the difference between the generated ${\vl}$'s and those
used for training.

\subsection{Annealed importance sampling}\label{sec:annealed-imp-sample}
As discussed in the previous section, training the RBM corresponds to maximising
the log-likelihood on the input training set, with respect to parameters of the
model. Therefore, in order to monitor the training progress it is helpful to obtain
an estimate of $\mathbb{E}_{\rm model}[\log p(\theta)]$.
This in turns requires the partition function for the model to be computed,
which is intractable in most cases. However, there are methods to estimate the
partition function. A simple method is to use a proposal distribution $p_0({\bf
v})=\frac{1}{Z_0}p_0^*({\bf v})$ whose partition function is tractable and can
be measured exactly. The asterisk here indicates that the probability is
unnormalised, with its corresponding partition function being the normalisation
factor. Suppose we wish to estimate an intractable partition function $Z_1$ for
$p_1({\bf v})=p^*_1({\bf v})/Z_1$. We can write
\begin{align}
    Z_1=\int p_1^*({\bf v}) d{\bf v} = Z_0 
    \int p_0({\bf v}) \frac{p_1^*({\bf v})}{p_0^*({\bf v})} d{\bf v}\ ,
\end{align}
and measure its estimator
\begin{align}
    \label{eq:Z1-estimate-MC}
    \hat{Z}_1=\frac{Z_0}{M}\sum_{m=1}^M
    \frac{p_1^*({\bf v}^{(m)})}{p_0^*({\bf v}^{(m)})} \, , 
\end{align}
where the sample $\{\vl^{(1)}, \ldots, \vl^{(M)}\}$ is drawn according to
$p_0({\bf v})$. One can readily observe that if the two distributions $p_0$ and $p_1$ have large overlap,
states that are drawn with a high probability from $p_0$, also have a high
probability in $p_1$, and the sum is close to the original integral. In
contrast, if the overlap is poor, the drawn states make negligible
contribution to the sum, making it a biased estimator for $\hat{Z}_1$. The issue
can be further quantified by computing the variance of $\hat{Z}_1$:
\begin{align}
    \text{Var}[\hat{Z}_1] = \frac{Z_0^2}{M^2} 
    \text{Var}\left[\sum_{m=1}^M
    \frac{p_1^*({\bf v}^{(m)})}{p_0^*({\bf v}^{(m)})}\right]
    =\frac{Z_0^2}{M^2} \sum_{m=1}^M
    \text{Var}\left[\frac{p_1^*({\bf v}^{(m)})}{p_0^*({\bf v}^{(m)})}\right]
    =\frac{Z_0^2}{M}\text{Var}
    \left[\frac{p_1^*({\bf v})}{p_0^*({\bf v})}\right],
\end{align}
where in the second equality, it is assumed that the samples are drawn
independently. Expanding the variance of the ratio, we obtain
\begin{align}
    \text{Var}\left[\frac{p_1^*({\bf v})}{p_0^*({\bf v})}\right]
    =\int p_0({\bf v})\left[\frac{p_1^*({\bf v})}{p_0^*({\bf v})}
    -\frac{Z_1}{Z_0}\right]^2 d{\bf v} 
    = \frac{Z^2_1}{Z^2_0}\int p_0({\bf v})
    \left[\frac{p_1({\bf v})}{p_0({\bf v})}-1\right]^2d{\bf v}\, .
\end{align}
Hence, the estimate for the variance of $\hat{Z}_1$ becomes
\begin{align}
    \label{eq:est-var-est-z1}
    \hat{\text{Var}}[\hat{Z}_1]=\frac{\hat{Z}_1^2}{M^2}
    \sum_{m=1}^M 
    \left[\frac{p_1({\bf v}^{(m)})}{p_0({\bf v}^{(m)})}-1\right]^2 \ \ \ 
    \text{s.t. : }{\bf v}^{(m)}\sim p_0\, .
\end{align}
If the two distributions $p_0$ and $p_1$ are not close to each other, in the
Monte Carlo simulation, there will be many samples drawn from $p_0$ with a high
probability which have a corresponding low probability with respect to $p_1$,
leading to a large contribution to the sum in Eq.~\ref{eq:est-var-est-z1}. In
fact, it has been discussed in
Refs.~\cite{MacKay:2002:ITI:971143,Salakhutdinov2008UTMLT2} that the above
variance can become large and even  infinite, leading to very poor estimates. \\

In most cases, as it is with the RBM, $p_1$ is a multimodal distribution over a
high dimensional space. As a result, it is difficult to propose a simple
distribution $p_0$ with a tractable partition function that gives a good
estimate for the original $Z_1$. In such situations a method known as annealed
importance sampling is adopted
\cite{Kirkpatrick1983OptimizationBS,PhysRevLett.78.2690,Neal:2001:AIS:599243.599401,Salakhutdinov2008UTMLT2}.
This method tries to bridge the original distribution $p_1$ and the simple
distribution $p_0$, by introducing intermediate closer distribution
$p_{\beta_0},p_{\beta_1},\cdots,p_{\beta_n}$ such that
$0=\beta_0<\beta_1<\cdots<\beta_{n-1}<\beta_n=1$. We therefore try to estimate
$Z_1/Z_0$ via:
\begin{align}
\frac{Z_1}{Z_0}=\frac{Z_{\beta_1}}{Z_0}\frac{Z_{\beta_2}}{Z_{\beta_1}}\cdots\frac{Z_{\beta_{n-2}}}{Z_{\beta_{n-1}}}\frac{Z_1}{Z_{\beta_{n-1}}}=\prod_{j=0}^{n-1}\frac{Z_{\beta_{j+1}}}{Z_{\beta_j}}\ ,
\end{align}
where $\frac{Z_{\beta_{j+1}}}{Z_{\beta_j}}$ can be estimated using the simple
importance sampling Eq.~\ref{eq:Z1-estimate-MC} described above. These estimates
are reliable as the corresponding distributions are close. The intermediate
distribution can be chosen such that they are proportional to the geometric
average of $p_1$ and $p_0$, \ie,
\begin{align}
p_{\beta}\propto p^*_1({\bf v})^{\beta}p^*_0({\bf v})^{1-\beta} \, .
\end{align}
To see why this is a sensible choice, notice that 
\begin{align}
p^*_1({\bf v})^{\beta} p^*_0({\bf v})^{1-\beta} = e^{-\beta E_1({\bf v})}
e^{-(1-\beta) E_0({\bf v})} = e^{-E_0(\vl)} e^{-\beta\left[E_1(\vl)-E_0(\vl)\right]} \, .
\end{align}
In other words, when $\beta$ is very small, $p_\beta$ is also very close to the
simple distribution $p_0$. Increasing $\beta$ gradually, until it is close to
one, will have $p_\beta$ being close to $p_1$, as required. Here, we make the
choice $E_0=0$, \ie, we sample from a uniform starting distribution $p^*_0({\bf
v})$ for $\beta_{j=0}=0$. This implies that, for $\beta_1,\cdots,\beta_n=1$, 
\begin{align}
    \label{eq:pbj}
    p^*_{\beta_j}(\vl) \propto e^{-\beta_j E_1({\bf v})}\, ,
\end{align}
where $e^{-E_1}({\bf v})$ is the numerator term in Eq.~\ref{eq:prbm}. In the
first step ${\bf v}$ is sampled from a uniform distribution and $p^*_{\beta_1}$
is computed according to Eq.~\ref{eq:pbj}, for a value of $\beta$ close to zero.
A set of configurations ${\bf v}$ is sampled according to that probability using
Gibbs sampling, after having taken into account the presence of $\beta$ in
Eqs.~\ref{eq:h-from-v} and \ref{eq:v-from-h}, where it enters as a
multiplicative factor in the argument of the logistic functions. Contrastive
divergence is taken to be $k_\text{CD}=1$, as described in the previous section.
We chose to increase $\beta$ by 0.0001 in each step and continue until
$\beta_{n}=1$ is reached. The process is repeated for several states ${\bf v}$
and the average is taken at the end. Multiplying the result by $Z_0$ gives the
estimate for $Z_1$, as required. As a final note, in order to avoid numerical
overflow, it is recommended to compute the logarithm of the ratio of the
partition functions, leading to a sum of logarithms of the bridging ratios.

\subsection{1D and 2D Ising model simulations}\label{sec:Ising_model}
We train the RBM on Ising spin configurations in 1D and 2D, distributed
according to the Boltzmann weight: 
\begin{align}
  P(s|J) = \frac{e^{-H_J(s)}}{Z_\text{Ising}} 
  = \frac{e^{-J/(k_BT)\sum_{\langle i,j\rangle}s_is_j}}{Z_\text{Ising}}\ ,
\end{align}
where $\langle i,j\rangle$ indicates that the sum is over nearest neighbour
spins, $T$ is the temperature of the system and $k_B$ is the Boltzmann constant.
The partition function is the sum over all possible configurations of the
system,
\begin{align}
  Z(J)=\sum_s e^{-H_J(s)} \cdot
\end{align}
Each Ising spin can take a binary value $-1$ or $1$, and the external field is
set to zero. The 2D Ising configurations were generated using {\tt Magneto}
\cite{Magneto}, a fast parallel C++ code available online. Magneto uses the
Swendsen-Wang Monte Carlo algorithm to generate the configurations minimising the autocorrelation. For
simplicity we set the combination $J/k_B=1$, and the temperature becomes the
defining feature in differentiating the behaviour of various Ising systems in a
given number of dimensions. Note that spins $\pm 1$ are converted to $0,1$
before setting them as input to the RBM. 

The expectation value for an Ising observable at a given temperature, 
\begin{align}
  \left\langle O \right\rangle|_T=\sum_{\bf s}p({\bf s}|T)O({\bf s}) \, ,
\end{align}
is estimated by the average of the measured observable over a finite set of
independent configurations:
\begin{equation}
	\label{eq:MCintegral}
  \hat{\left\langle O \right\rangle}|_T = 
  \frac{1}{N} \sum_{n=1}^N O \left({\bf s}{(n)},T\right) \ \ \ , \ \ \ \text{s.t.} \ \ \ {\bf s}^{(N)}\sim p({\bf s}|T) \ ,
\end{equation}
where $N $ is the total number of sampled states. The observables under
considerations are the average magnetisation, susceptibility, energy and heat
capacity: 
\begin{equation}
	\begin{split}
  &\left\langle m\right\rangle = \frac{1}{L^2}
    \left\langle\left|\sum_{i=1}^{L^2}s_i\right|\right\rangle ,\\
  &\left\langle\chi\right\rangle = \frac{L^2}{T}
    \left\langle\left\langle m^2 \right\rangle -
    \left\langle m\right\rangle^2 \right\rangle ,\\
  &\left\langle E \right\rangle = -\frac{1}{L^2}
    \left\langle \sum_{\left\langle i,j \right\rangle} s_i s_j\right\rangle , \\
  &\left\langle c_v\right\rangle = \frac{L^2}{T^2}
    \left\langle\left\langle E^2 \right\rangle - 
    \left\langle E \right\rangle^2\right\rangle\, .
	\end{split}
\end{equation}

Once the training process is completed we generate configurations according to
the probability encoded by the RBM. We then compare the results with the
corresponding values from Magneto. The methods for generating new configurations
from the RBM distribution, as well as measurements of observables in the case of
the 2-dimensional Ising model are given in detail in Sec.~\ref{sec:2d-ising}. 


\section{Validation in one dimension}
\label{sec:valid-1d}

To validate our code, we trained a machine on a set consisting of 1-dimensional
Ising states with 6 sites, sampled from a distribution with $J=T=1$, similar to
Ref.~\cite{PhysRevB.94.165134}. The training was divided into three steps of
2000, 1000 and 1000 epochs respectively, with a decreasing learning rate at each
step, as summarised in Table~\ref{tab:rbm-params-1d}. In this case the
analytical form of the probability distribution is relatively easy to compute so
it was used to validate our implementation of both the training code and the
annealing algorithm, similar to Ref.~\cite{PhysRevB.94.165134}.

\begin{table}[hb!]
	\begin{center}
		\begin{tabular}{c c c c | c c c | c c c |c c c | c }
			\hline
			Visible & Hidden  & $N_\text{train}$ & Batch size & Epochs 1 & LR $\alpha_1$ & CD $k_1$ & Epochs 2 & LR $\alpha_2$ & CD $k_2$ & Epochs 3 & LR $\alpha_3$ & CD $k_3$ & $T$ \\\hline\hline
			6    & 6   &  100000 &  200  & 2000  & 0.01 & 1  & 1000 & 0.001 &1  & 1000 & 0.0001   & 1 & 1  \\
			\hline 
		\end{tabular}
	\end{center}
	\caption{Parameters for training an RBM on the 1-dimensional Ising model with six spins, corresponding to the visible nodes. With this choice, there are a total of 64 possibles states, making it possible for the partition function to be measured exactly. We start the training with $\alpha=0.01$ and gradually fine-tune it to smaller values as the training progresses. In this cases keeping $k=1$ throughout the training is sufficient for learning the distribution of the data accurately. }
	\label{tab:rbm-params-1d}
\end{table}

\begin{figure}[!htb]
	\minipage{0.50\textwidth}
	\includegraphics[width=\linewidth]{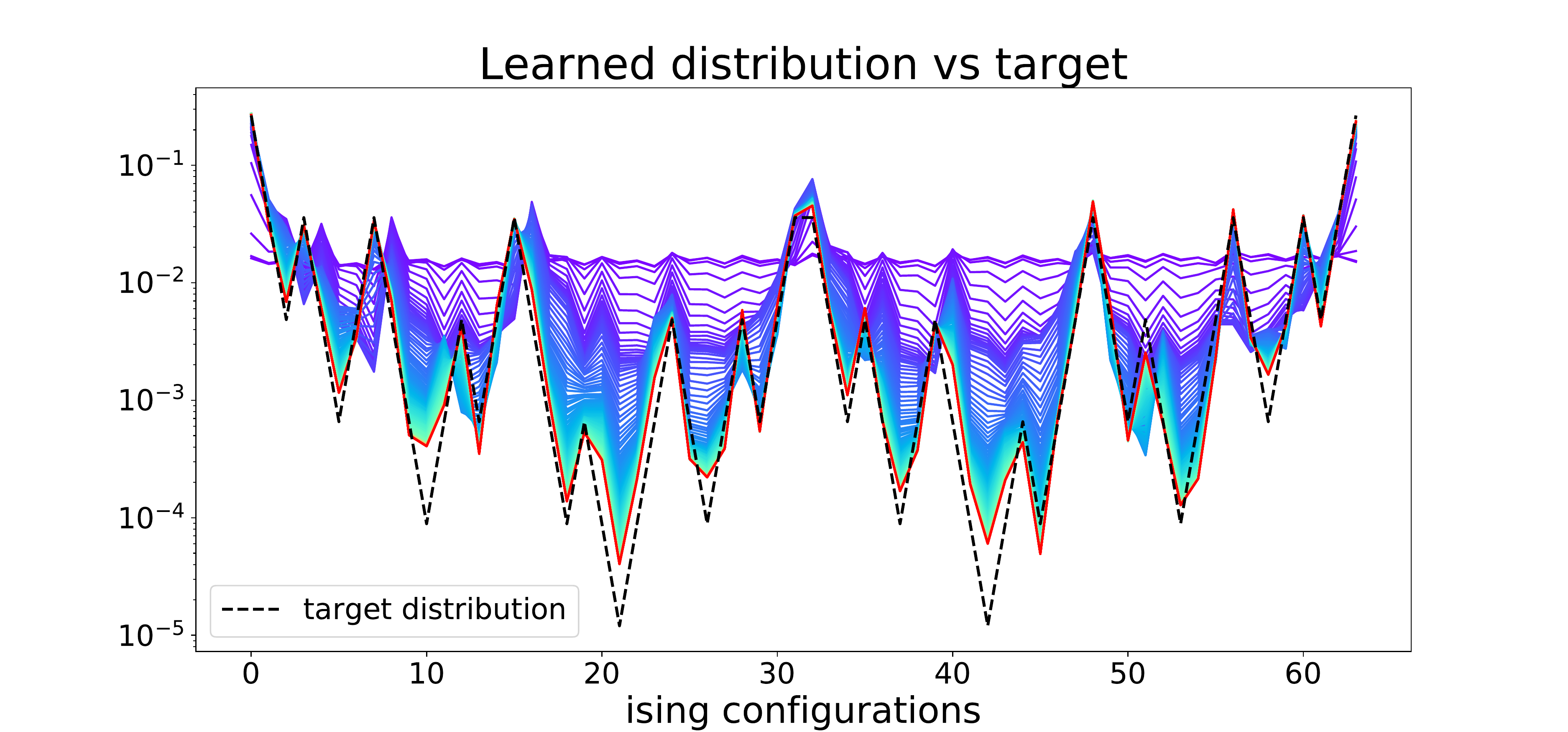}  
	\endminipage\hfill
	\minipage{0.50\textwidth}
	\includegraphics[width=\linewidth]{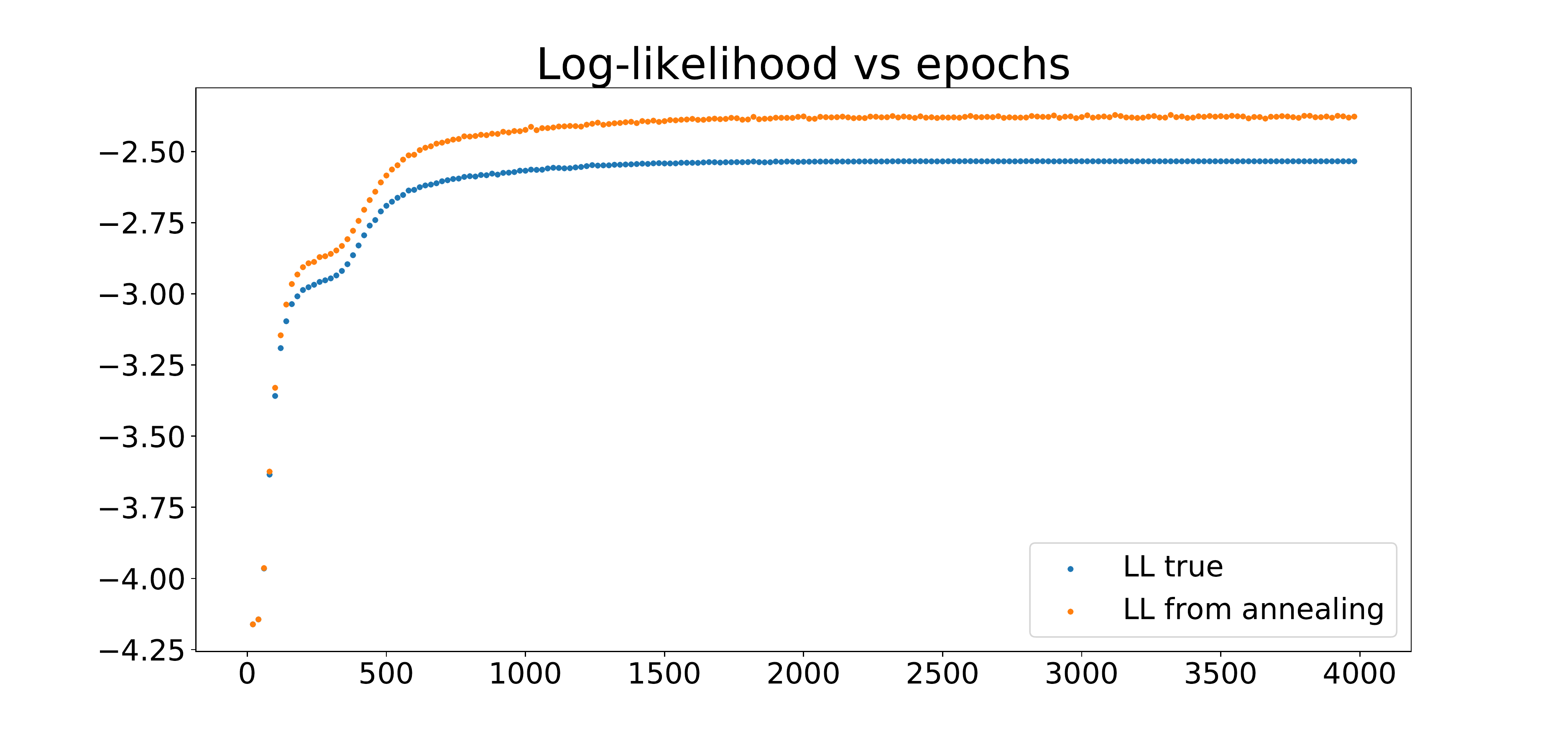}  
	\endminipage\hfill
	\caption{The left hand plot shows the RBM distributions obtained at different number of training epochs, from 20 (blue) to 4000 (red) compared to the target distribution (dotted black line). The right hand plot presents is a comparison between the exact log-likelihood (blue) and that obtained from the annealing algorithm (orange). The uncertainty for the estimated log-likelihood is smaller than the $10\%$.}
	\label{check}
\end{figure}

The results for this test are shown in Fig.~\ref{check}. For a 1D chain with 6
sites, there are 64 different possible states. The $x$-axis of the plot on the
left hand side of Fig.~\ref{check} indicates these states, with their
corresponding probabilities indicated on the $y$-axis. The true probability of
these states, measured exactly, has been plotted with a black dashed line. The
prediction for the underlying distribution improves as the number of training
epochs increases, with the colour blue corresponding to the learned distribution
at 20 epochs and the colour red indicating the learned distribution at epoch
4000, which is very close to the true distribution. The plot on the right hand
side of Fig.~\ref{check} shows, in blue, how the true log-likelihood increases
vs epochs, while the annealing algorithm prediction, plotted in orange, is able
to reproduce the exact log-likelihood with approximately $10\%$ error, keeping
the same overall functional form. We will assume that this is the typical order of
magnitude for accuracy of the log-likelihood resulting from the annealing
algorithm; and will use the log-likelihood computed with the annealing procedure
as our measure for the effectiveness of a training procedure, particularly for
more complicated systems where measurement of the true partition function is
intractable. Notice that the annealing underestimates $Z$, thus resulting in an
upper bound for the true log-likelihood~\cite{Salakhutdinov2008UTMLT2}.

\begin{figure}[!htb]
	\minipage{0.50\textwidth}
	\includegraphics[width=\linewidth]{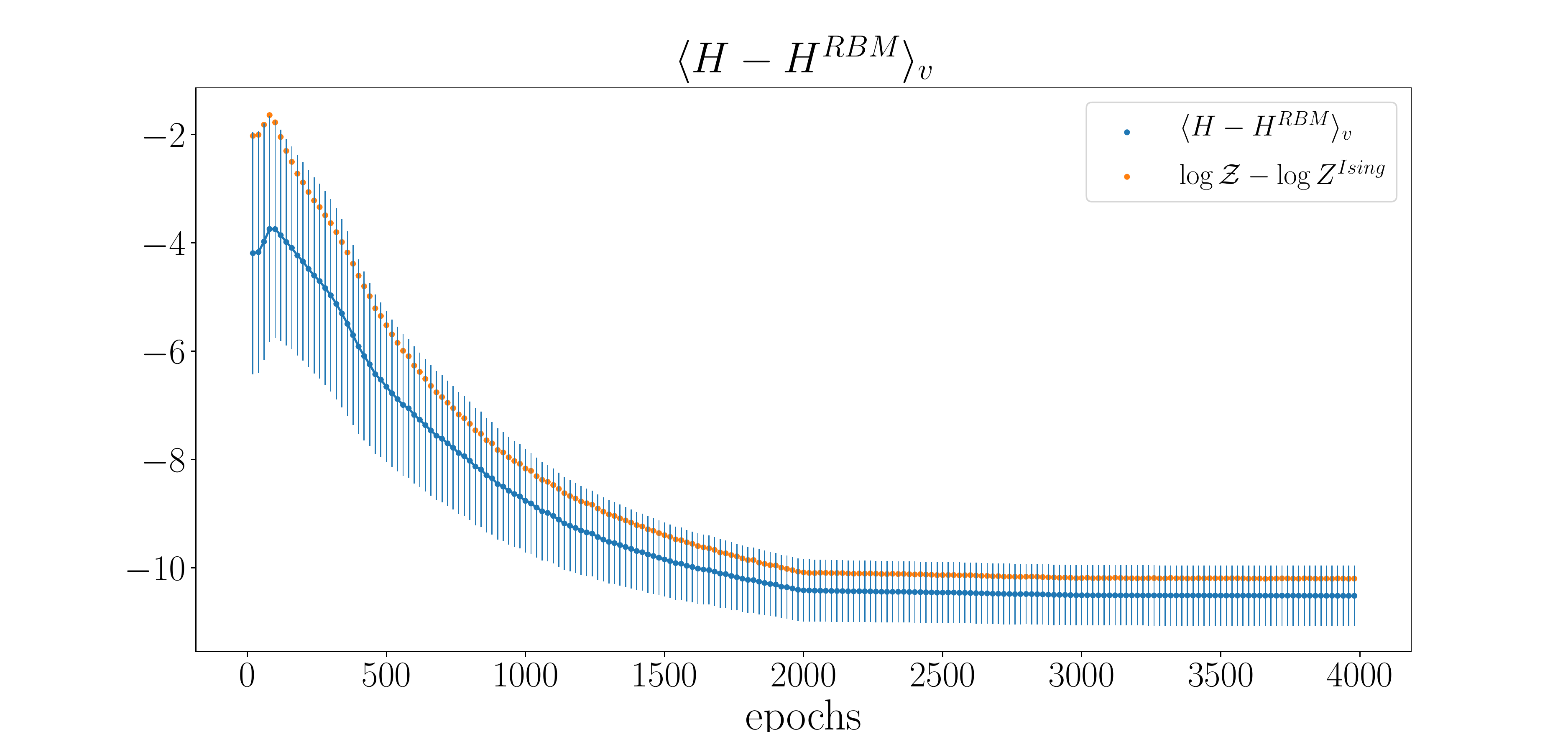}  
	\endminipage\hfill
	\minipage{0.50\textwidth}
	\includegraphics[width=\linewidth]{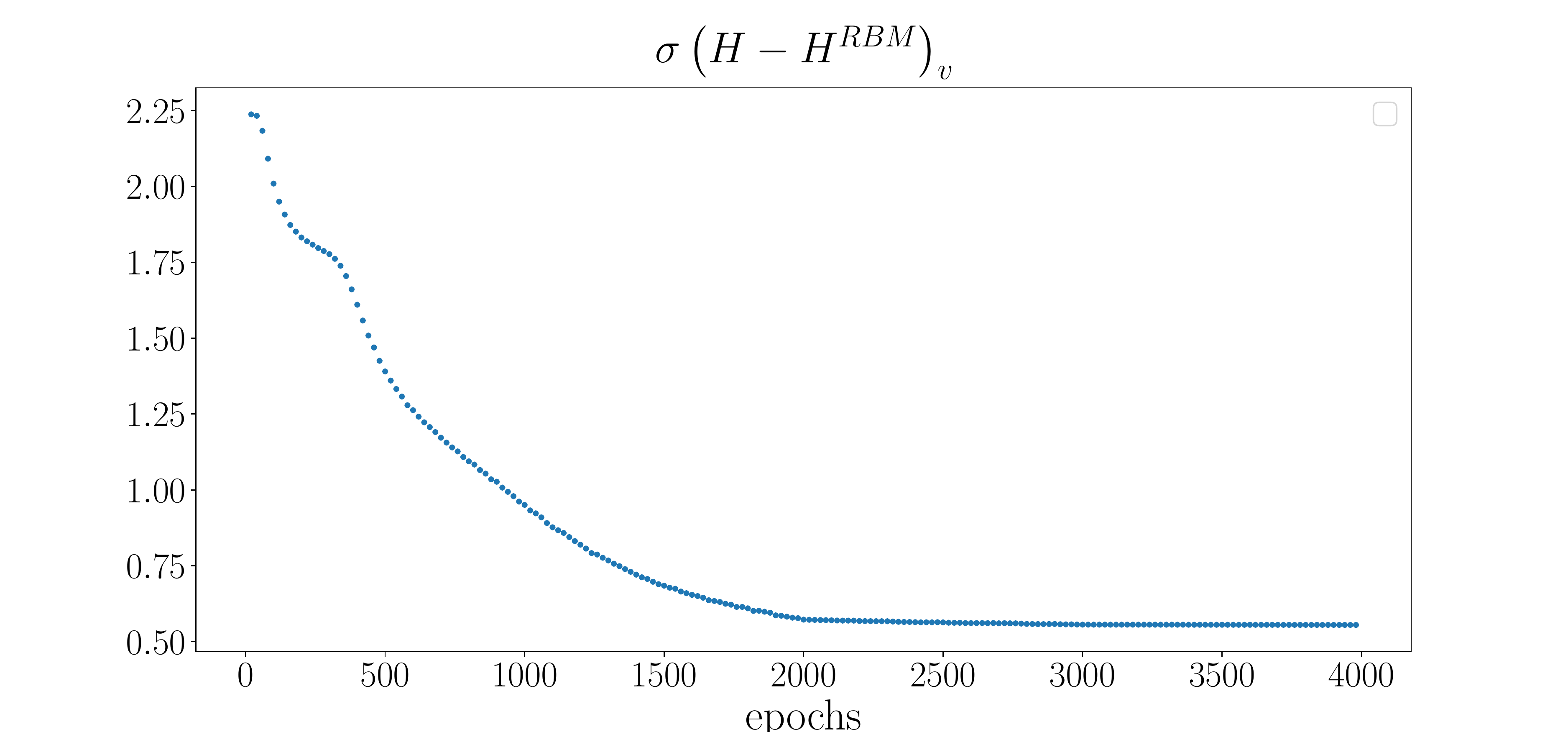}  
	\endminipage\hfill
	\caption{The left hand plot shows $\langle H-H^\text{RBM}  \rangle_{v} $ (blue) and $\log{\mathcal{Z}}-\log{Z^\text{Ising}}$ (orange) per epoch. The right hand plot shows the standard deviation on $\langle H-H^\text{RBM}  \rangle_{v} $ per epoch.}
\label{fig:difference}
\end{figure}

At the end of the training we expect
\begin{align}
\label{eq:H-Z-equal}
\frac{e^{-H\left(v\right)}}{Z^\text{Ising}} = \frac{e^{-H_\lambda^\text{RBM}(v)}}{\mathcal{Z}} ,
\end{align}
which, for every v,  implies
\begin{align}
  \label{eq:deltaH-deltaZ}
  H(v)- H_\lambda^\text{RBM}(v) = 
  \log{\mathcal{Z}}-\log{Z^\text{Ising}} = \text{constant}.
\end{align}
We can reformulate this statement in the following way: minimising the KL
divergence only requires the difference between the Ising probability
distribution and that of the RBM to go to zero
\begin{equation}
  \label{eq:KL-deltaH-deltaZ}
  \begin{split}
    D_\text{KL}\Big[q_\text{Ising}({\bf v})||p_{\theta}({\bf v})\Big] &= 
    \sum_{\vl} \Big( q_\text{Ising}({\bf v}) 
    \log\left(q_\text{Ising}({\bf v})\right) - 
    q_\text{Ising}({\bf v})
    \log\left( p_{\theta}({\bf v}) \right) \Big)  \\
    &= \sum_\vl q_\text{Ising}({\bf v})
    \Bigg[(\log{\mathcal{Z}}-\log{Z^\text{Ising}}) -  
    (H({\bf v})- H_\lambda^\text{RBM}({\bf v})) \Bigg] ,
  \end{split}
\end{equation}
which in turn implies that the Hamiltonians and the corresponding partition
functions may differ by a constant, getting cancelled in the above equation.
Notice also, that this statement has to be true term-by-term once the machine
has learned, \ie for any state $\{\bf v\}$. This is observed numerically, where 
\begin{align}
  \label{eq:eng-diff-state-by-state}
  H({\bf v}^{(1)})-H^\text{RBM}({\bf v}^{(1)}) = \cdots = 
  H({\bf v}^{(64)})-H^\text{RBM}({\bf v}^{(64)}) = 
  \log{\mathcal{Z}}-\log{Z^\text{Ising}}=\text{constant},
\end{align}
up to numerical errors, for all the 64 possible states.
In order to verify if this is in indeed the case in the training the machine, 
on the left hand side of Fig.~\ref{fig:difference}, 
we plot the average $\langle H-H^\text{RBM}  \rangle_{v} $ of the 64 energy differences 
in Eq.~\ref{eq:eng-diff-state-by-state}, 
together with their corresponding standard deviation $\sigma(H-H^\text{RBM})_{v} $, 
as a function of the training epochs. The plot on the right hand side of Fig.~\ref{fig:difference}, 
shows only the values of the standard deviation per epoch which decreases until it reaches a constant. 
Notice that both the average and the standard deviation converge to constant values, $-10.56$ and $0.55$ respectively, 
showing how the differences of Eq.~\ref{eq:eng-diff-state-by-state} indeed converge to a common constant value of $-10.56$ up to a numerical error of 0.55.
We have also checked that this numerical values are consistent with $\log{\mathcal{Z}}-\log{Z^\text{Ising}}$, 
which is plotted, in orange, together with $\langle H-H^\text{RBM}  \rangle_{v} $ in the left hand side of 
Fig.~\ref{fig:difference}: the former lies within 1 sigma from the latter, showing how the KL divergence in 
Eq.~\ref{eq:KL-deltaH-deltaZ} indeed decreases towards zero along the training, converging to a situation 
where $ H(v^{(i)}) - H^\text{RBM}(v^{(i)}) $ is constant.

\section{RBM for the 2D Ising model}\label{sec:2d-ising}
We now consider training sets consisting of spin configurations of 2-dimensional Ising systems at various temperatures.
The configurations are generated using Magneto, setting $J/k_B = 1$, with temperatures above, at and below the critical temperature, 
starting from $T=1.8$ to $T=3.0$ in steps of $\Delta T=0.1$. The training of each RBM on these configurations 
has been studied independently, with the aim of obtaining the set of parameters which provide the correct value of observables 
at each temperature, as compared with Magneto values. The observables for each trained RBM have been computed as described in 
Sec.~\ref{sec:Ising_model}, for different checkpoints along the training, and plotted as a function of the number of epochs. 
This allows us to monitor the training procedure, and to assess the number of epochs required to converge to the underlying 
distribution of the training data. This procedure has been performed for the case of an $8^2$ lattice, as well as a $16^2$ lattice. 
We highlight the increasing difficulties in training when the volume is increased.

\subsection{Ising simulation and RBM training parameters}

The simulation parameters for generating Ising spins using Magneto~\cite{Magneto} and training the RBM are presented in Tables~\ref{tab:ising-params} and \ref{tab:rbm-params-2d} respectively. In the following we present detailed results for three systems of $8^2$ spin configurations, at temperatures $T=1.8$, $T=2.2$ and $T=3.0$ respectively. The procedure for training a system of $16^2$ spins is reported in Appendix.~\ref{app:training-1616}.

In order to train the RBMs we used $10^5$ Ising configurations with batch-size 200. Although the training procedure is specific to each dataset and RBM architecture, we give a general prescription for the case where the number of hidden and visible nodes are chosen to be the same. Most our trainings were performed in three steps:
\begin{enumerate}
\item We start the training with a large learning rate $\alpha_1$, the value of which has to be tested for. The idea for such a choice is to speed up the process of maximising the log-likelihood at the beginning. The CD parameter $k_1$, at this stage, is set equal to 1. These settings were kept as long as they were enough to ensure an increasing log-likelihood. As the number of epoch increases, we observe either the log-likelihood stabilising at a fixed value or starting decreasing. An example of this behaviour is reported in the Appendix, where, in the case of $T=1.8$, it is observed that between 3000 and 4000 epochs, the log-likelihood starts to decrease. This behaviour is observed elsewhere in the literature, \eg see Ref.~\cite{Fischer2012AnIT}.
\item In order to prevent the decrease in the log-likelihood, we decrease the variance in the estimate of the gradient of the log-likelihood, namely we increase the CD parameter to a larger value $k_2$. Again, the value is chosen according to whether or not it results in an increase in the log-likelihood. At this stage, we also decrease the learning rate $\alpha$,  in order to fine-tune the training as we approach the maximum of the log-likelihood. 
\item A single fine-tuning such as the one described above may not be enough to fully train the machine. In that case, we iterate step 2 with a higher value of $k$ and lower value of $\alpha$. This has been done for all our trained machines. 
\item At each of the above steps, we monitor the moments of the learned distribution and compare the values to the expected ones, obtained directly from the training set. This allows us to monitor if and how the training is progressing.
\end{enumerate}
Therefore, the training of each machine in our case is characterised by three different phases with specific number of epochs, learning rates, and values of $k$, as well as comparing the moments extracted from the machine with the expected values obtained from the training data. Our choices for the hyperparameters are summarised in Table~\ref{tab:rbm-params-2d}. \\

\begin{table}
  \begin{center}
    \begin{tabular}{c c c c c c c c c c}
      \hline
      $L^2$ & $N_\text{therm}$  & $N_\text{measure}$ & Binning & algorithm & $T$ & Steps $\Delta T$\\\hline\hline
       $8^2$    & 50000 &  100000  & 1  & SW  & 1.8-3.0 \& 2.27 & 0.1 \\ \hline
       $16^2$  & 50000 &  100000  & 1  & SW  & 1.8-3.0 \& 2.27  & 0.1\\
      \hline 
    \end{tabular}
  \end{center}
	\caption{2D Ising model parameters for data generation. Using the Swendsen-Wang (SW) algorithm, 
	the autocorrelation drops below 1 therefore no binning was required. $N_\text{therm}$ denotes the number of MC iterations used for thermalisation, while $N_\text{measure}$ denotes the measurement taken after thermalisation, which are saved to be used as training examples for the RBM.}
  \label{tab:ising-params}
\end{table}

\begin{table}
  \begin{center}
    \begin{tabular}{c c c c | c c c | c c c |c c c | c c}
      \hline
      Visible & Hidden  & $N_\text{train}$ & Batch size & Epochs 1 & LR $\alpha_1$ & CD $k_1$ & Epochs 2 & LR $\alpha_2$ & CD $k_2$ & Epochs 3 & LR $\alpha_3$ & CD $k_3$ & $T$ & Steps $\Delta T$ \\\hline\hline
       $8^2$    & $8^2$   &  $10^5$ &  200  & 3000  & 0.1 & 1  & 1000 & 0.01 &5  & 1000 & 0.001   & 10 & 1.8-2.1 & 0.1 \\
       $8^2$    & $8^2$   &  $10^5$ &  200  & 2000  & 0.01  &1  &1000 & 0.001 &5  & 1000 &0.0001 & 5 & 2.2-3.0 & 0.1 \\ \hline
       $16^2$  & $16^2$ &  $10^5$ &  200  & 8000  & 0.01  & 10   & 2000  & 0.01 & 20 & $//$& $//$  &$//$ & 1.8 & $//$\\
      \hline 
    \end{tabular}
  \end{center}
  \caption{Parameters for training the RBMs. The number of visible and hidden nodes for each case is presented. $N_\text{train}$ denotes the number of examples used to train the machine. $T$ indicates the temperatures of the given systems, with intervals $\Delta T=0.1$. The learning rate (LR), $\alpha$, is set fixed for Epochs 1 for the initial training and is later fine-tuned until the observables results for the last several epochs converge within statistics. As advised in Ref.~\cite{Fischer2012AnIT}, whenever the increase in the log-likelihood plateaus and then starts to decrease, we reduce the value of $\alpha$ and increase $k$. The number of epochs in each phase depends on the size and temperature of each system. Our tests indicate that a higher value of $k$, \ie $k=10$, needs to be used to train the machine on the larger configuration $v^2=16\times16$, which is then increased to $k=20$ for the last phase of the training. }
  \label{tab:rbm-params-2d}
\end{table}

\begin{figure}[!htb]
	\minipage{0.50\textwidth}
	\includegraphics[width=\linewidth]{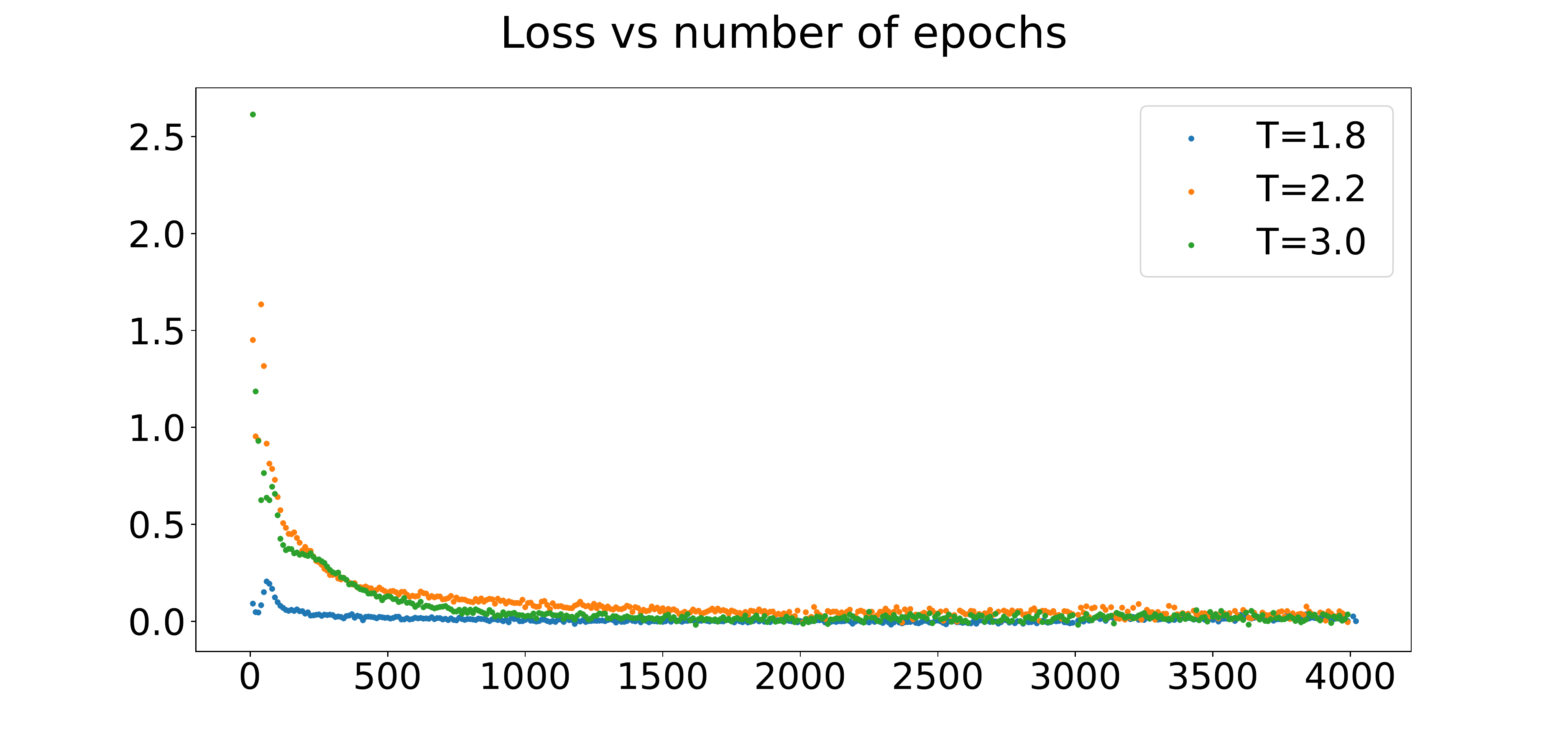}  
	\endminipage\hfill
	\minipage{0.50\textwidth}
	\includegraphics[width=\linewidth]{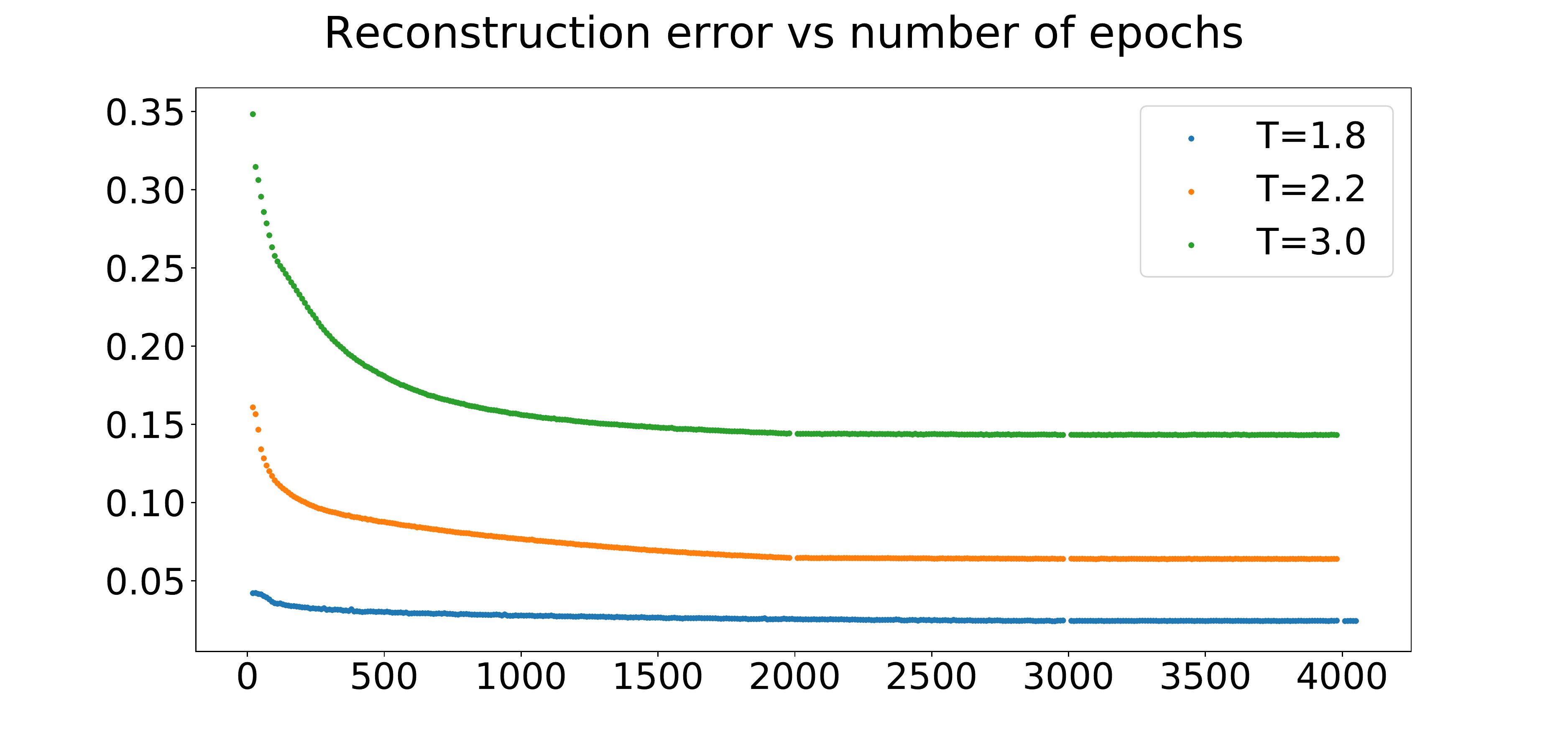}  
	\endminipage\hfill
	\minipage{0.50\textwidth}
	\includegraphics[width=\linewidth]{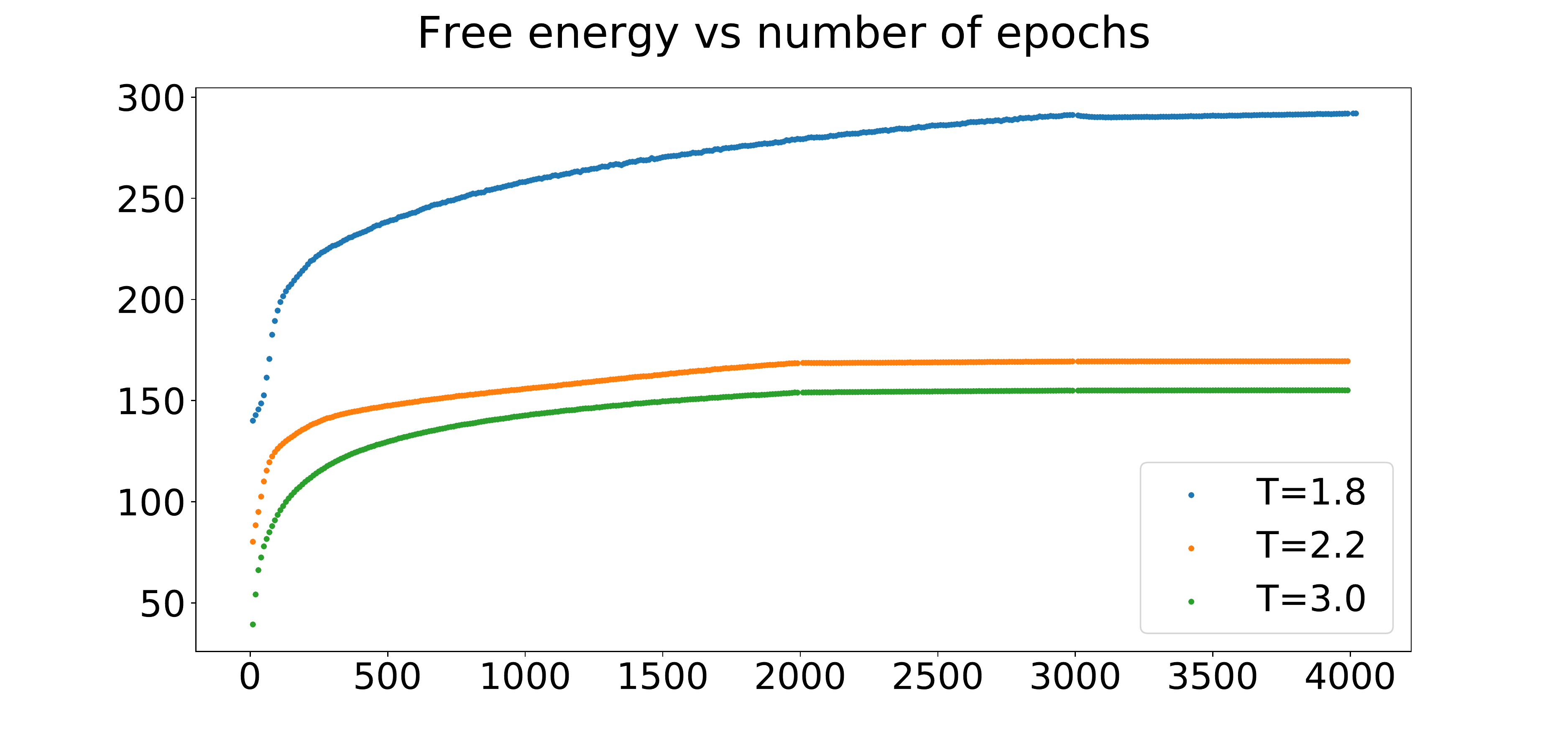}  
	\endminipage\hfill
	\minipage{0.50\textwidth}
	\includegraphics[width=\linewidth]{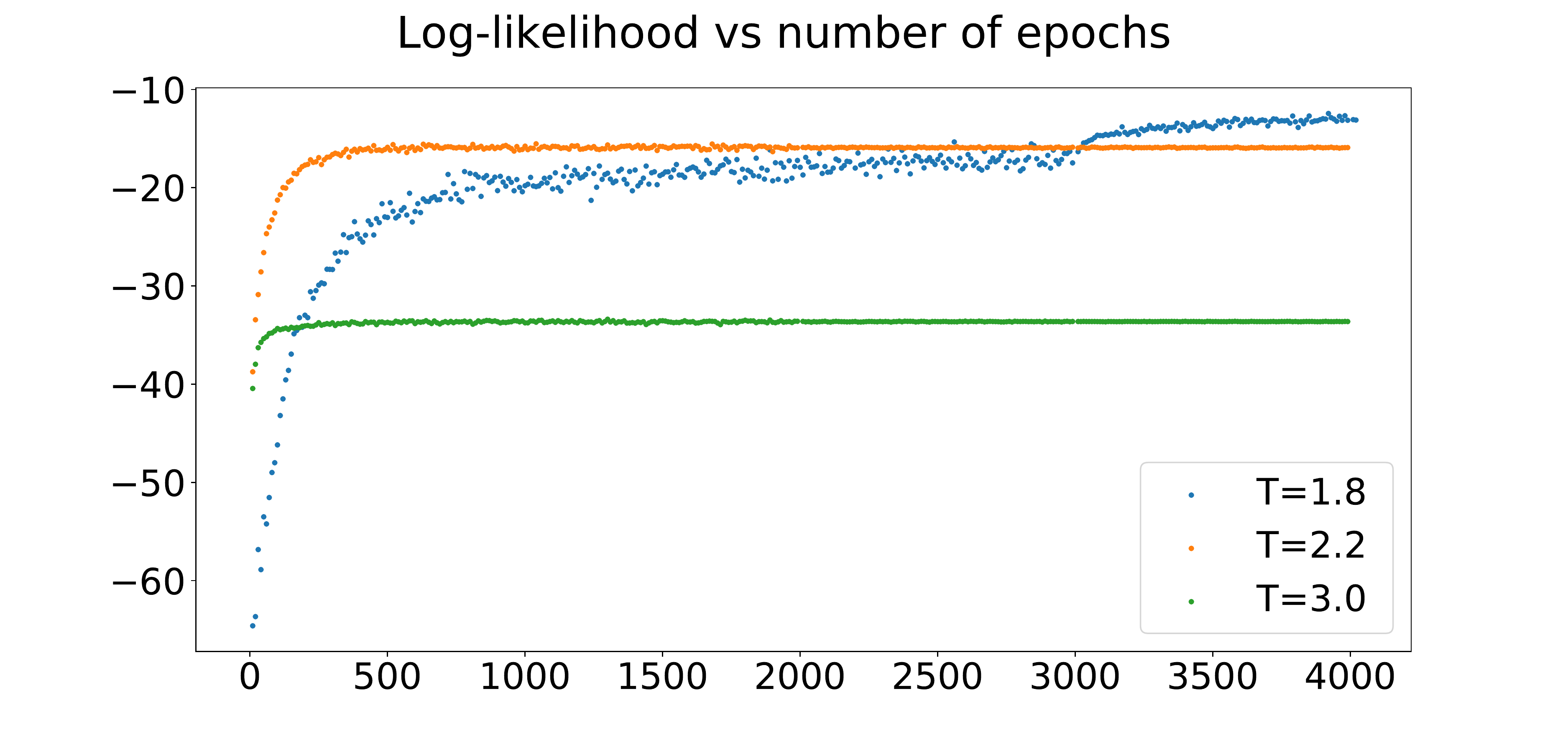}  
	\endminipage\hfill
	\caption{From left to right, loss function and reconstruction error in the first line, free energy and log-likelihood in the second one. The results are presented for three different $T=1.8$, $T=2.2$ and $T=3.0$. }
	\label{fig:allestimators}
\end{figure}

The four quantities used to monitor training, \ie the log-likelihood, free energy, loss function and reconstruction error are reported as functions of the training epoch in Fig.~{\ref{fig:allestimators}}, for three different temperatures. They all have the expected behaviour, as described in Sec.~\ref{sec:rbm-formalism}. We stop the training process when there is no further increase in the log-likelihood and, after fine-tuning with a higher value of $k$ and smaller $\alpha$, the first and second moments of distributions generated by machines at different epochs are statistically equivalent. We can conclude that the training algorithm, so far, has been successful. In the next section we assess how well the machine is able to predict the observables, as compared to the measured Magneto values. 

\subsection{Gibbs sampling}
As anticipated in Sec.~\ref{sec:Ising_model}, for each trained machine we have computed the average magnetisation, susceptibility, energy and heat capacity, together with their corresponding uncertainties. The results have been compared to the values obtained from the training set, in order to examine if our machines are able to reproduce the first and second moments of the underlying distribution.
In order to sample the states from the RBM distribution, required to implement Eq.~(\ref{eq:MCintegral}), we have used Gibbs Sampling, Ref.~\cite{Fischer2012AnIT}.
Block Gibbs sampling is a particular Markov Chain Monte Carlo method (MCMC). 
The general idea behind any MCMC method is that of setting up a Markov Chain that converges to the distribution we want to sample from, in this case Eq.~(\ref{eq:p joint rbm}), and then running the chain long enough to reach its stationary point. In the case of Gibbs sampling, this translates into building the chain
$${\bf v}^{(0)} \rightarrow {\bf h}^{(0)} \rightarrow {\bf v}^{(1)} \rightarrow {\bf h}^{(1)} \rightarrow {\bf v}^{(2)} \,\,\, \cdots \ ,$$ 
where ${\bf h}^{(l)} $ and ${\bf v}^{(l+1)} $ are sampled from the conditional probabilities given in Eq.~\ref{eq:h-from-v} and Eq.~\ref{eq:v-from-h} respectively. It can be shown that this chain converges to $P({\bf v},{\bf h};\theta) $, which implies that at a certain point in the chain, the states ${\bf h}^{(l)}$ and ${\bf v}^{(l)}$ will be sampled according to the joint probability distribution encoded in the RBM. The word {\it block} here, refers to the fact that we can simultaneously update all variables $h_i$ given ${\bf v}^{(l)}$, and vice versa, as the RBM is a restricted network which means that the variables within each layer are independent of each other. Note that the Contrastive Divergence algorithm, used to obtain estimates of the log-likelihood gradient as described in Sec.\ref{sec:RBMtrain}, is only a Gibbs sampling run with a single step, to get ${\bf v}^{(1)}$ from ${\bf v}^{(0)}$ through ${\bf v}^{(0)} \rightarrow {\bf h}^{(0)} \rightarrow {\bf v}^{(1)} $.
One of the main advantages of Gibbs sampling over other MCMC methods (see the next section on Metropolis sampling for an example) is that it converges relatively fast. This allowed us to measure and plot the observables at many epochs along the training, together with their corresponding uncertainties, as an indication of progress in learning. The details concerning the Gibbs sampling parameters are reported in Table~(\ref{tab:gibbs-metro-params}). After a thermalisation step of $2\times10^4$ configurations, we measure magnetisation and energy on every 100th configuration, allowing the Markov chain to run for $2\times10^6$ configurations. 
\begin{table}
  \begin{center}
    \begin{tabular}{c c c c c c c c c c}
      \hline
     Algorithm  & $N_\text{therm}$  & $N_\text{measure}$ & Binning   \\\hline\hline
       Gibbs &  20000 & $2\times 10^6$   & 100   \\ \hline
       Metropolis &  $50000$ & $1\times10^6$  & 50   \\
      \hline 
    \end{tabular}
  \end{center}
  \caption{Gibbs and Metropolis sampling parameters, for MC simulations on an already trained RBM. The measurements are made on configurations generated after the initial thermalisation step $N_\text{therm}$. The binning factor indicates the number of successive measurement binned to ensure the remaining are indeed independent.   }
  \label{tab:gibbs-metro-params}
\end{table}
\begin{figure}[!htb]
	\minipage{0.50\textwidth}
	\includegraphics[width=\linewidth]{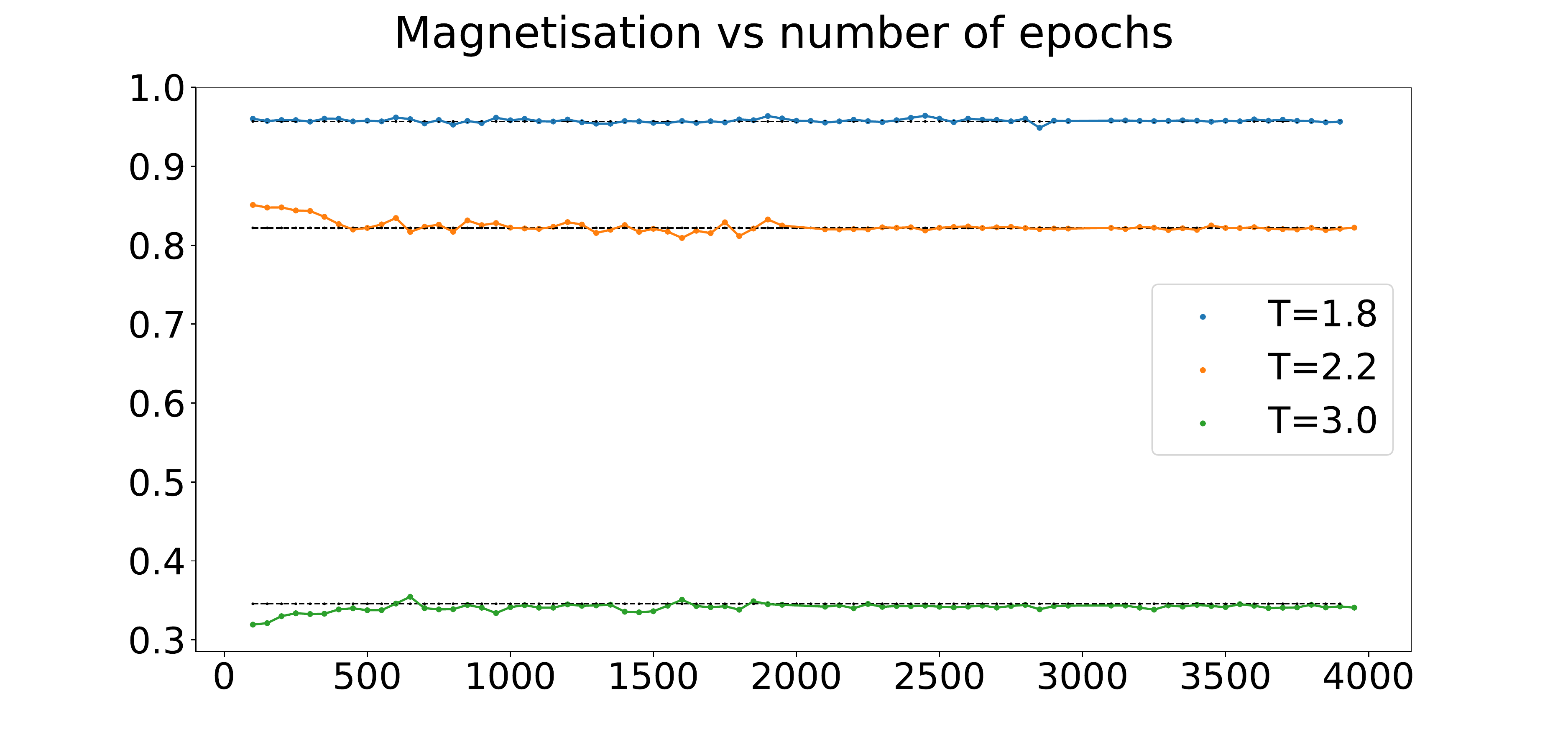}  
	\endminipage\hfill
	\minipage{0.50\textwidth}
	\includegraphics[width=\linewidth]{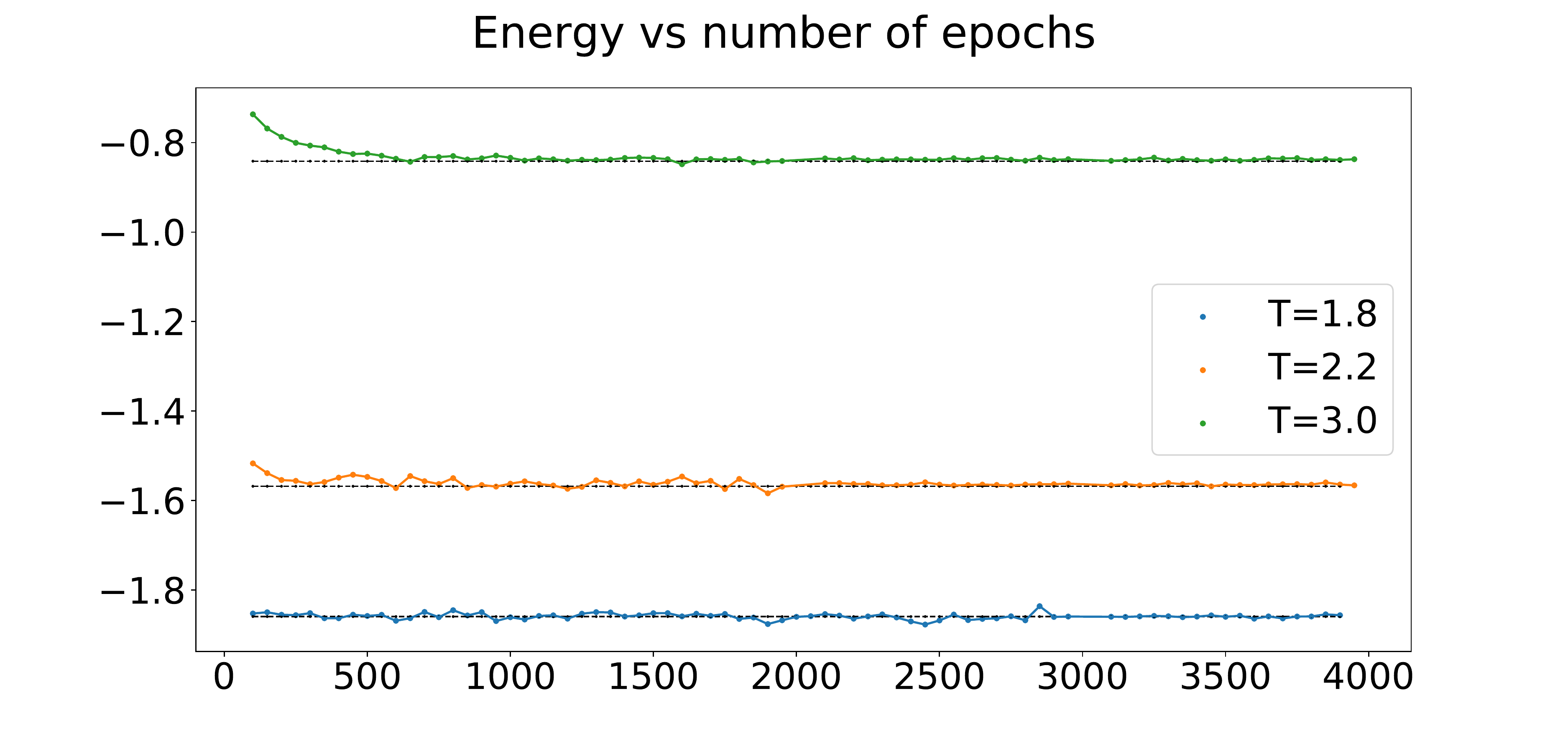}  
	\endminipage\hfill
	\minipage{0.50\textwidth}
	\includegraphics[width=\linewidth]{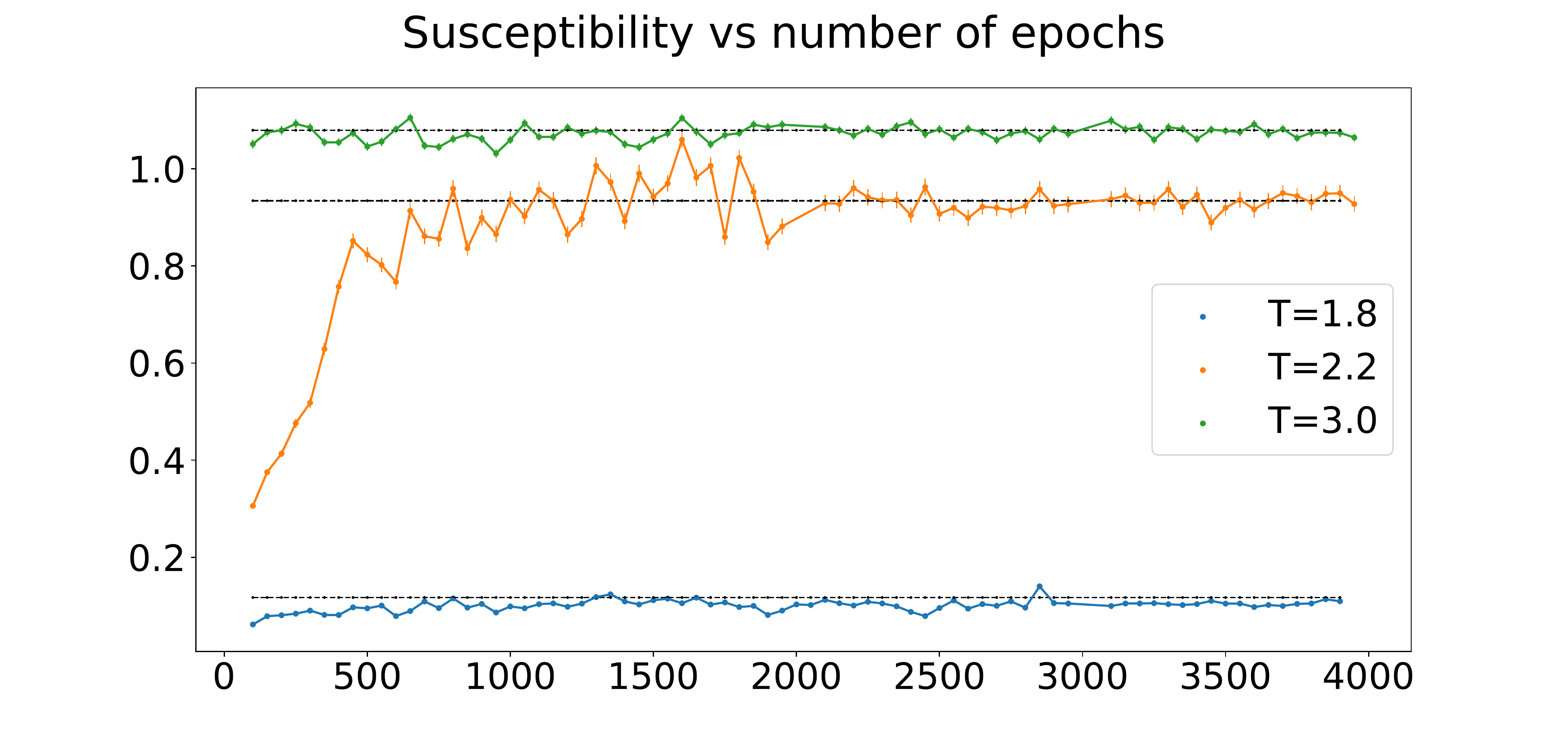}  
	\endminipage\hfill
	\minipage{0.50\textwidth}
	\includegraphics[width=\linewidth]{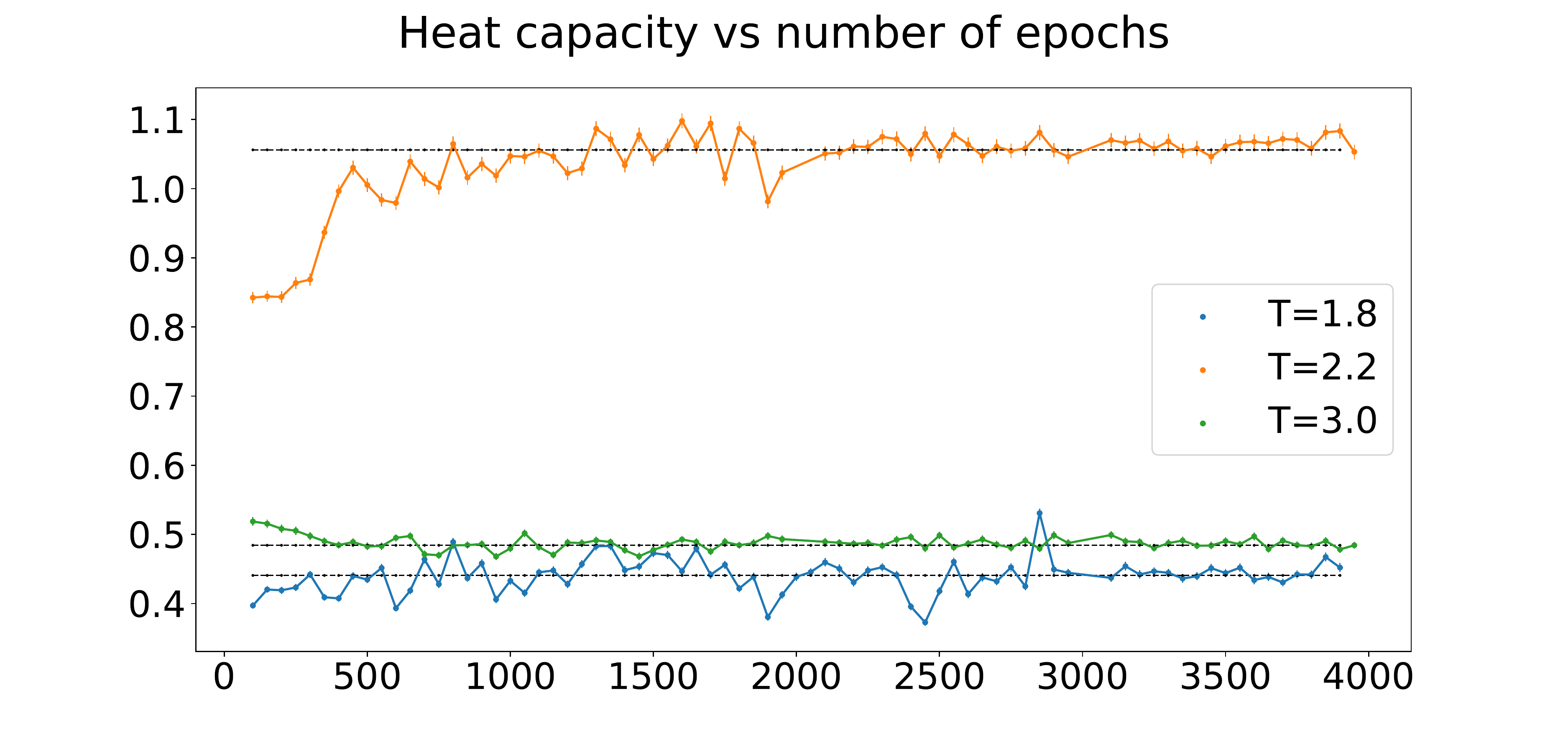}  
	\endminipage\hfill
	\caption{From left to right, magnetization and energy in the first line, susceptibility and heat capacity in the second one. The results are presented for three different T=1.8, T=2.2 and T=3.0, as function of the training epoch. }
	\label{fig:allobservables}
\end{figure}

\subsection{Metropolis sampling}
In order to have an independent measurement of the observables, we implemented the Metropolis algorithm. The algorithm takes, as input, the parameters of a trained RBM, $\theta = \{\bf w, \bf b, \bf c\}$. Starting from a random configuration, a new spin configuration is proposed by randomly flipping a spin using the transformation $v_i=1-v_i$. Letting $u$ to be a random number between zero and one from the uniform distribution, if,
\begin{align} 
\min\left(1, \frac{p_\text{RBM}({\bf v}_\text{new})}{p_\text{RBM}({\bf v}_\text{old})}\right) > u \ , 
\end{align}
we accept the proposed configuration. Otherwise we reject the new configuration and return to the original one. At each Monte Carlo sweep, this process is repeated for all spin variables. Since we take the ratio between the new and the old probabilities, the partition functions cancel and so it does not need to be estimated. The procedure is repeated until the thermalisation is reached. These can be checked by checking that the Monte Carlo history of an observable e.g. magnetisation, follows a normal distribution. We chose to discard the first $5\times10^5$ configurations for the thermalisation step. See \eg Fig.~\ref{fig:hist-metro-18} in the appendix. Note that the generated configurations in the $0,1$ basis are mapped to those of the Ising model in the $-1,1$ according to $v_\text{ising}=2v_{RBM}-1$. \\

In order to take independent measurements of the observables, successive number of measurements of $|m|$ and $E$ were binned. To determine the bin size, the error on \eg $|m|$ vs bin size was plotted in Fig.~\ref{fig:metro-bin-size}. To confirm independence, the autocorrelation time defined as:
\begin{align}\label{eq:auto-corr-time}
A(T)=\frac{1}{N_\text{meas}-t}\left[\sum_{i=1}^{N_\text{meas}-t}m^{(i)}m^{(i+t)}-(\overline{m}^{(i)})^2\right]/\left(\overline{(m^{(i)})^2}-(\overline{m}^{(i)})^2\right) \ ,
\end{align}
where $N$ is the total number of Monte Carlo steps, was also computed. The autocorrelation time vs MC steps is plotted in Fig.~\ref{fig:metro-auto-corr}. Based on these two plots, choosing bin size equal to 50, seems sufficient for independent measurements to be obtained. We present the parameters of the Metropolis sampling in Table.~\ref{tab:gibbs-metro-params}. The magnetisation $\langle |m|\rangle$, energy $\langle E \rangle$, susceptibility $\chi$ and heat capacity $C_V$, are then measured. The error on these quantities have been computed using the statistical bootstrap procedure. \\

We also tried to extract the third and fourth moments, however, the corresponding uncertainties turned out to be of the same order as the moments themselves.

\subsection{Observable predictions at all temperatures}
\begin{figure}[!htb]
	\minipage{0.50\textwidth}
	\includegraphics[width=\linewidth]{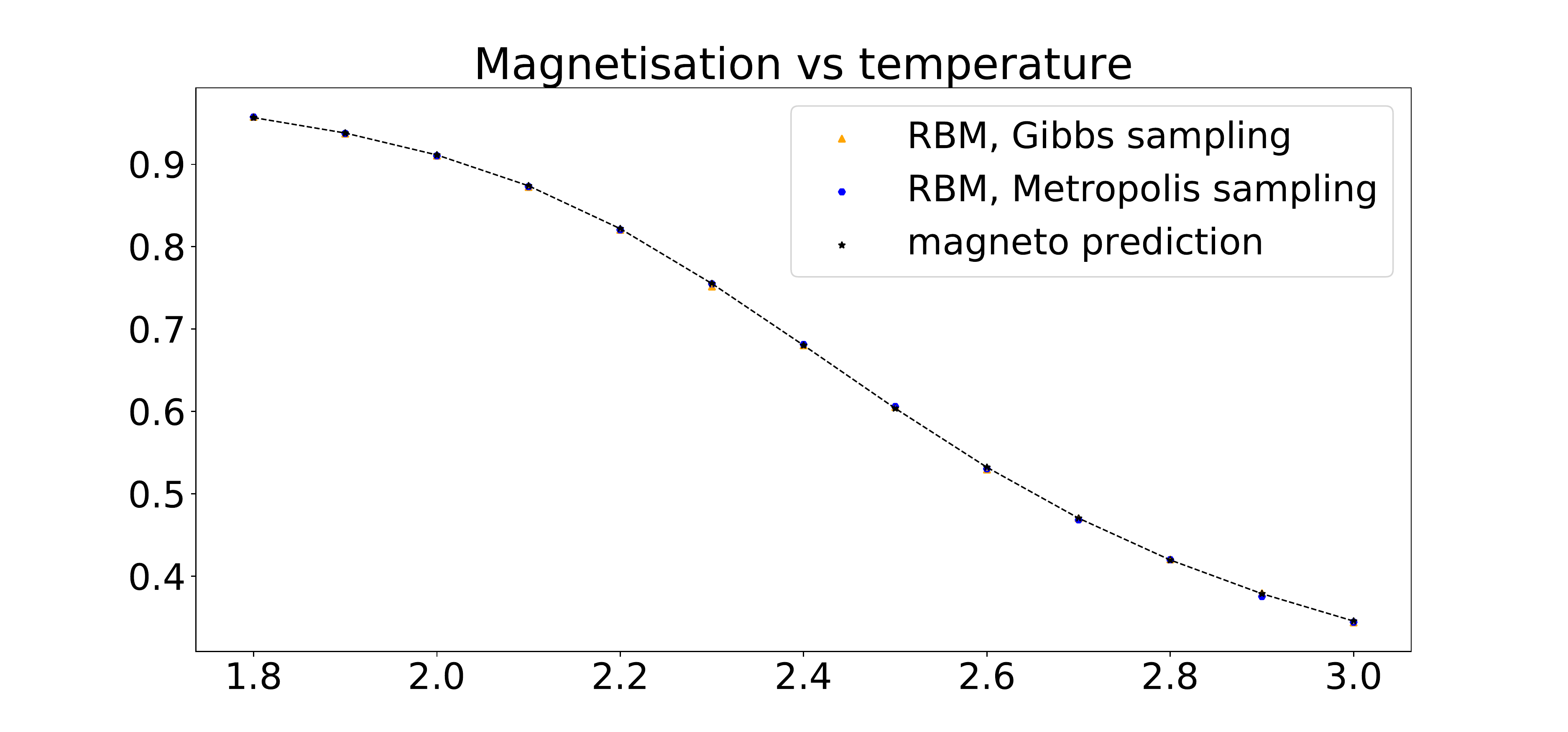}  
	\endminipage\hfill
	\minipage{0.50\textwidth}
	\includegraphics[width=\linewidth]{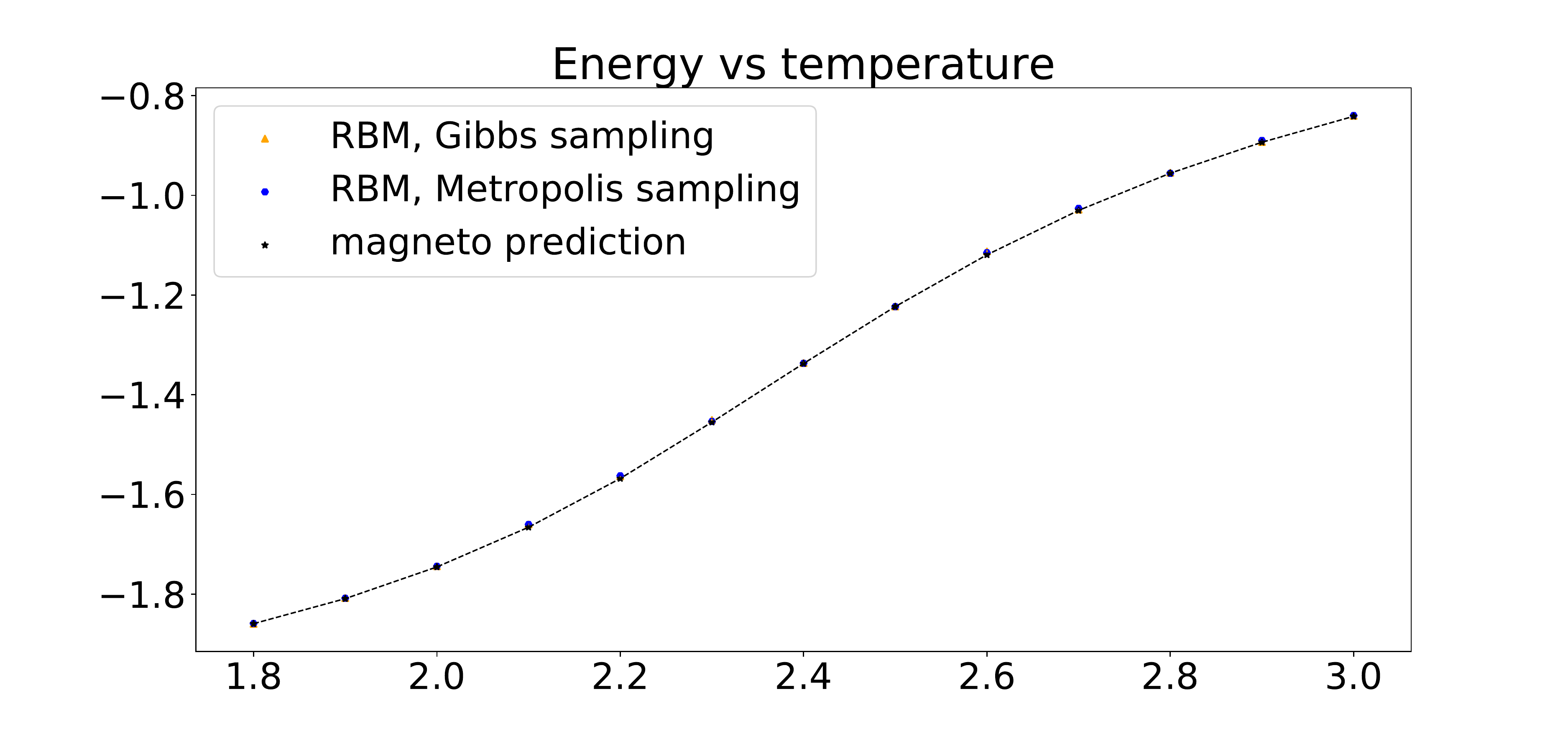}  
	\endminipage\hfill
	\minipage{0.50\textwidth}
	\includegraphics[width=\linewidth]{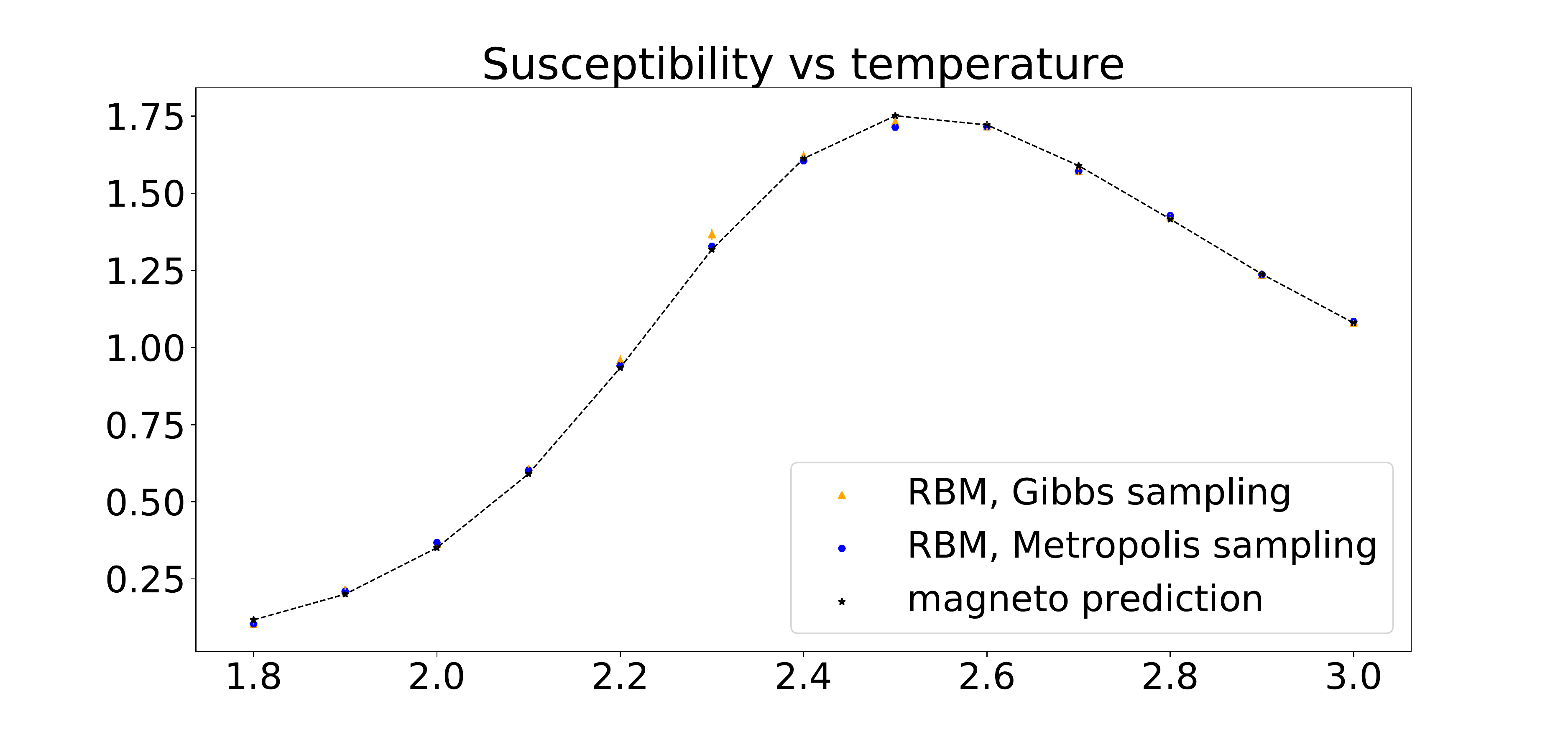}  
	\endminipage\hfill
	\minipage{0.50\textwidth}
	\includegraphics[width=\linewidth]{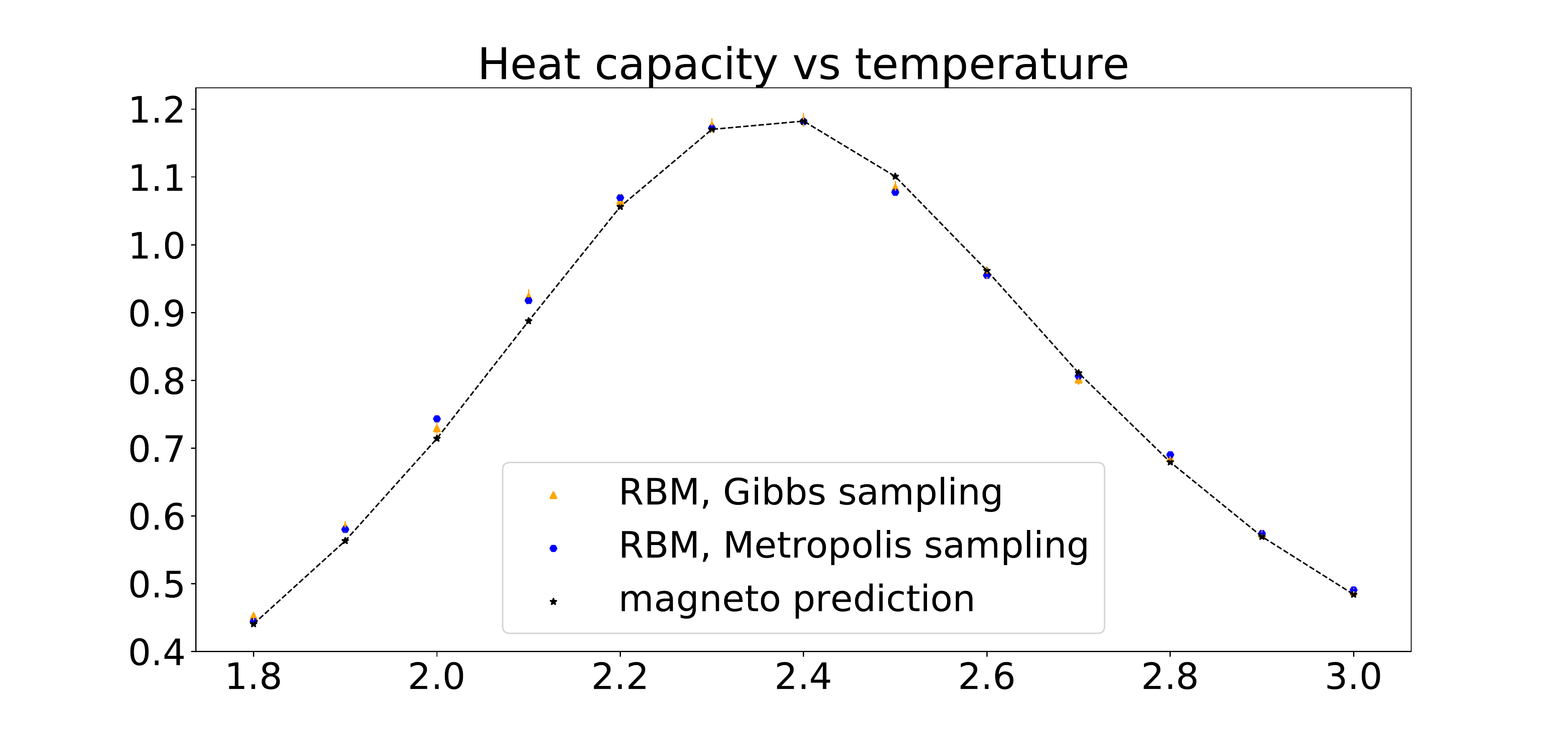}  
	\endminipage\hfill
	\caption{Observables vs temperature. Results coming from Gibbs and Metropolis samplings are compared to the expected values from Magneto.}
	\label{fig:allobservablest}
\end{figure}

\begin{figure}[!htb]
	\minipage{0.50\textwidth}
	\includegraphics[width=\linewidth]{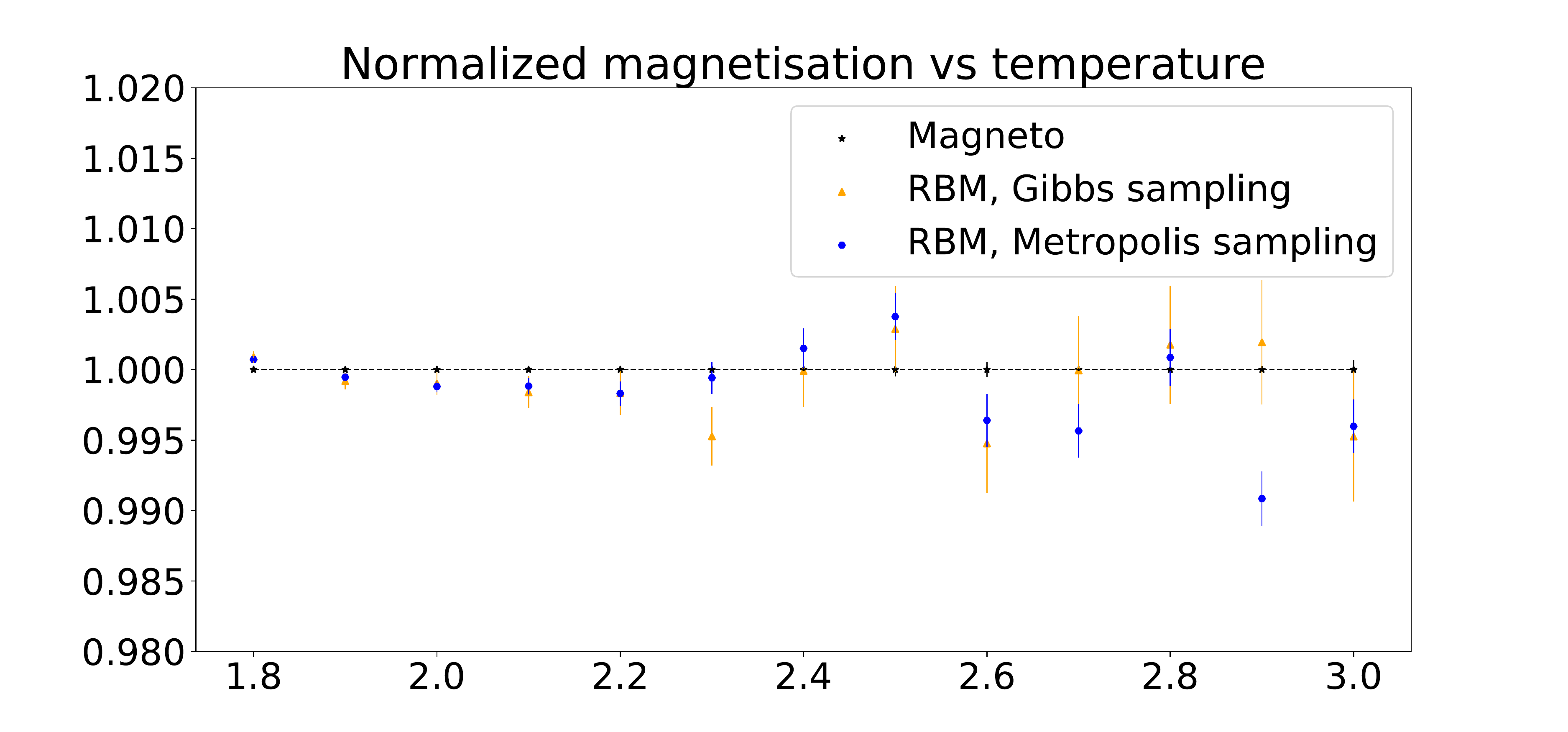}  
	\endminipage\hfill
	\minipage{0.50\textwidth}
	\includegraphics[width=\linewidth]{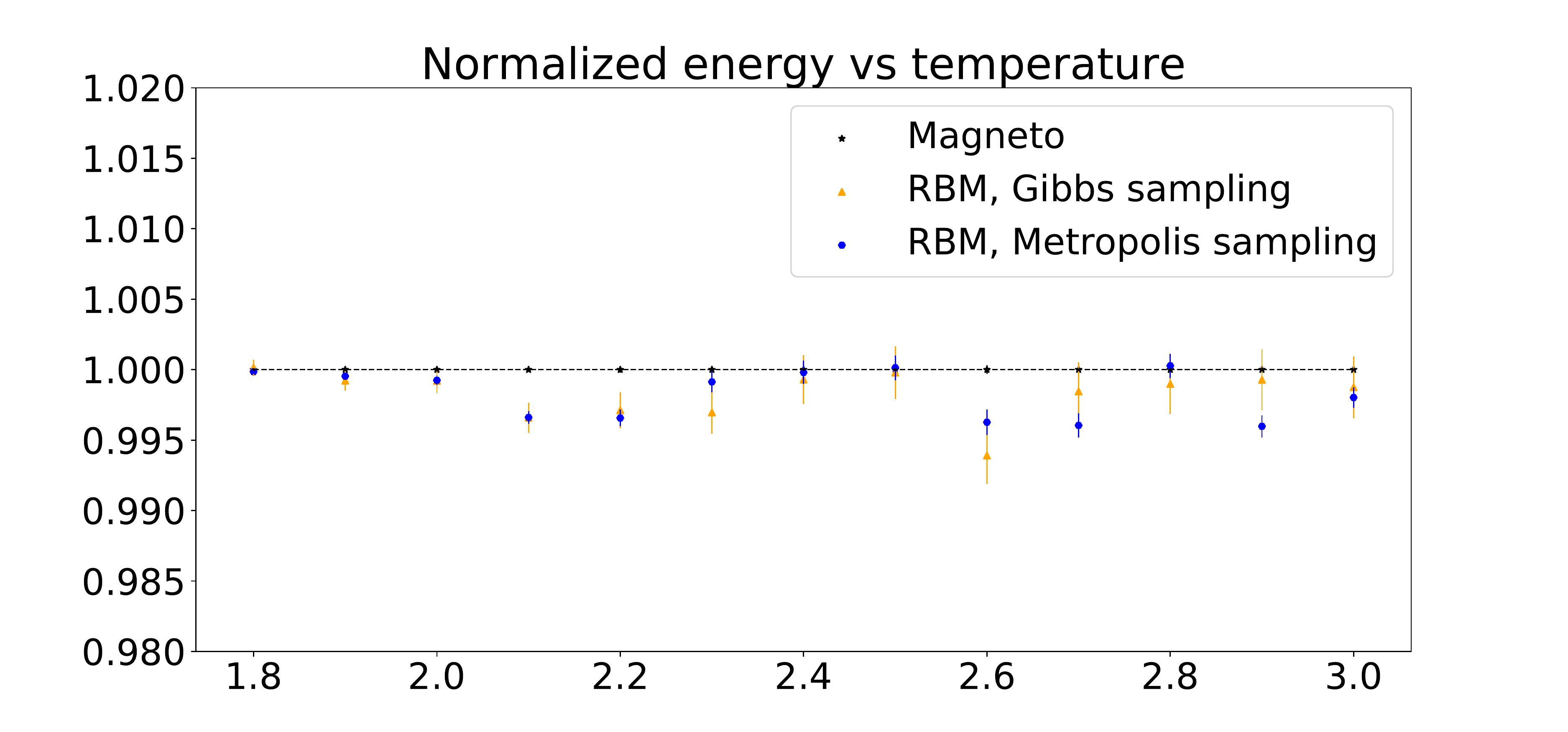}  
	\endminipage\hfill
	\minipage{0.50\textwidth}
	\includegraphics[width=\linewidth]{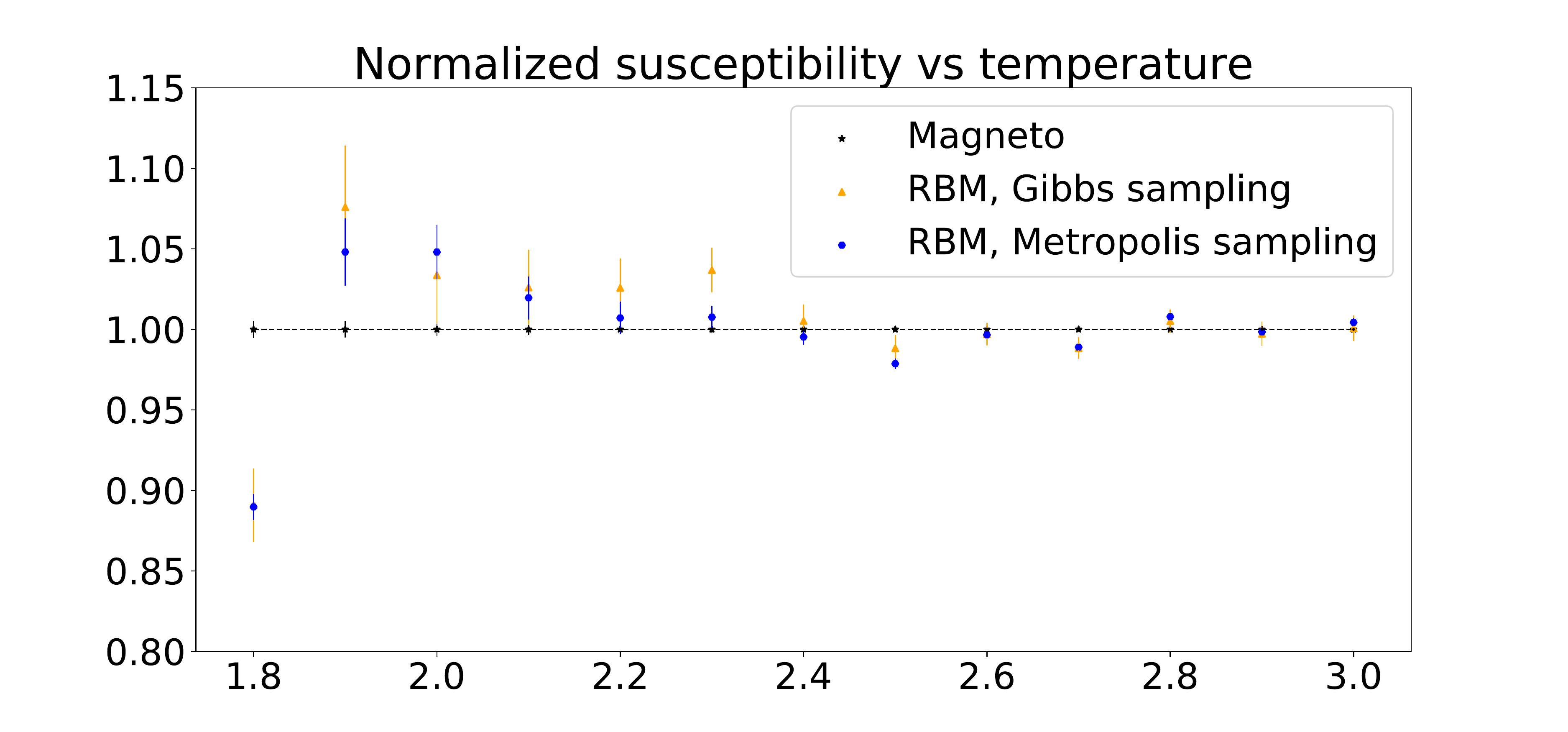}  
	\endminipage\hfill
	\minipage{0.50\textwidth}
	\includegraphics[width=\linewidth]{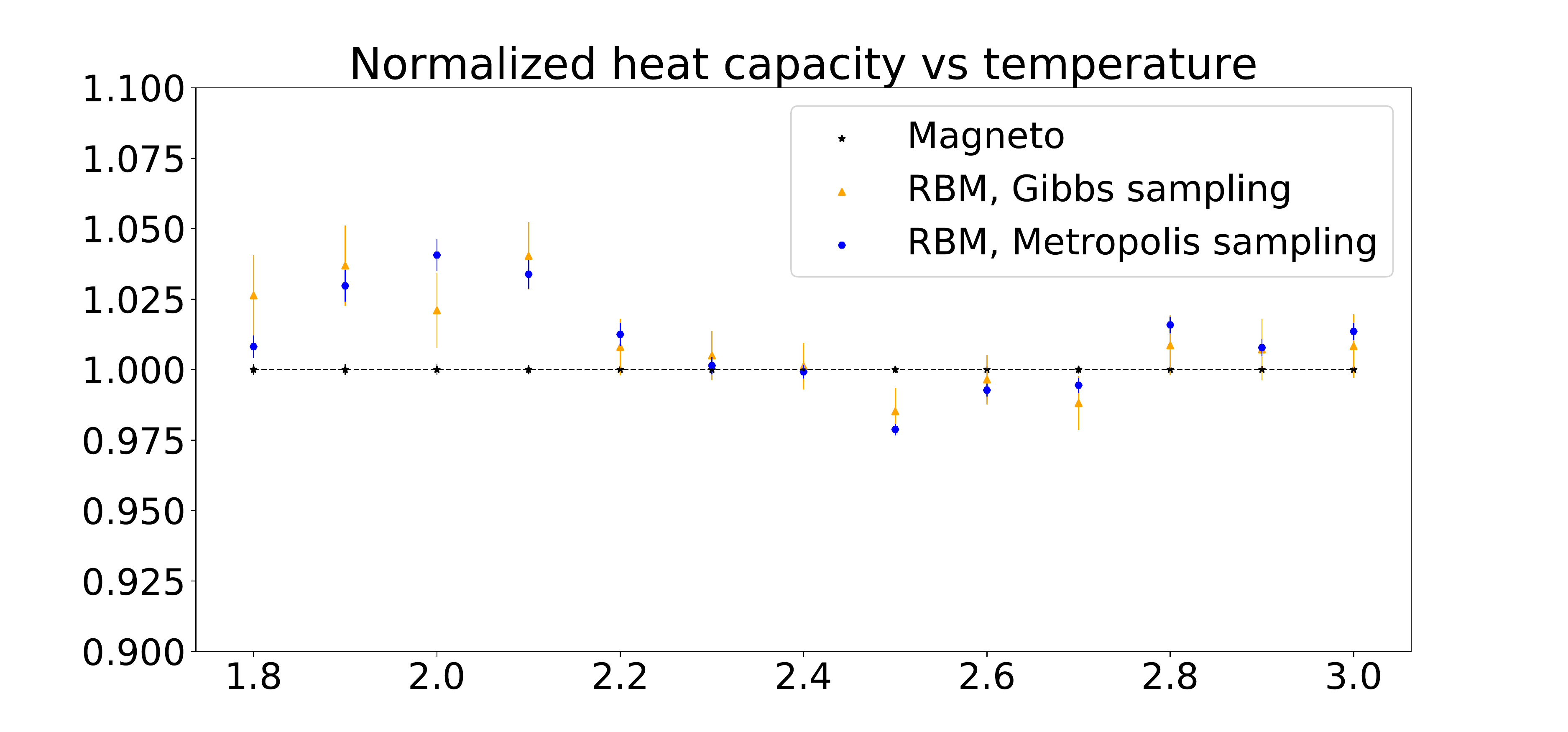}  
	\endminipage\hfill
	\caption{Observables, normalised by the expected magneto value, vs temperature. }
	\label{fig:allobservablesNt}
\end{figure}

In the following we report on the results of Ising observable measurements, as extracted from a trained RBM, for 13 different values of temperature. In Fig.~\ref{fig:allobservablest} the values for the different observables are shown, obtained independently from Gibbs and Metropolis sampling. The expected values from magneto are also plotted. We can observe resulting curves reproduce the expected values for magnetisation, energy, susceptibility and heat capacity.
However in these plots it is not possible to discern between the results obtained by different sampling methods and expected magneto values. To indicate more clearly the precision with which the RBM is able to reproduce the moments, we plot the observables normalised over their expected values, with their  corresponding error bars in Fig.~(\ref{fig:allobservablesNt}). Notice that, as expected, the first moments \ie magnetisation and energy, agree better with the corresponding magneto results as compared to the second moments. Moreover, the measurements using the two different sampling methods always agree within 2$\sigma$. For the first moments, the values measured from the RBM distribution are almost always compatible with the expected values within 2$\sigma$, while in the case of the second moments some discrepancy is observed, \eg see susceptibility at $T=1.8$.
We conclude that the RBM is in general able to obtain precise values for the first moments of the underlying distribution at all temperature studied, however, for the second moments the agreement is not always as precise.


\section{Extracting couplings from the RBM}
In the previous sections we described how to train an RBM for a 2D Ising model and assess the quality of the training by comparing the derived first and second moments, from the RBM, to the corresponding observables measured using the target distribution. This is a strong check that the machine has trained correctly, as first and second moments can always be measured from the training data directly. The predictive power of the machine, however, is in capturing the interactions between visible units, at every temperature. In this section we address how to extract the couplings between the visible units from the trained RBM, in the case of a binary system \eg the Ising model. Starting from an observation made in Chapter~16 of Ref.~\cite{Mehta2018AHL}, we derive a closed form expression for the two-point coupling between the visible units, as a function of the parameters of the trained RBM, and compare the results with the two-points interactions expected from the Ising model. We then generalise the computation to higher order interactions, presenting a closed form expression for the tensor describing the $n$-point coupling between visible units learned by the machine, as a function of the learned RBM parameters. \\

\begin{figure}
\minipage{0.50\textwidth}
\includegraphics[width=\linewidth]{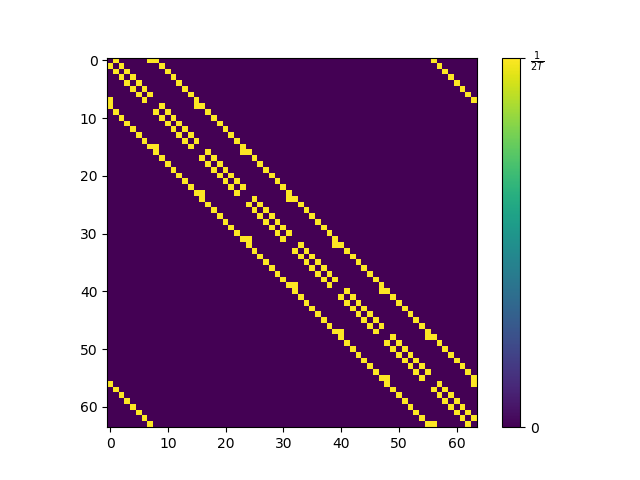}  
\endminipage\hfill
\minipage{0.50\textwidth}
\includegraphics[width=\linewidth]{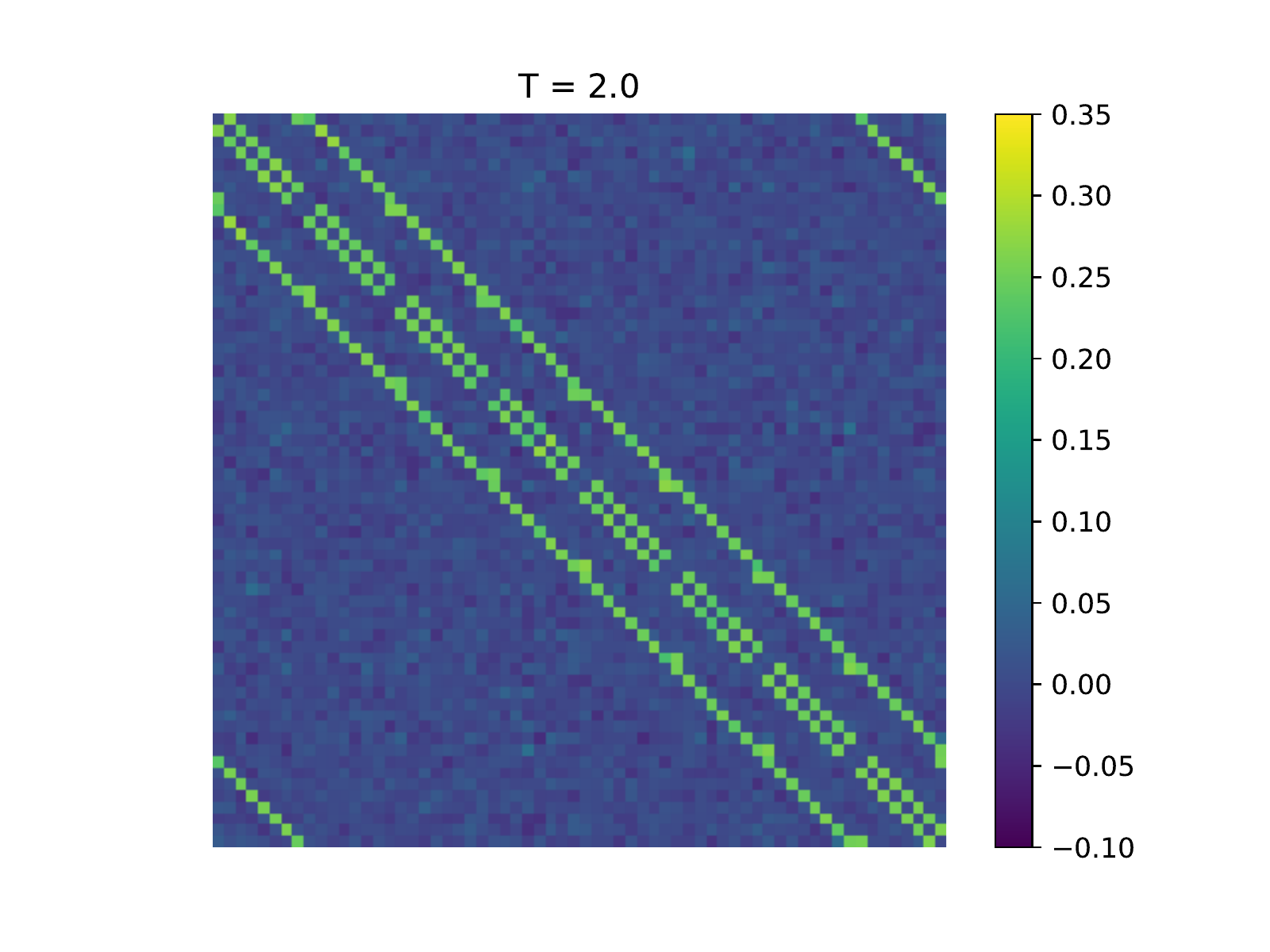}  
\endminipage\hfill
\caption{The matrix of interactions, $H_{j_1j_2}$, for an $8\times8$ Ising lattice with periodic boundary conditions as used in the generation of the training set (left) and the one learnt by the RBM at the end of the training (right). In this example T=2.0. The spins are labelled from 0 to 64 and the nearest neighbour structure is evident. }
\label{fig:hising}
\end{figure}

In order to extract the coupling from the weight matrix of the trained RBM, $W_{ij}$, we expand the generating function in powers of $v_{j_{1}}$. Marginalising over the hidden units, the energy of the RBM as a function of the visible units can be written as:
\begin{equation}\label{eq:E1}
\begin{split}
E(\vl) &= \ln \sum_{\hl} e^{\Evh} \\
&= \sum_{i}\ln \sum_{h_{i}} e^{- \sum_{j} b_{j} v_{j} - \sum_{i} c_{i} h_{i} - \sum_{i, j} h_{i} W_{i j} v_{j}}.
\end{split}
\end{equation}
Since the first term in $\Evh$ is independent of the hidden layer, it can be brought outside the sum. This is the \emph{visible bias} term. $E(\vl)$ then becomes,
\begin{equation}\label{eq:E2}
\begin{split}
E(\vl) &= - \sum_{j} b_{j}v_{j} - \sum_{i} \ln \sum_{h_{i}} e^{c_{i} h_{i}} e^{ \sum_{j} h_{i} W_{i j}  v_{j}} \\
& = - \sum_{j} b_{j}v_{j} - \sum_{i} \ln \sum_{h_{i}} q(h_{i}) e^{th_{i}} \ ,
\end{split}
\end{equation}
where we have defined $t \equiv \sum_{j} W_{i j} v_{j}$ and $q(h_{i}) \equiv e^{c_{i} h_{i}}$. This allows us to introduce the cumulant generating function, similar to what was done in Ref.~\cite{Mehta2018AHL}: 
\begin{equation}\label{eq:Kdef}
K_{i}(t) \equiv \ln \sum_{h_{i}} q(h_{i}) e^{th_{i}} = \sum_{n} \frac{\kappa_{i}^{(n)} t^{n}}{n!} \ ,
\end{equation}
where the $n$th cumulant $\kappa_{i}^{(n)} = \partial_{t}^{n} K_{i}(t)|_{t=0}$. Expanding the generating function as a power series in $n$ gives,
\begin{equation}\label{eq:E3}
\begin{split}
E(\vl) &= - \sum_{j} b_{j}v_{j} - \sum_{i}\kappa_{i}^{(0)} - \sum_{i}\kappa_{i}^{(1)} t -  \sum_{i}\frac{\kappa_{i}^{(2)} t^{2}}{2!} - \ldots \\
&= - \sum_{i}\kappa_{i}^{(0)} - \sum_{j} \left( b_{j} + \sum_{i} \kappa_{i}^{(1)} W_{i j} \right) v_{j} - \frac{1}{2!}\sum_{j_{1}, j_{2}} \left( \sum_{i} \kappa_{i}^{(2)} W_{i j_{1}} W_{i j_{2}} \right)v_{j_{1}} v_{j_{2}} - \ldots \ ,
\end{split}
\end{equation}
where in the final line we have used the definition of $t$ to rewrite $E(\vl)$ in terms of $n$-point interactions between the visible units. When the RBM has learned the physics of the Ising model, we would expect that $E(\vl) = H_{\mathrm{ising}}(\vl)$ up to a constant. Since the two energies can differ by an overall constant without affecting the physical observables, we are not interested in $- \kappa_{i}^{(0)}$. However, the two energies are equal across states and we would expect that $E(\vl)$ and $H_{\mathrm{ising}}(\vl)$ to be equal, order-by-order in $\vl$. For the standard Ising interactions, with no external field,
\begin{equation}
H_{\mathrm{Ising}}(\tilde{\vl}) = \sum_{j_{1},j_{2}} H_{j_{1}j_{2}} \tilde{v}_{j_{1}} \tilde{v}_{j_{2}}\ ,
\end{equation}
where $\tilde{\vl}$ describes the spin states, with each spin taking the value $+1$ or $-1$. $H_{j_{1}j_{2}}$ is the appropriate Ising matrix with nearest neighbour interactions. In this Ising model, the nearest neighbour coupling is set to $J = -1/T$ where $T$ is the temperature of the model. Therefore, $H_{j_{1}j_{2}}$ is zero except for components which correspond to nearest neighbour interactions, which are equal to $\frac{1}{2T}$. This is shown pictorially in Fig.~\ref{fig:hising} for the case of an $8\times8$ lattice. The $x$ and $y$ axes represent spins on this two-dimensional lattice, numbered from 0 to 64. Notice the structure of couplings due to periodic boundary conditions.
\\

States in the \plusmb basis can be related to those in the \zob basis by the simple transformation $\tilde{\vl} = 2\vl - 1$. This allows us to rewrite the Ising Hamiltonian in the \zob basis as:
\begin{equation}\label{eq:hzob}
H_{\mathrm{ising}}(\vl) = 4\sum_{j_{1}, j_{2}} H_{j_{1}j_{2}} v_{j_{1}} v_{j_{2}} - 4 \sum_{j_{1}} \left( \sum_{j_{2}} H_{j_{1}j_{2}} \right) v_{j_{1}} + \left( \sum_{j_{1}, j_{2}}H_{j_{1}j_{2}} \right) ,
\end{equation}
where the symmetric form of $H_{j_{1}j_{2}}$ has been exploited to simplify the expression. The 2-point interaction in Eq.~\ref{eq:hzob} and Eq.~\ref{eq:E3} can now be compared to extract the coupling, $\frac{1}{2T}$, from the trained RBM.

There is however a subtlety which arises in the expansion of the cumulant generating function. We can see that in \zob basis:
\begin{equation}\label{eq:vrel}
v_{j}^n = v_{j} \ \ , \ \ n \in \mathbb{Z}^+ \ \cdot
\end{equation}
This implies that higher order terms in $n$ also contribute to the $N_v$-point function, where $N_v$ is the number of {\it distinct} vertices. To make this statement more clear, let us write Eq.~\ref{eq:E3} in a more general form,
\begin{equation}
\label{eq:Eexpansion}
E(\vl) = - \kappa_{i}^{(0)} - \sum_{j} \left( b_{j} + \sum_{i} \kappa_{i}^{(1)} W_{i j} \right) v_{j} - \sum_{n > 1}\frac{1}{n!} \sum_{j_{1} \ldots j_{n}} \left( \sum_{i} \kappa_{i}^{(n)} W_{i j_{1}} \ldots W_{i j_{n}} \right)v_{j_{1}} \ldots v_{j_{n}} \ \cdot
\end{equation}
In other words, $n$ can be thought of as counting the powers of $W_{ij}$. For each $n$, the contributions from the sum over the visible vertices, $\sum_{j_{1} \ldots j_{n}}$, can be grouped by the number $N_v$ of distinct vertices being multiplied, \ie the $N_v$-point functions. For example, using Eq.~\ref{eq:vrel}, we see that the numerical contributions to the 2-point interactions are terms proportional to $v_{j_{1}}^k v_{j_{2}}^{n-k} = v_{j_{1}} v_{j_{2}}$ where the vertices are distinct, \ie, $j_{1} \neq j_{2}$. The number of combinations of powers of $v_{j_{1}}$ and $v_{j_{2}}$ for a given $n$ are simply the binomial coefficients. Therefore, 2-point contributions from all orders in $n$ can be written as:
\begin{equation}\label{eq:resum1}
\sum_{n > 1} \frac{1}{2(n!)} \sum_{0<k<n} \sum_{j_{1} \neq j_{2}} \left( \sum_{i} \kappa_{i}^{(n)} \binom{n}{k} W_{i j_{1}}^{k} W_{i j_{2}}^{n-k} \right)v_{j_{1}} v_{j_{2}} \ ,
\end{equation}
where the factor of two has been included to account for the double counting arising from the symmetry under the exchange of $j_{1}$ and $j_{2}$. Performing the sum over $k$ gives,
\begin{equation}\label{eq:resum2}
\frac{1}{2} \sum_{n > 1}\frac{1}{n!} \sum_{j_{1}\neq  j_{2}} \left( \sum_{i} \kappa_{i}^{(n)} \left[ \left( W_{i j_{1}} + W_{i j_{2}} \right)^{n} - (W_{i j_{1}})^{n} - (W_{i j_{2}})^{n} \right] \right) v_{j_{1}} v_{j_{2}} \ \cdot
\end{equation}
Comparing this result to the Ising 2-point interaction in Eq.~\ref{eq:hzob}, provided that the RBM is properly trained, the Ising pair-wise coupling can be written as:
\begin{equation}
H_{j_{1}j_{2}} = \frac{1}{8} \sum_{n > 1}\frac{1}{n!} \sum_{i} \kappa_{i}^{(n)} \left[ \left( W_{i j_{1}} + W_{i j_{2}} \right)^{n} - (W_{i j_{1}})^{n} - (W_{i j_{2}})^{n} \right]\cdot
\end{equation}
On the other hand, $\kappa_{i}^{(n)} = \partial_{t}^{n}K_{i}(t)|_{t = 0}$, giving
\begin{equation}
\begin{split}
H_{j_{1}j_{2}} =& \frac{1}{8}\sum_{n} \left( \frac{1}{n!} \sum_{i} \left[ \left( W_{i j_{1}} + W_{i j_{2}} \right)^{n} - (W_{i j_{1}})^{n} - (W_{i j_{2}})^{n} \right]\partial_{t}^{n}K_{i}(t)|_{t=0} \right) + \frac{1}{8}\sum_{i} K_{i}(0) \\
=& \frac{1}{8}\sum_{i} \left( e^{\left( W_{i j_{1}} + W_{i j_{2}} \right) \partial_{t}} - e^{W_{i j_{1}} \partial_{t}} - e^{ W_{i j_{2}} \partial_{t}}  + 1\right)K_{i}(t)|_{t=0} \ \cdot
\end{split}
\end{equation}
We recognise the shift operators $e^{a \partial_{x}}f(x) = f(a+ x)$ in the second line of the above equation. Therefore, the expression can be simplified further as: 
\begin{equation}\label{eq:closedform}
\begin{split}
H_{j_{1}j_{2}} =& \frac{1}{8}\sum_{i} \Big( K_{i}(W_{i j_{1}} + W_{i j_{2}}) - K_{i}(W_{i j_{1}}) - K_{i}( W_{i j_{2}}) + K_{i}(0) \Big) ,
\end{split}
\end{equation}
which is the closed form expression for $H_{j_{1}j_{2}}$, including all order contributions in $n$. More explicitly, using Eq.(\ref{eq:Kdef}) we have 
\begin{equation}
K_{i}(t) = \log\left(1 + e^{c_i + t}\right)
\end{equation}
so that
\begin{equation}\label{eq:exp2}
H_{j_{1}j_{2}} = \frac{1}{8} \sum_{i} \ln \frac{(1 + e^{c_{i} + W_{i j_{1}} + W_{i j_{2}}})(1 + e^{c_{i}})}{(1 + e^{c_{i} + W_{i j_{1}}})(1 + e^{c_{i} + W_{i j_{2}}})}
\end{equation}

\begin{figure}
\centering
\includegraphics[width=\textwidth]{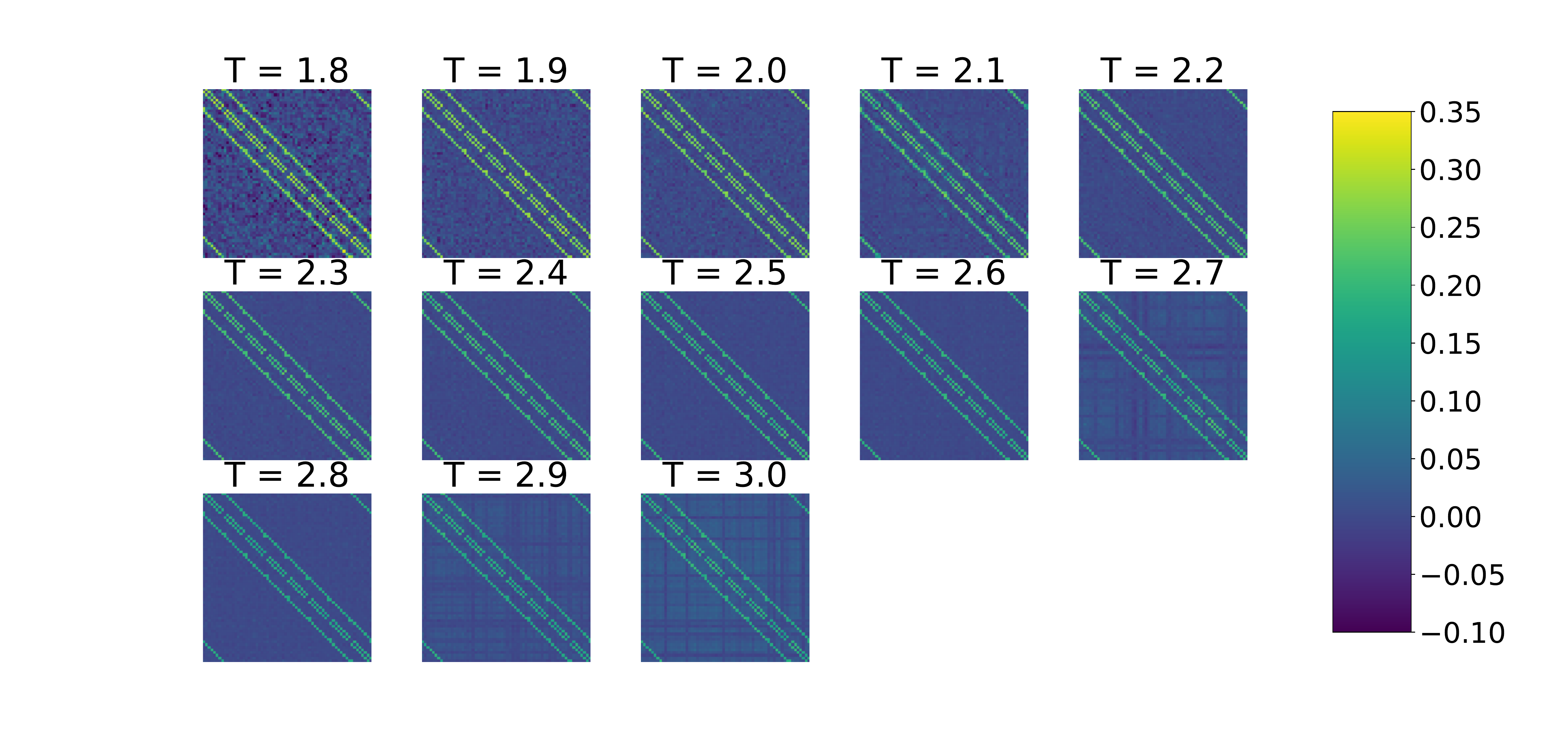}
\caption{The interaction matrix $H_{j_{1}j_{2}}$ extracted from RBMs, with $L^2=8\times8$ and $h^2=8\times8$, trained at a temperature indicated above each subplot. Again, the spins are labelled from 0 to 64 and show the same structure as for the generic Ising training set in Fig.~\ref{fig:hising}.}
\label{fig:rbmweights}
\end{figure}

\begin{figure}
\centering
\includegraphics[width=\textwidth]{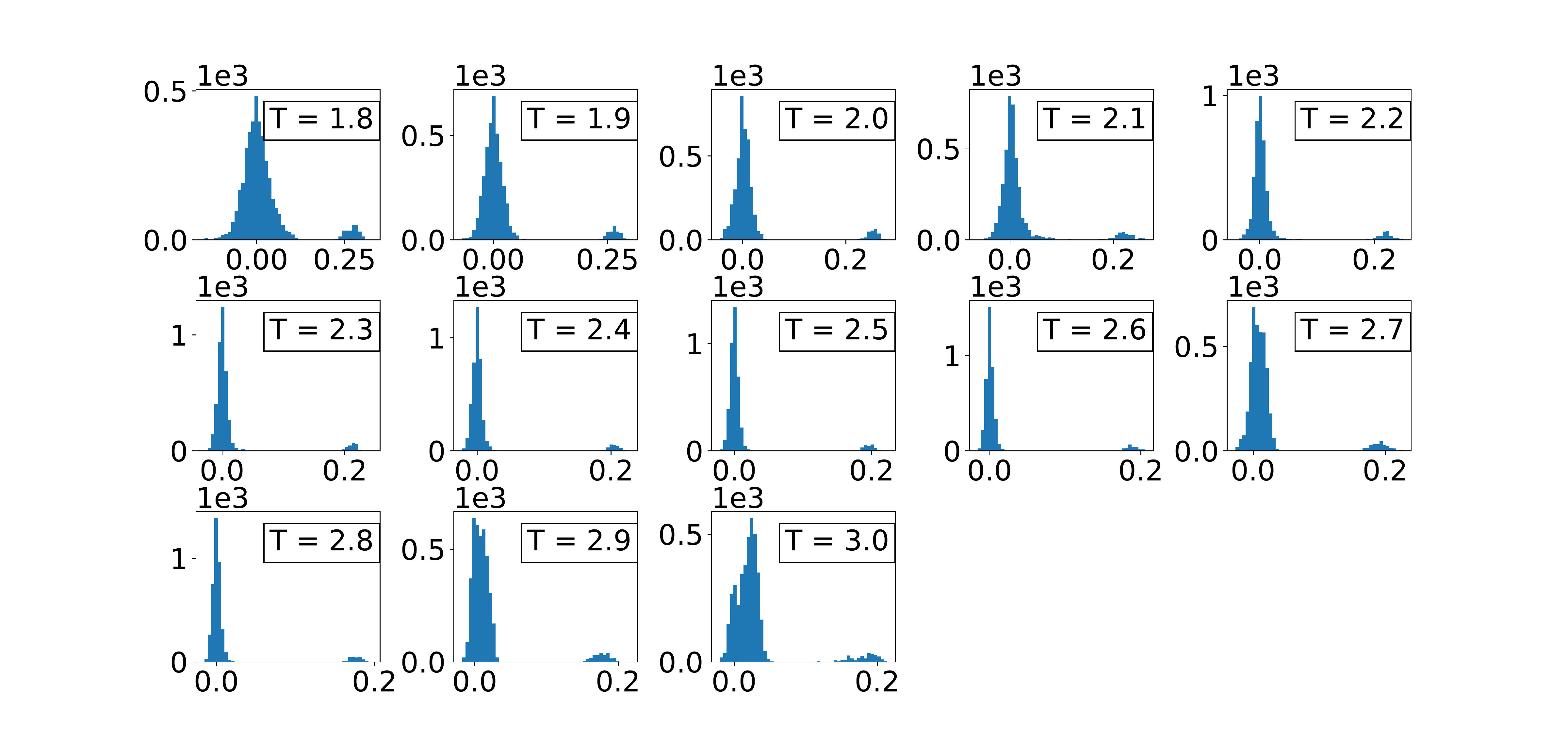}
\caption{Histograms of the entries of $H_{j_{1}j_{2}}$ extracted from RBMs trained at a temperature indicated above each subplot. As it can be observed, there are always two peaks: The smaller peak, represents the number of nearest neighbour on the $y$-axis, with the value of the coupling indicated on the $x$-axis; the larger peak represent all other sites that are not nearest neighbours and are not expected to couple to each other, hence it being centred around zero on the $x$-axis.  }
\label{fig:rbm2phist}
\end{figure}

\begin{figure}
\centering
\includegraphics[width=\textwidth]{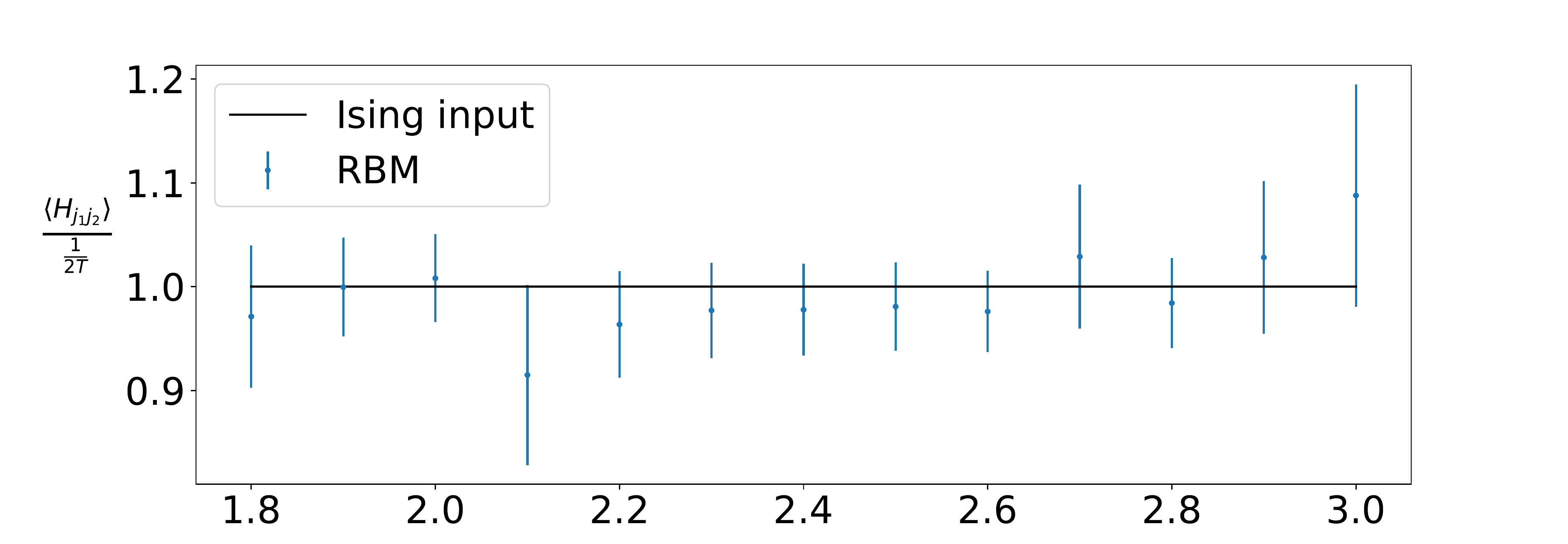}
\caption{The predicted 2-point interaction coupling, for $L^2=8\times8$ at different values of temperature, normalised by the corresponding true value $1/2T$. The coupling is extracted from the nearest neighbour diagonals observed in the interaction matrices, with its error bar computed by taking the standard deviation of the diagonal components. The predicted values agree with the expected ones within statistics.}
\label{fig:2pcoupling}
\end{figure}

Using the trained RBMs, the coupling given by Eq.~\ref{eq:closedform} was evaluated for temperatures ranging from $T=1.8$ to $T=3.0$. The results, presented in Fig.~\ref{fig:rbmweights}, have the same nearest neighbour structure as the Ising model used to train the machines, in Fig.~\ref{fig:hising} with generic temperature $T$.
\begin{figure}
	\centering
	\includegraphics[width=\textwidth]{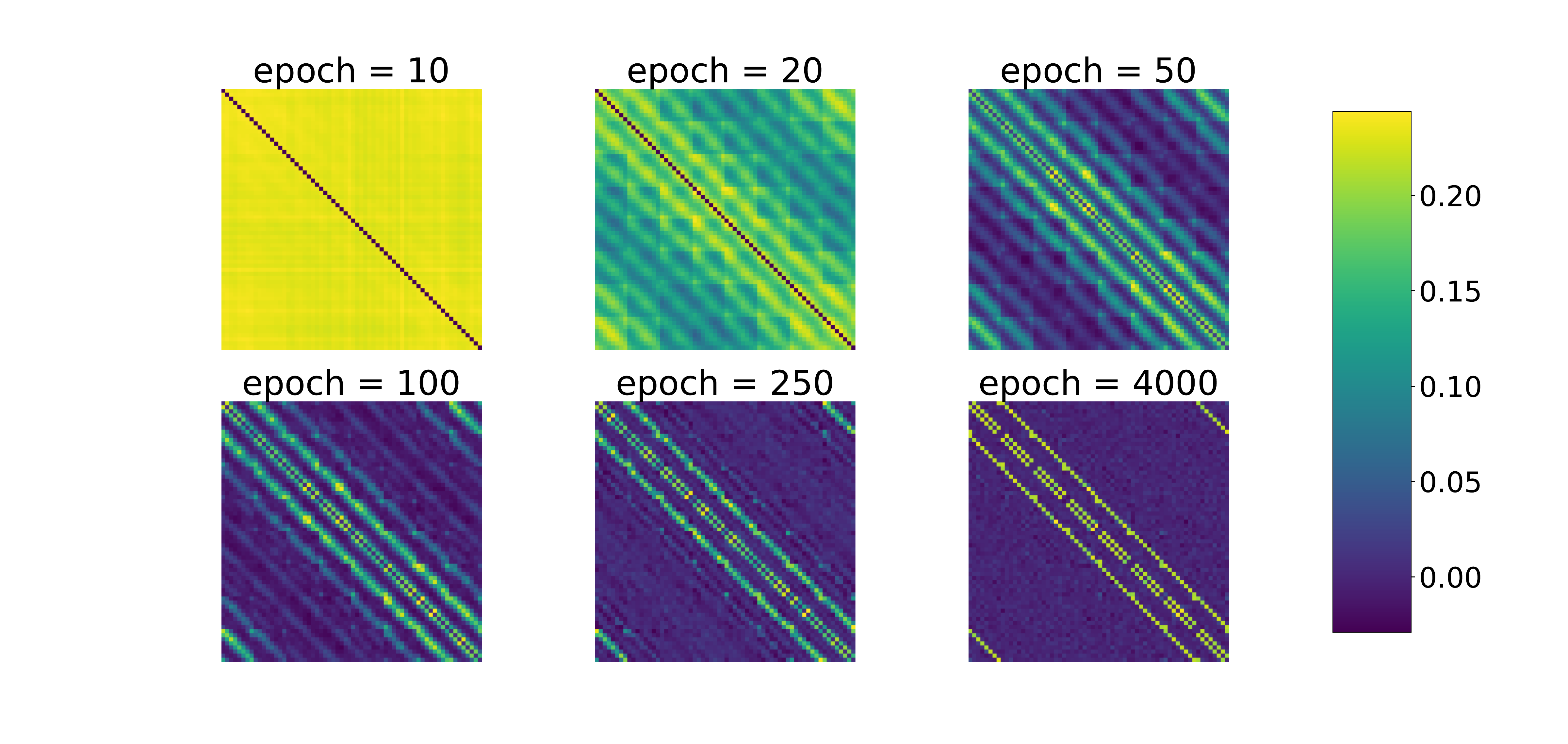}
	\caption{$H_{j_{1}j_{2}}$ extracted from RBMs at different stages of the training (10, 20, 50, 100, 250 and 4000 epochs), for $L^2=8\times8$, $h^2=8\times8$ and temperature $T=2.2$. As the machine approaches the end of the training, the expected structure of Fig.\ref{fig:hising} becomes more and more evident.}
	\label{fig:rbmweightsprogression}
\end{figure}
In Fig.~\ref{fig:rbm2phist}, we have plotted the corresponding histograms of the values $H_{j_1j_2}$ at various temperatures. The expected bimodal structure of coupling matrices is observed, 
with most of the entries, \ie those associated with the non-nearest neighbour spins interactions, being centred around zero, while the nearest neighbour interactions introduce a distinct second peak, around the value of the coupling. 
By taking the average and the standard deviation across the nearest neighbour diagonals we can compare the couplings extracted from the RBM against the exact values from the Ising model, as shown in Fig.~\ref{fig:2pcoupling}. We see that the RBM predictions agrees with the analytical results within statistics. \\

It is also interesting to look at the change in the couplings along the training: this is shown in Fig.~\ref{fig:rbmweightsprogression}, where the 2-points interaction matrix for $T=2.2$ is plotted at different stages of the training. The training algorithm starts by initialising the RBM visible-hidden interaction terms $W_{ij}$ from a normal distribution centred around zero. The bias terms are also initialised to zero. In the early epochs, we observe no particular structure in the $H_{j_1,j_2}$ interaction matrix. As learning progresses, more of the nearest neighbour structure appears, together with further non nearest neighbour correlations. Towards the final phases of the training, only the nearest neighbour structure remains, with the rest of the interactions being almost zero. \\

\begin{figure}
	\centering
	\includegraphics[width=\textwidth]{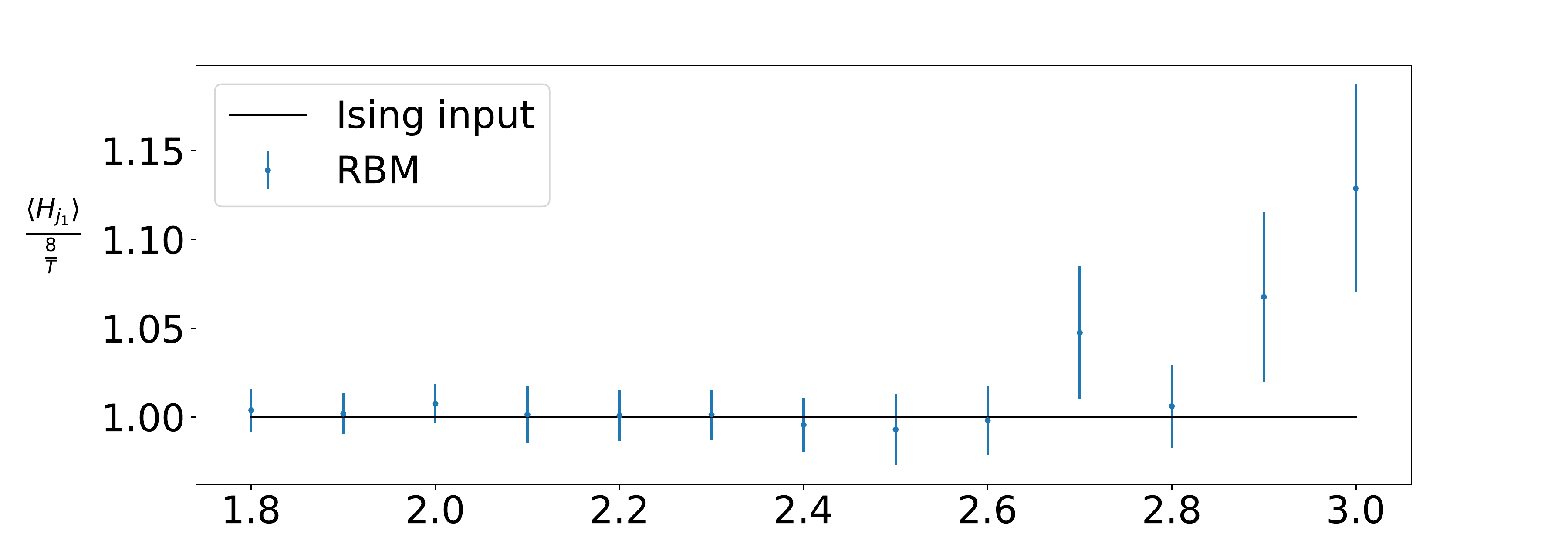}
	\caption{The linear terms extracted by the trained machine, normalised by the corresponding true value $8/T$. All predicted values are compatible with the true ones within $2\sigma$.}
	\label{fig:linearterm}
\end{figure}

Let us now return to Eq.~\ref{eq:Eexpansion} and collect terms  linear in $v_j$. The terms contributing are those with $j_1 = j_2 = \ldots = j_n $, giving,
\begin{equation}\label{eq:coup-lin-term}
\sum_j J_j v_j = - \sum_{j} \left( b_{j} + \sum_{i} \kappa_{i}^{(1)} W_{i j} \right) v_{j} - \sum_{n > 1}\frac{1}{n!} \sum_{j} \left( \sum_{i} \kappa_{i}^{(n)} \left(W_{i j}\right)^n  \right)v_{j} .
\end{equation}
The vector $J_i$ can be read from the linear term in Eq.~\ref{eq:hzob},
\begin{equation}
J_i = 4\sum_j H_{ij} = \frac{8}{T} \ ,
\end{equation}
where we have used $\sum_j H_{ij} = 2/T$, for every component $i$ in the Ising 2-point interaction matrix $H_{ij}$.
The second term on the right hand side of the Eq.~\ref{eq:coup-lin-term} can be treated in the same way as the 2-point interaction described above above,\ie ,
\begin{equation}
\sum_{n > 1}\frac{1}{n!} \sum_{j} \left( \sum_{i} \kappa_{i}^{(n)} \left(W_{i j}\right)^n  \right)v_{j} = \sum_j \sum_i\left[K_i\left(W_{ij}\right) - K_i\left(0\right)\right]v_j - \sum_j \sum_i\kappa_{i}^{(1)} W_{i j}v_j \ \cdot
\end{equation}
It then follows that,
\begin{equation}
\label{eq:linearterm}
J_j = -\left(b_j + \sum_i K_i\left(W_{ij}\right) - \sum_i K_i\left(0\right)\right) =-\left(b_j + \sum_i \log\left(\frac{1+e^{c_i+ W_{ij}}}{1+e^{c_i}}\right) \right) .
\end{equation}
This allows us to extract the value of an external magnetic field, if present, from the parameters of the trained RBM. The original training set was generated without any external magnetic fields, therefore, if working in the \plusmb basis, we do not expect to observe any linear term from the trained machine. However, we train the machine in the \zob basis, which implies the existence of a linear term $\vec{J}$ such that $J_i = 8/T$ for every component $i$. To extract this value from the trained machine we have taken the average across all the entries of the vector $J_j$ defined in Eq.~\ref{eq:linearterm}. The linear term extracted from the trained machine are plotted in Fig.~\ref{fig:linearterm}, where the error bar is obtained by computing the standard deviation of the entries $J_j$. Again, we see how these values are compatible with the expected ones within $2\sigma$.\\

The 3- and 4-point interactions can also be calculated and plotted. This is mainly used as a sanity check to confirm that the RBM has not incorrectly learned unexpected additional interactions. For the 3 point-interactions, we want to include all combinations which leave three distinct indices, similar to Eq.~\ref{eq:resum1} we get,

\begin{equation}\label{eq:resum3}
\sum_{n > 2} \frac{1}{6(n!)} \sum_{0<l<(n-k)} \sum_{0<k<(n-1)} \sum_{j_{1} \neq j_{2} \neq j_{3}} \left( \sum_{i} \kappa_{i}^{(n)} \binom{n}{k} \binom{n-k}{l} W_{i j_{1}}^{k} W_{i j_{2}}^{l} W_{i j_{3}}^{n-k-l} \right)v_{j_{1}} v_{j_{2}} v_{j_{3}} \ \cdot
\end{equation}

Summing over $k$ and $l$ gives

\begin{equation}
\begin{split}
\sum_{n > 2}^{\infty} \frac{1}{6(n!)} \sum_{j_{1} \neq j_{2} \neq j_{3}} \Big( & \sum_{i} \kappa_{i}^{(n)} \big[  (W_{i j_{1}} + W_{i j_{2}} + W_{i j_{3}})^{n}  - 
(W_{i j_{1}} + W_{i j_{2}})^{n} -
(W_{i j_{1}} + W_{i j_{3}})^{n} -
(W_{i j_{2}} + W_{i j_{3}})^{n} \\
& + (W_{i j_{1}})^{n} + (W_{i j_{2}})^{n} + (W_{i j_{3}})^{n} \big] \Big) v_{j_{1}} v_{j_{2}} v_{j_{3}} \ \cdot
\end{split}
\end{equation}
If we add an subtract terms corresponding to $n = 0, 1, 2$ then the sum over $n$ can be evaluated as before, giving
\begin{equation}
\begin{split}
\frac{1}{6}  \sum_{j_{1} \neq j_{2} \neq j_{3}} \sum_{i} \Big( & K_{i}(W_{i j_{1}} + W_{i j_{2}} + W_{i j_{3}}) - K_{i}(W_{i j_{1}} + W_{i j_{2}}) - K_{i}(W_{i j_{1}} + W_{i j_{3}}) - K_{i}(W_{i j_{2}} + W_{i j_{3}}) \\
& + K_{i}(W_{i j_{1}}) +  K_{i}(W_{i j_{2}}) + K_{i}(W_{i j_{3}}) + K_{i}(0) \Big)v_{j_{1}} v_{j_{2}} v_{j_{3}} \ \cdot
\end{split}
\end{equation}
We can then write the interaction tensor explicitly as
\begin{equation}\label{eq:exp3}
\frac{1}{6} \sum_{i} \ln \frac{(1 + e^{c_{i} + W_{i j_{1}} + W_{i j_{2}} + W_{i j_{3}}})(1 + e^{c_{i} + W_{i j_{1}}})(1 + e^{c_{i} + W_{i j_{2}}})(1 + e^{c_{i} + W_{i j_{3}}})(1 + e^{c_{i}})}{(1 + e^{c_{i} + W_{i j_{1}} + W_{i j_{2}} + })(1 + e^{c_{i} + W_{i j_{1}} + W_{i j_{3}}})(1 + e^{c_{i} + W_{i j_{2}} + W_{i j_{3}}})} \ \cdot
\end{equation}
One can observed the pattern emerging from \eqref{eq:exp2} and \eqref{eq:exp3} and write down a general form for the $N$-point interaction tensor
\begin{equation}\label{eq:cfall}
\frac{1}{N!}\sum_{l=0}^{N} (-1)^{l} \sum_{\alpha_{1} < \ldots < \alpha_{N-l}} K_{i}(W_{i, j_{\alpha_{1}}} + \ldots + W_{i, j_{\alpha_{N-l}}})
\end{equation}
where $\{ \alpha_{i} \} = [1..N]$ and the product simply multiplies the $\binom{N}{N-l}$ combinations of choosing $N-l$ unique indices from $\{ \alpha_{i} \}$. Using Eq.~\ref{eq:cfall}, histograms for the tensors for the 3- and 4-point interaction can then be plotted for the trained RBMs, to check if any higher order interactions were learned. It is expected that the histograms have a single peak at zero. These histograms are presented in Fig.~\ref{fig:3hist} and Fig.~\ref{fig:4hist} of App.~\ref{app:3-4-point-interaction}, respectively.

\section{Conclusions}\label{sec:conclusions}
We have trained several RBMs on 1- and 2-dimensional Ising models at various values of temperature. The training procedure in each case has been discussed in detail. We have used five different criteria to test whether the learning process has been successful. The first four are measurements of the loss function, reconstruction error, free energy and log-likelihood throughout the training procedure. The measurement of the latter, which is the quantity that is being maximised in the algorithm, involves the estimation of the partition function of the model, which is computed using annealed importance sampling. The fifth and last criterion, is the measurements of the first and second moments of the distribution, as given by the RBM, and comparing them to those obtained directly from the training set. These five criteria are essential to the RBM being trained correctly. Hence, we have provided a generic prescription in training an RBM on a binary model, from our experience. Moreover, we have used the RBM to predict the interaction couplings between the Ising spins by re-summing the cumulant generating function. The re-summations can be performed for any model with binary $\{0,1\}$ states. The ability to extract the couplings exactly can be useful in studying the relation between the RBM and the renormalization group (RG), as first noted by Ref.~\cite{DBLP:journals/corr/MehtaS14}. In this paper, we have also demonstrated the difficulty in training an RBM with smaller number of hidden nodes as compared to the visible. The numerical verification of RG is therefore left for future work. Another application of predicting the couplings is for the case where the interactions between the visible nodes of the RBM are unknown. In general, there couplings are more complicated than simple pair-wise interactions and as such are not extractible from the data directly with conventional methods. The RBM has no such assumption, as there are indirect all-to-all connections between the visible nodes, via the hidden layer. As the machine learns from the data, some of these connections turn-off while others increase in strength. This makes the RBMs a powerful tool for predicting complex connections between the visible nodes and their relative strength.

\acknowledgements
AK is most grateful to Chris Ponting, both for his academic and financial support of this project via the MRC Programme, grant number MC\_UU\_00007/15. GC acknowledges funding by Intel and an STFC IAA award, and is supported by STFC, grant ST/L000458/1 and ST/P002447/1. MW is supported by STFC, grant ST/R504737/1. TG is supported by The Scottish Funding Council, grant H14027. LDD is supported by an STFC Consolidated Grant, ST/P0000630/1 and a Royal Society Wolfson Research Merit Award, WM140078.

\clearpage

\appendix

\section{The training procedure in more detail: $L^2=8\times8$}
\label{sec:num-tests}
\begin{figure}[!htb]
	\minipage{0.50\textwidth}
	\includegraphics[width=\linewidth]{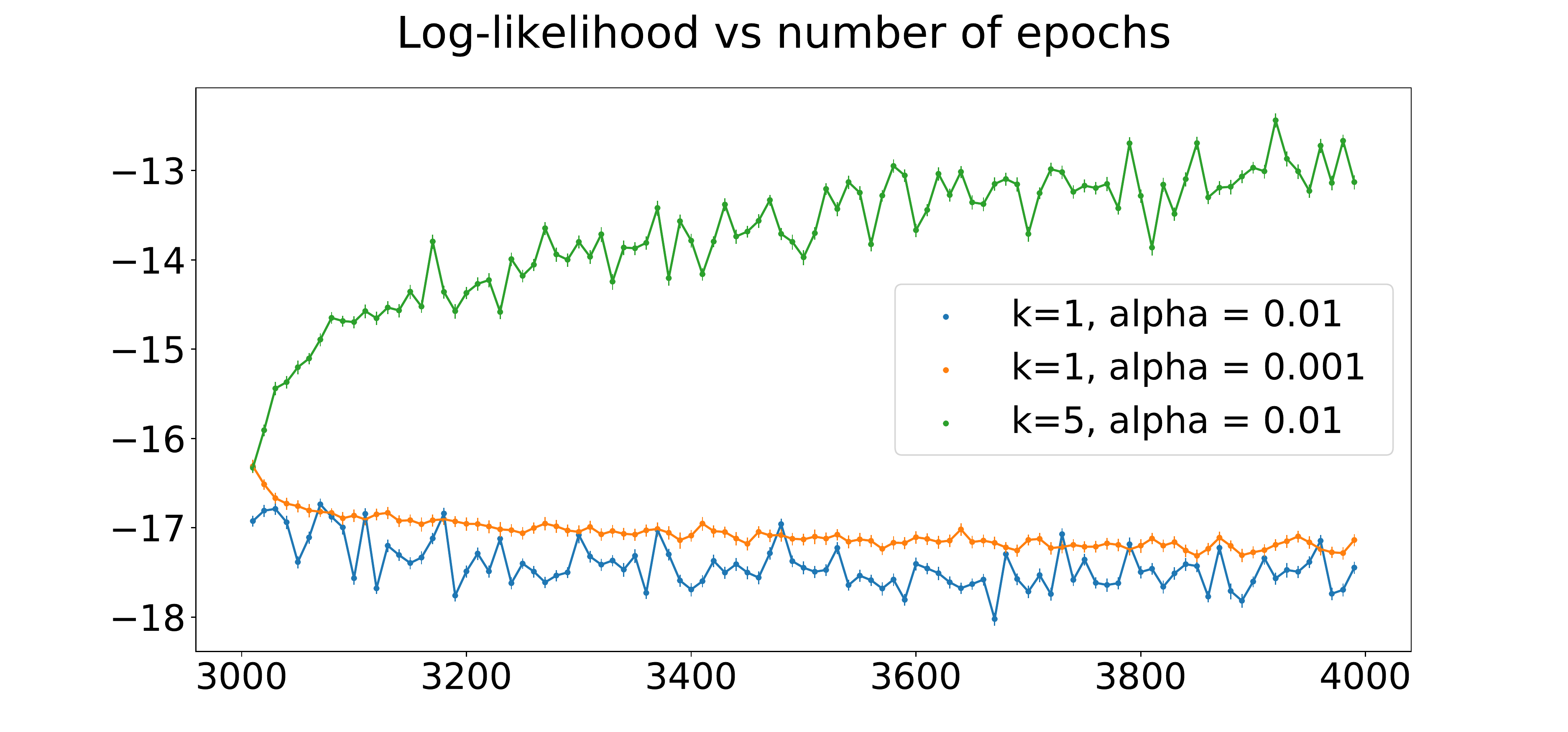}  
	\endminipage\hfill
	\minipage{0.50\textwidth}
	\includegraphics[width=\linewidth]{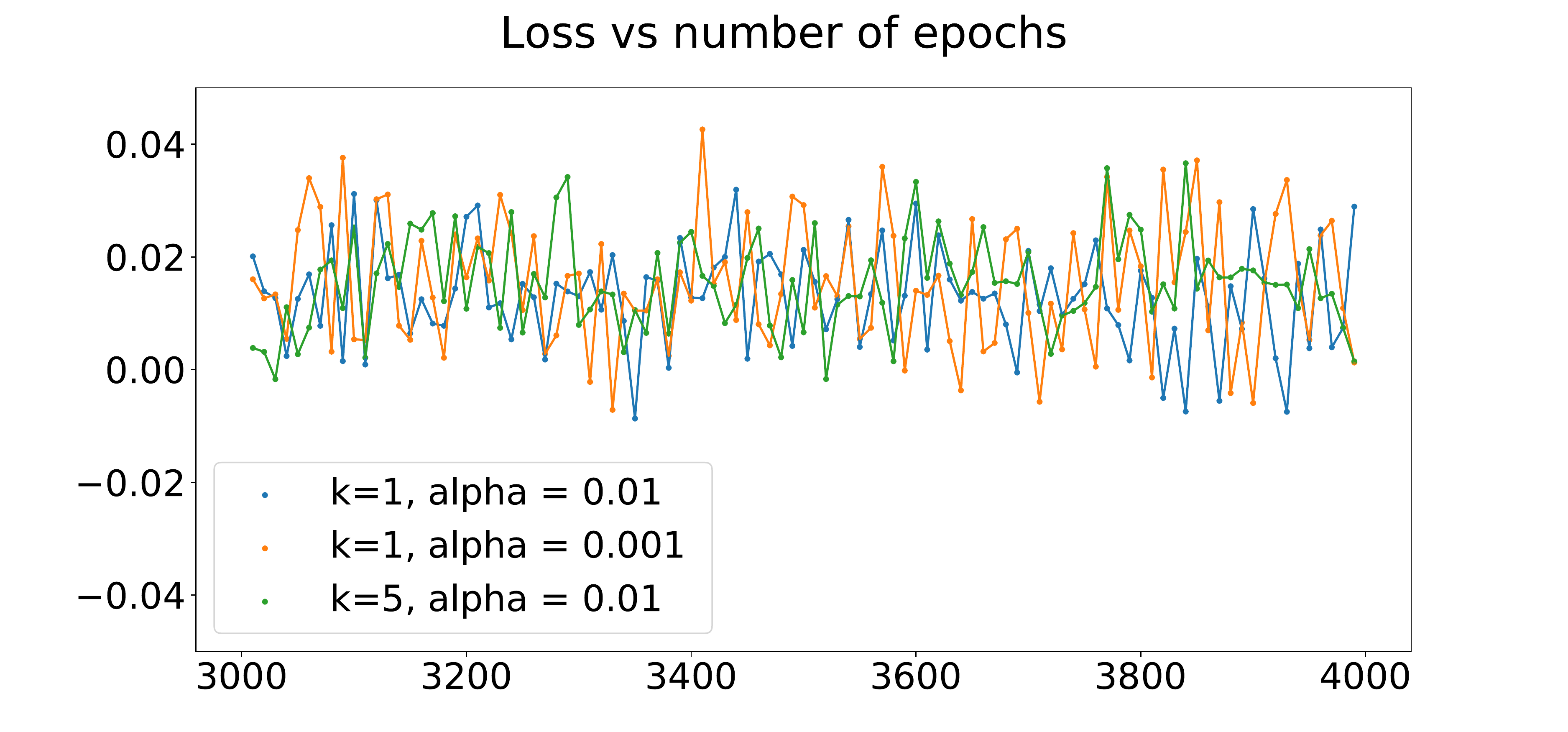}  
	\endminipage\hfill
	\caption{From left to right, log-likelihood and loss function between 3000 and 4000 epochs for three different values of k. While the former shows different behaviors, keeping an increasing trend just for the highest k value, the latter doesn't change at all, always remaining near zero.}
	\label{fig:ll_diff_k}
\end{figure}

Here we report further details regarding the training of the RBM on the 8$\times$8 Ising model at $T=1.8$.
The specific parameters used for the training are presented in Table ~\ref{tab:rbm-params-2d}.
As mentioned before, the CD parameter $k$ was increased during the training, while the learning rate $\alpha$ was decreased, splitting the training in three phases. The reason for the former prescription, is exemplified in Fig.~\ref{fig:ll_diff_k}, where the loss function and the log-likelihood between epoch 3000 and 4000 are plotted for different values of $k$. While the loss function is left unchanged with varying $k$, the log-likelihood displays a clear difference in shape, \ie it increases as $k$ is increased, while keeping $\alpha$ fixed at 0.01. If, instead, we choose to keep $k$ fixed at $k=1$, and decrease $\alpha$ from 0.01 to 0.001, we see no improvement in the log-likelihood as observed on the left hand side of Fig.~\ref{fig:ll_diff_k}. 
Therefore, it is the increase in the value of $k$ that leads to the increase in the log-likelihood, visible in the curves corresponding to $T=1.8$ of Fig.~\ref{fig:allestimators}.

\begin{figure}[!htb]
	\minipage{0.50\textwidth}
	\includegraphics[width=\linewidth]{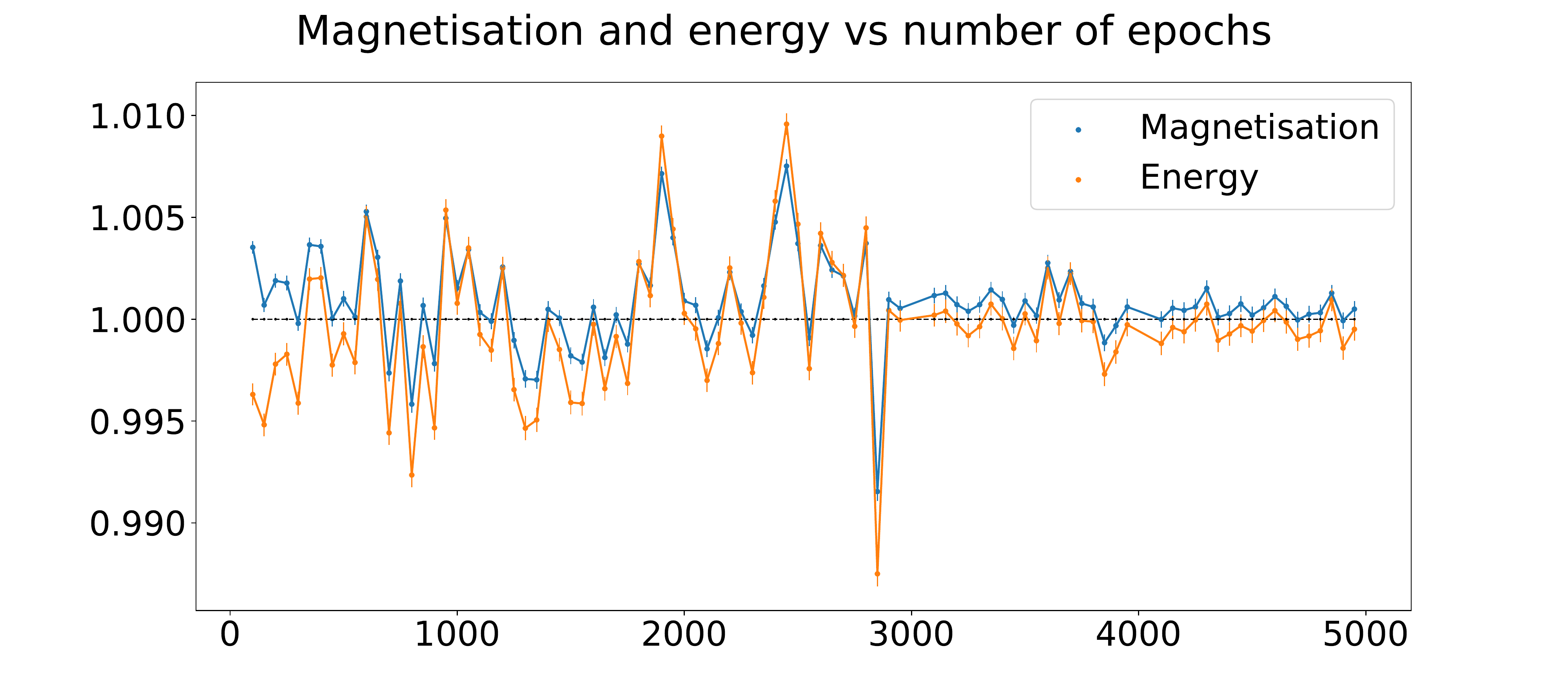}  
	\endminipage\hfill
	\minipage{0.50\textwidth}
	\includegraphics[width=\linewidth]{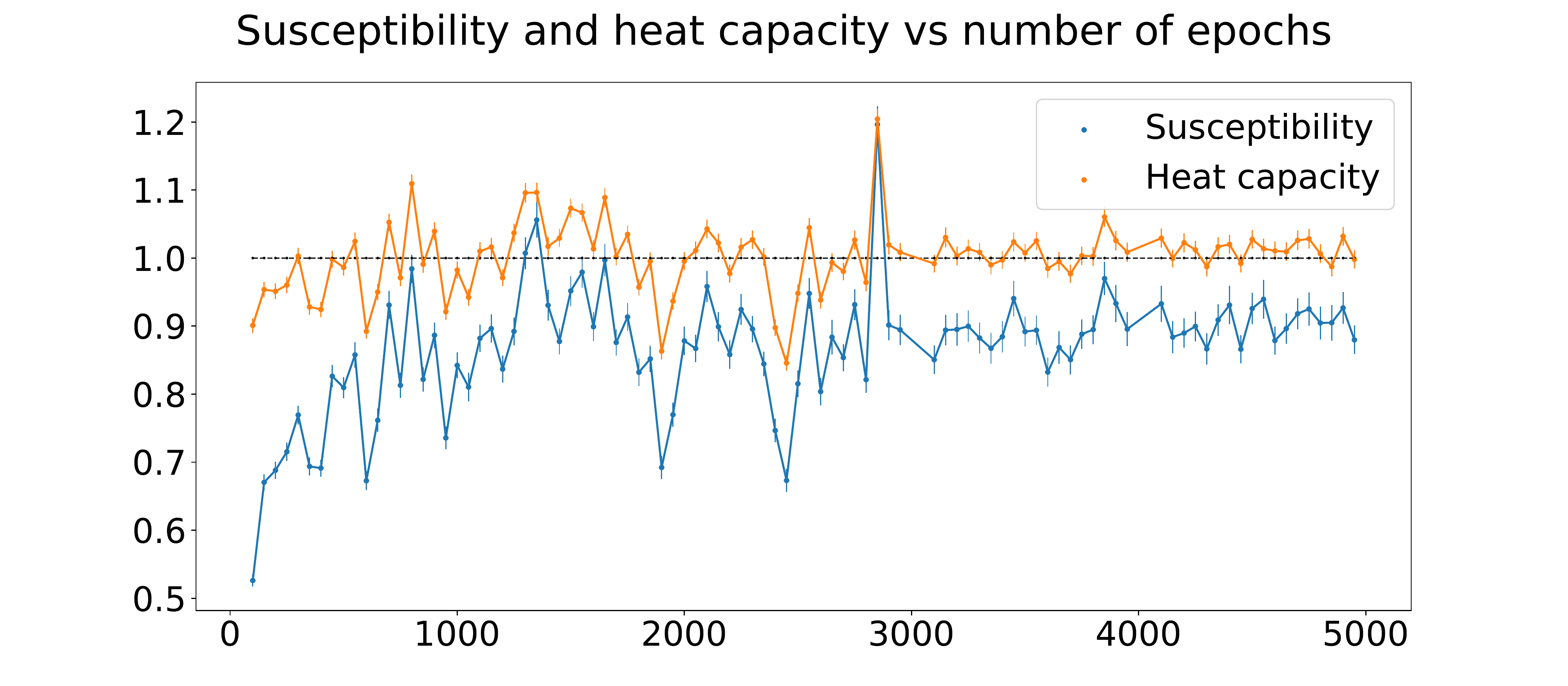}  
	\endminipage\hfill
	\caption{Observables for $T=1.8$ normalized by their expected values as a functions of the training epoch. Magnetization and energy are shown on the left, susceptibility and heat capacity are on the right.}
	\label{fig:obs_1.8}
\end{figure}

The reason for reducing the learning rate along the training is visible in Fig.~\ref{fig:obs_1.8}, where we have taken observables for  $T=1.8$ from Fig.~\ref{fig:allobservables}, magnified and normalized by their expected value from Magneto. We therefore observe that decreasing the learning rate implies a reduction of the fluctuations between successive epochs, resulting in a fine tuning of our predictions. This is more evident starting from epoch 3000, where the learning rate is decreased by factor 10. As shown in the plots, at the end of the training, the predictions from the RBMs are compatible, within statistics, with the expected value from Magneto. Also, when we approach the end of the training, all the machines shown in the plots are statistically equivalent to each other, since their predictions are all compatible within 2 sigma.

\section{Training on a larger system: $L^2=16\times16$}\label{app:training-1616}
\begin{figure}[!htb]
	\minipage{0.50\textwidth}
	\includegraphics[width=\linewidth]{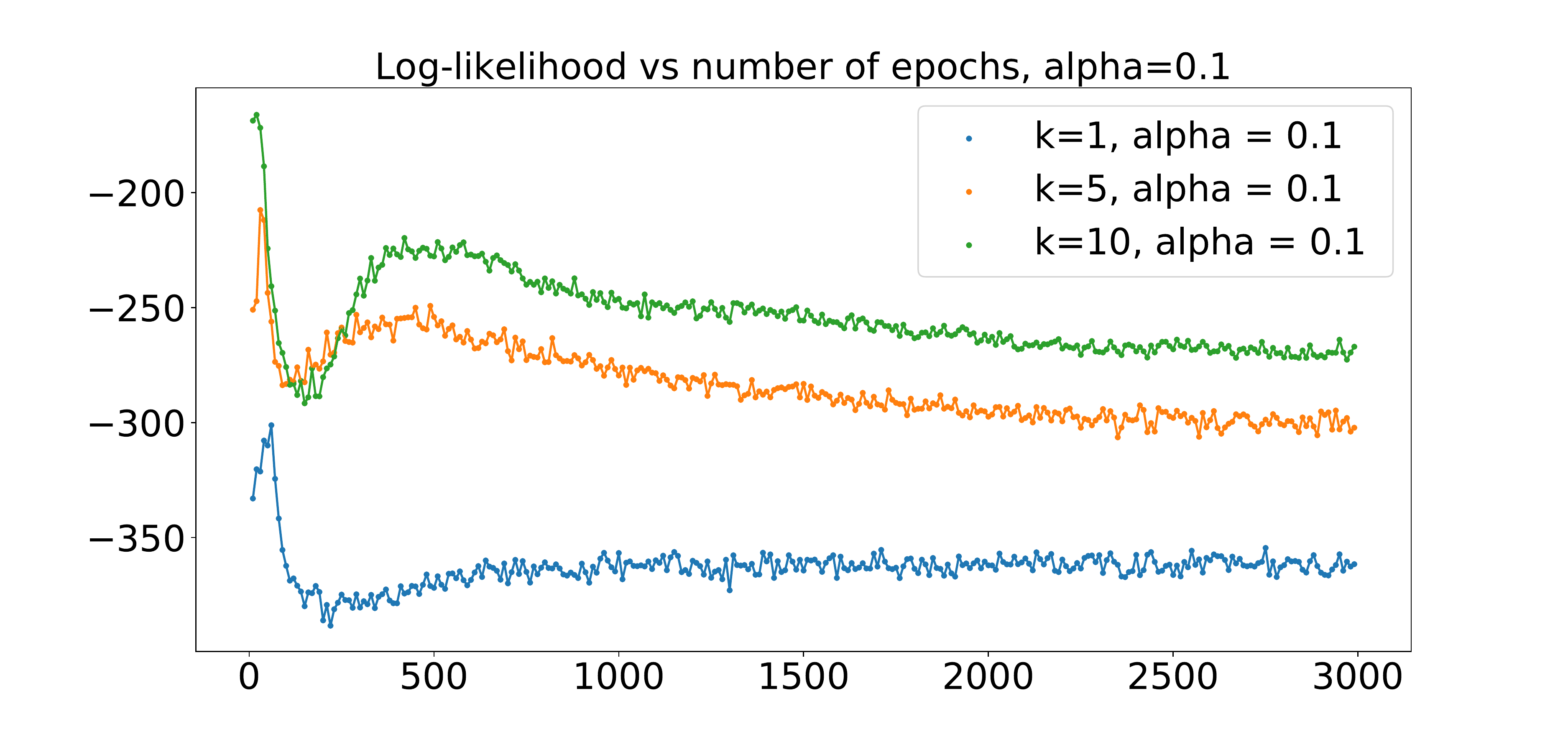}  
	\endminipage\hfill
	\minipage{0.50\textwidth}
	\includegraphics[width=\linewidth]{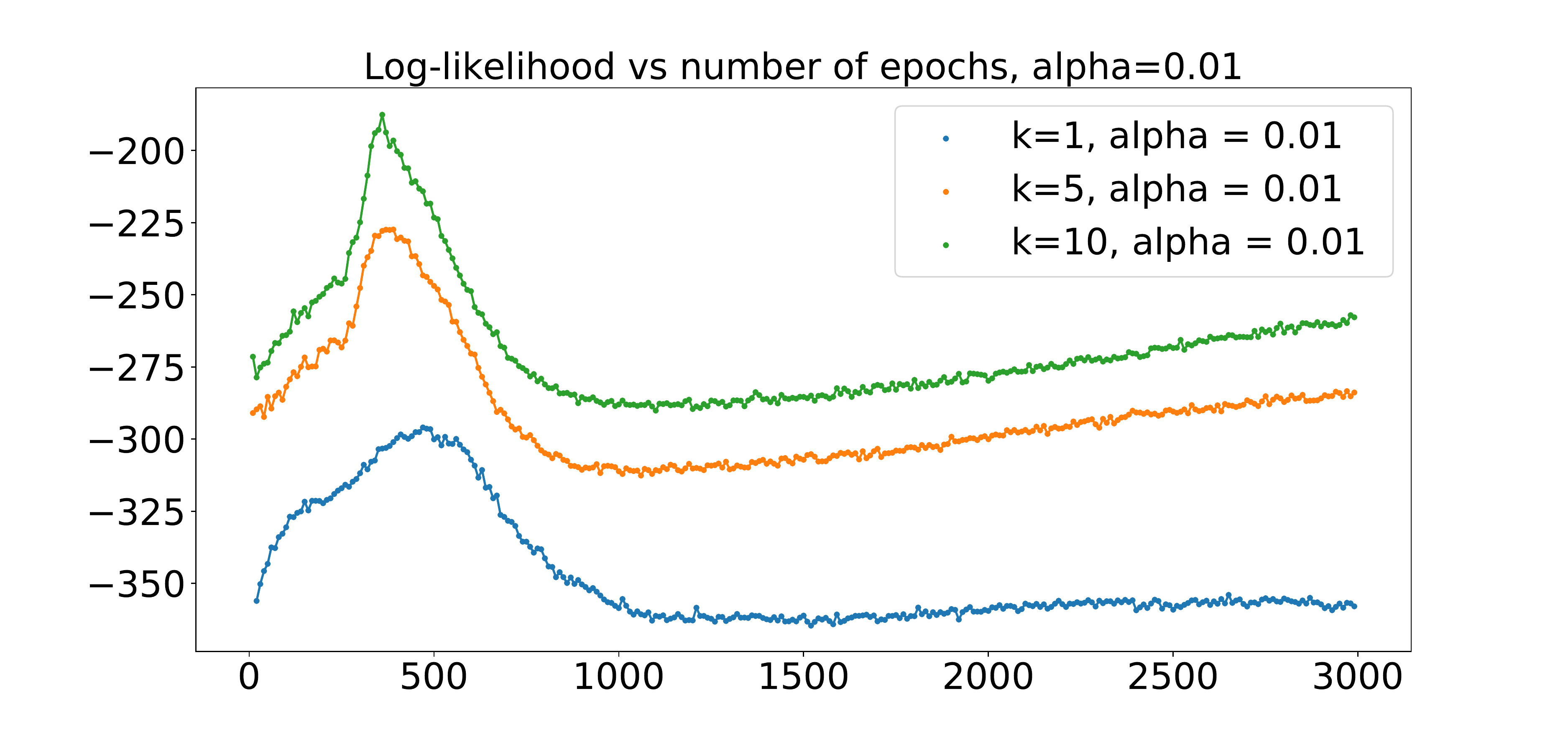}  
	\endminipage\hfill
	\caption{Log-likelihood for different values of $k$ and $\alpha$.}
	\label{fig:ll_16_diffk}
\end{figure}

\begin{figure}[!htb]
	\minipage{0.50\textwidth}
	\includegraphics[width=\linewidth]{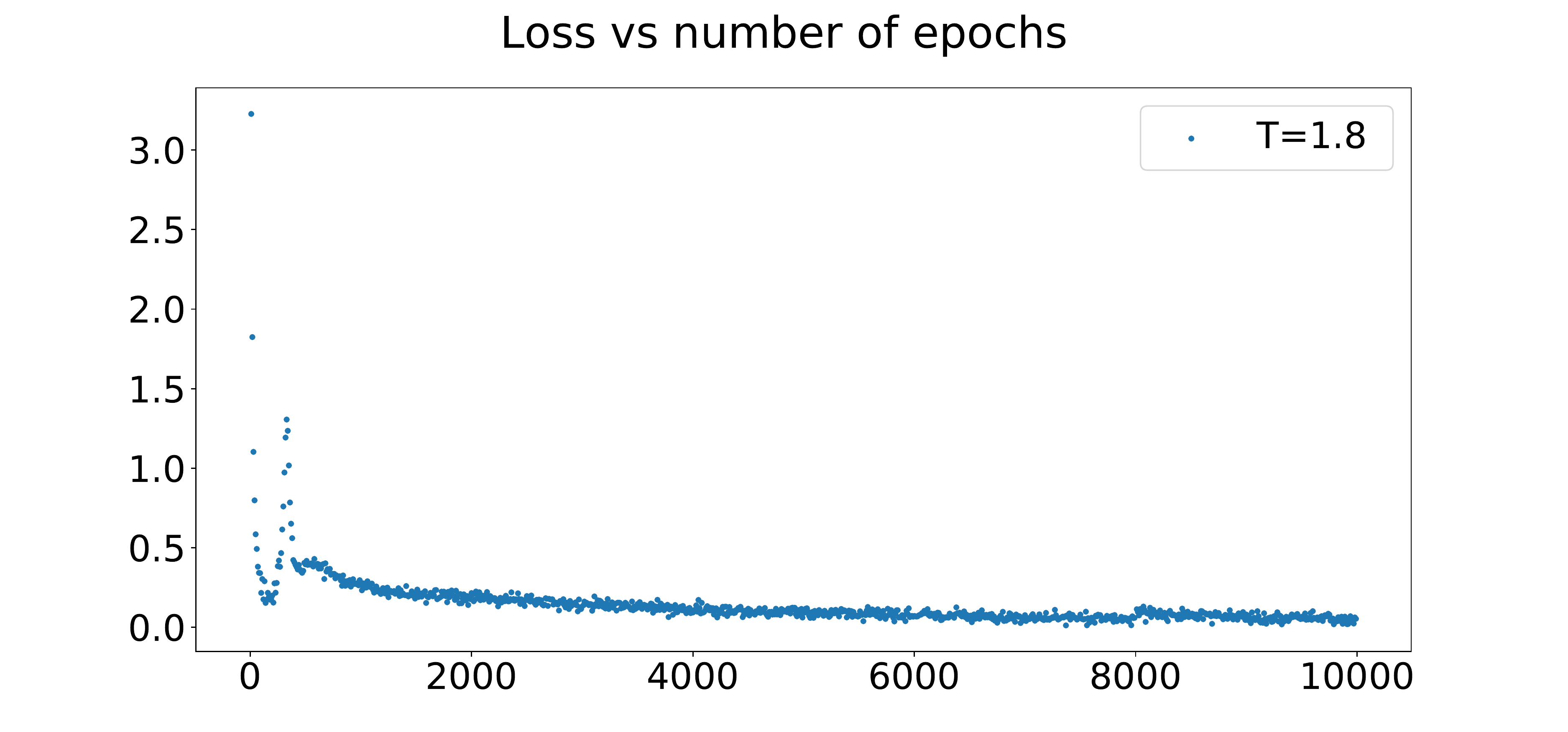}  
	\endminipage\hfill
	\minipage{0.50\textwidth}
	\includegraphics[width=\linewidth]{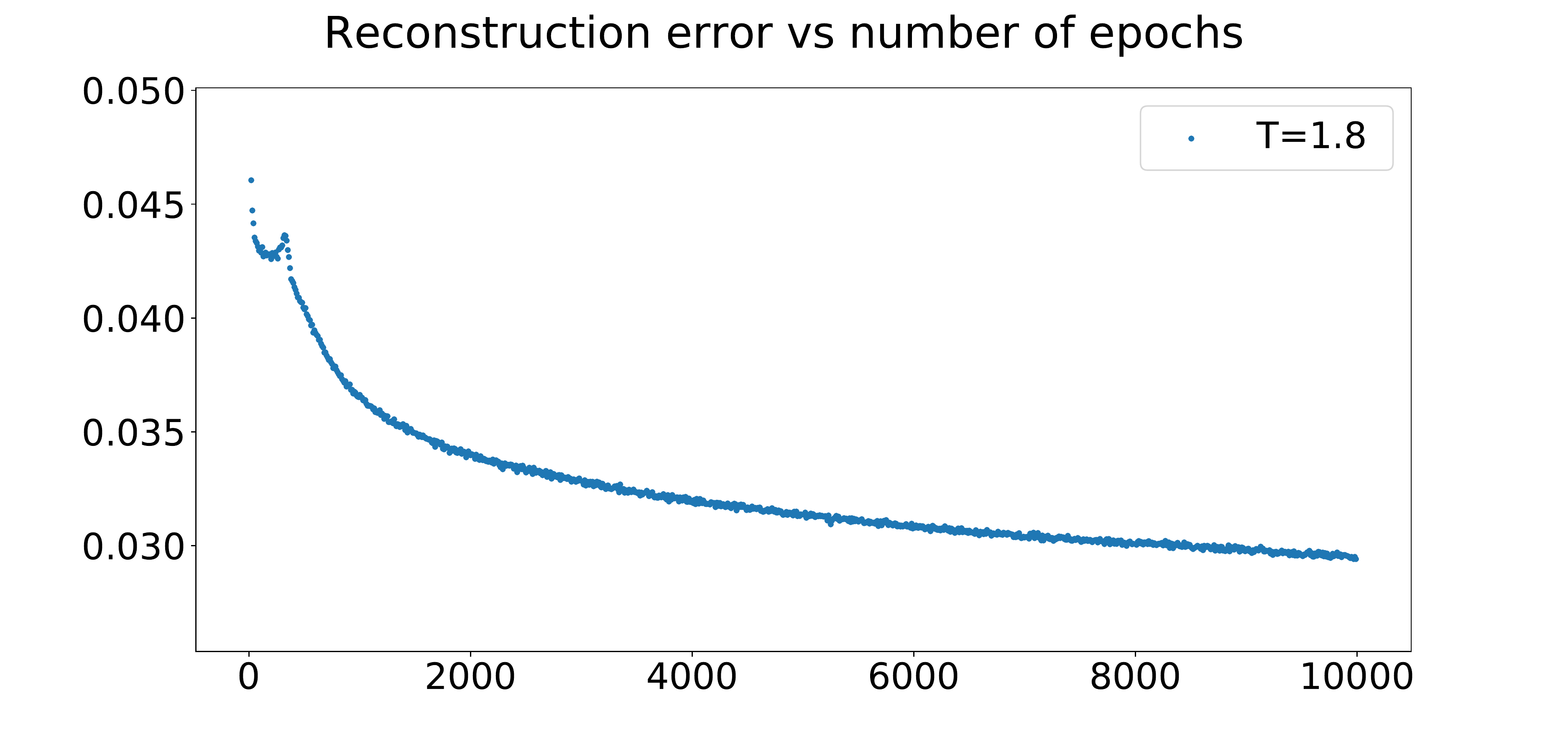}  
	\endminipage\hfill
	\minipage{0.50\textwidth}
	\includegraphics[width=\linewidth]{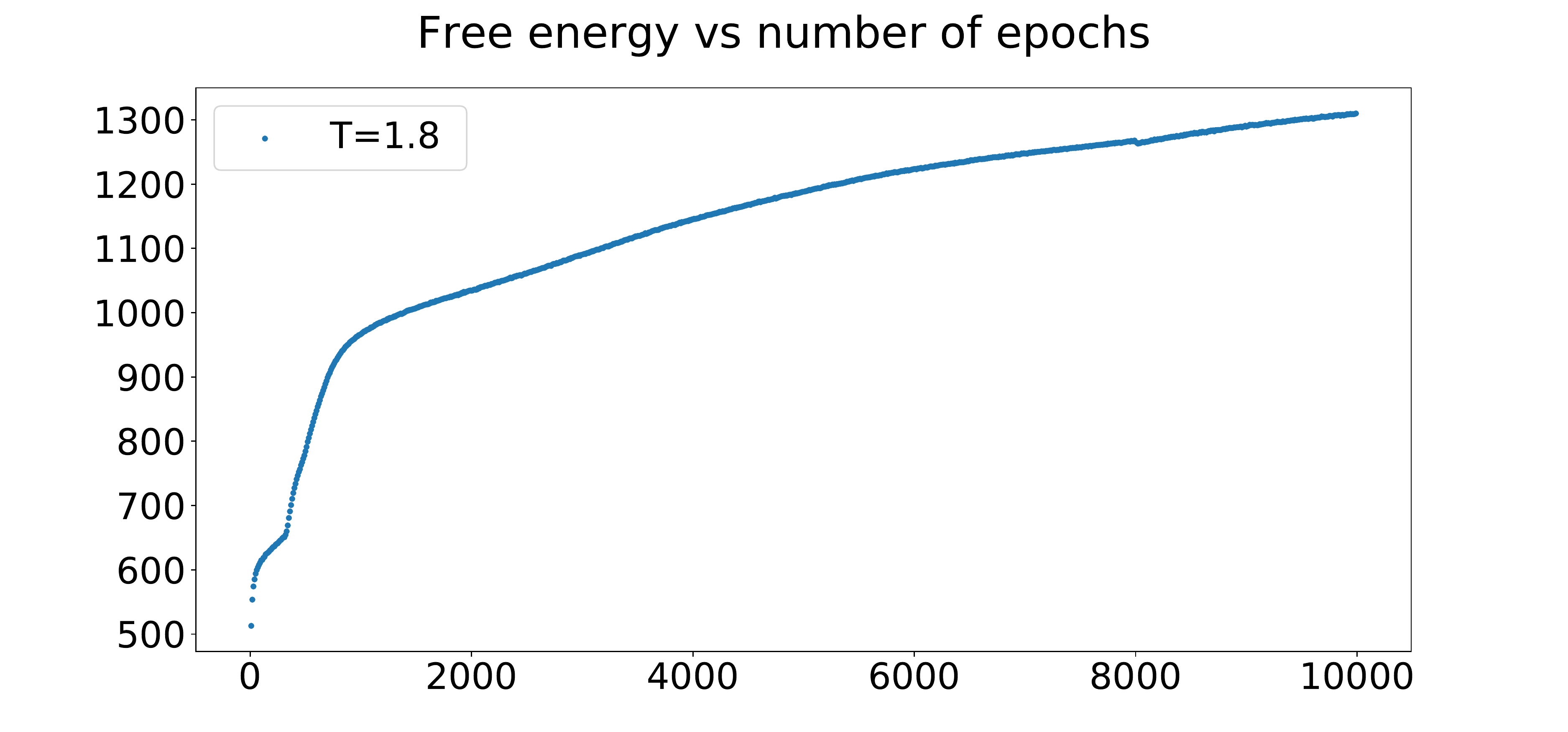}  
	\endminipage\hfill
	\minipage{0.50\textwidth}
	\includegraphics[width=\linewidth]{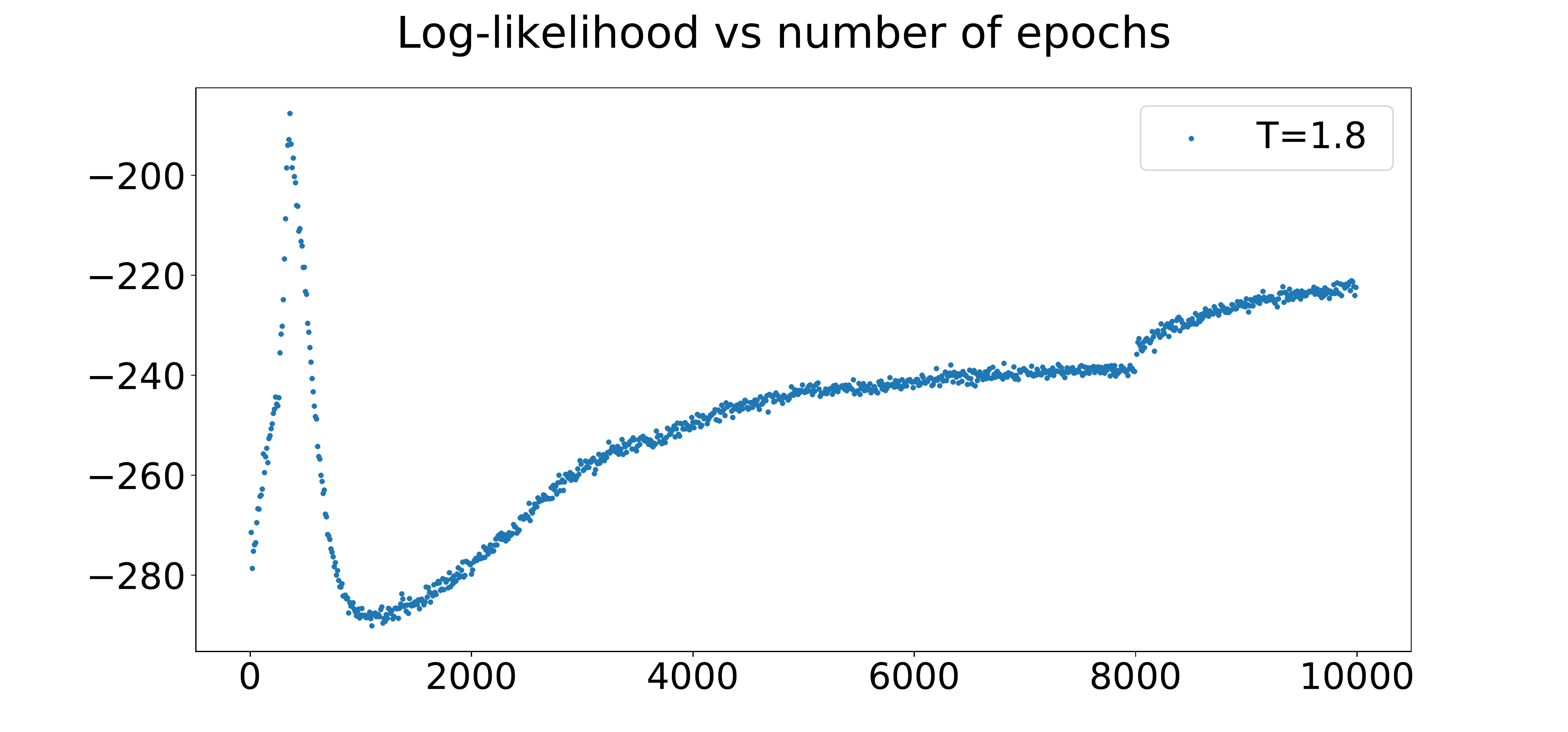}  
	\endminipage\hfill
	\caption{Here we observed the increase in the log-likelihood behaviour for our chosen values of $k$ and $\alpha$, as given in Table~\ref{tab:rbm-params-2d}. Both the loss function and reconstruction error decrease as the training progresses.}
	\label{fig:estimators16}
\end{figure}

\begin{figure}[!htb]
	\minipage{0.50\textwidth}
	\includegraphics[width=\linewidth]{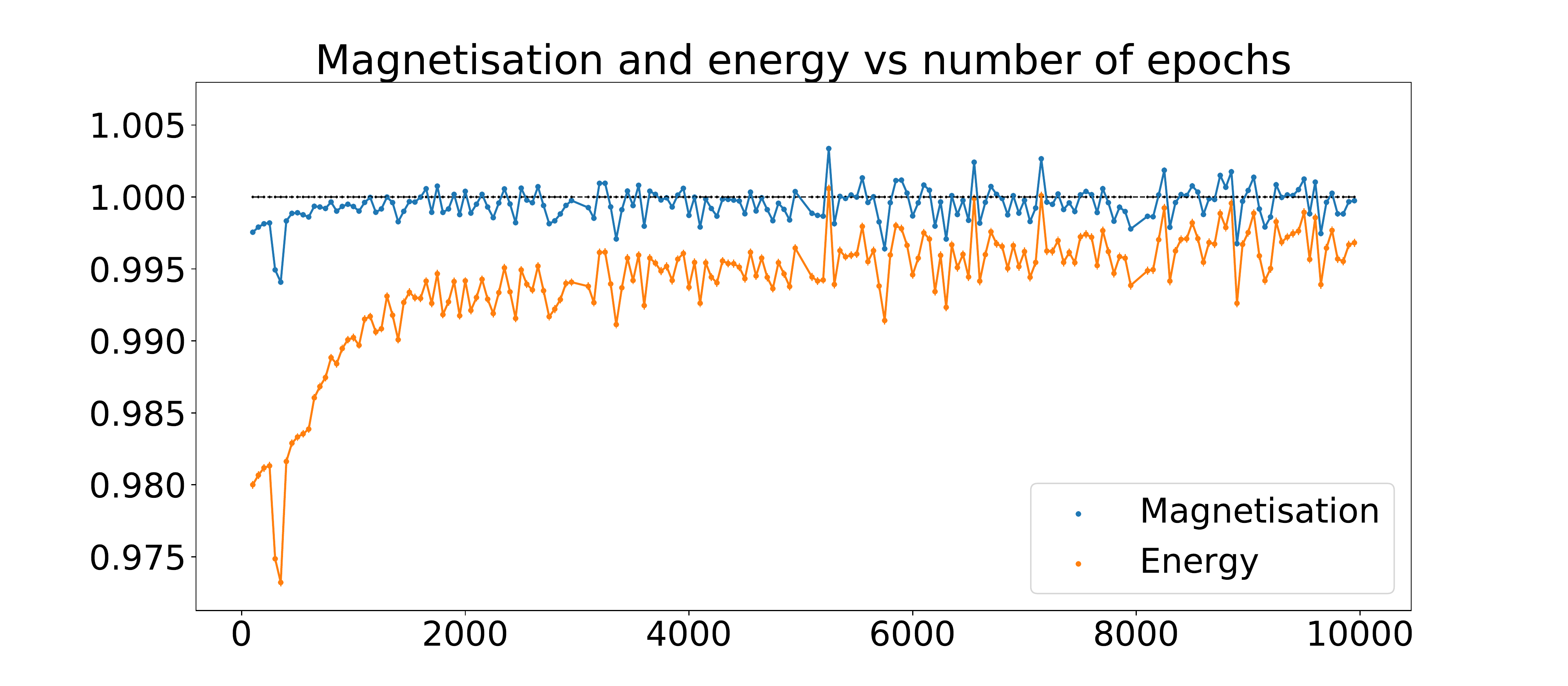}  
	\endminipage\hfill
	\minipage{0.50\textwidth}
	\includegraphics[width=\linewidth]{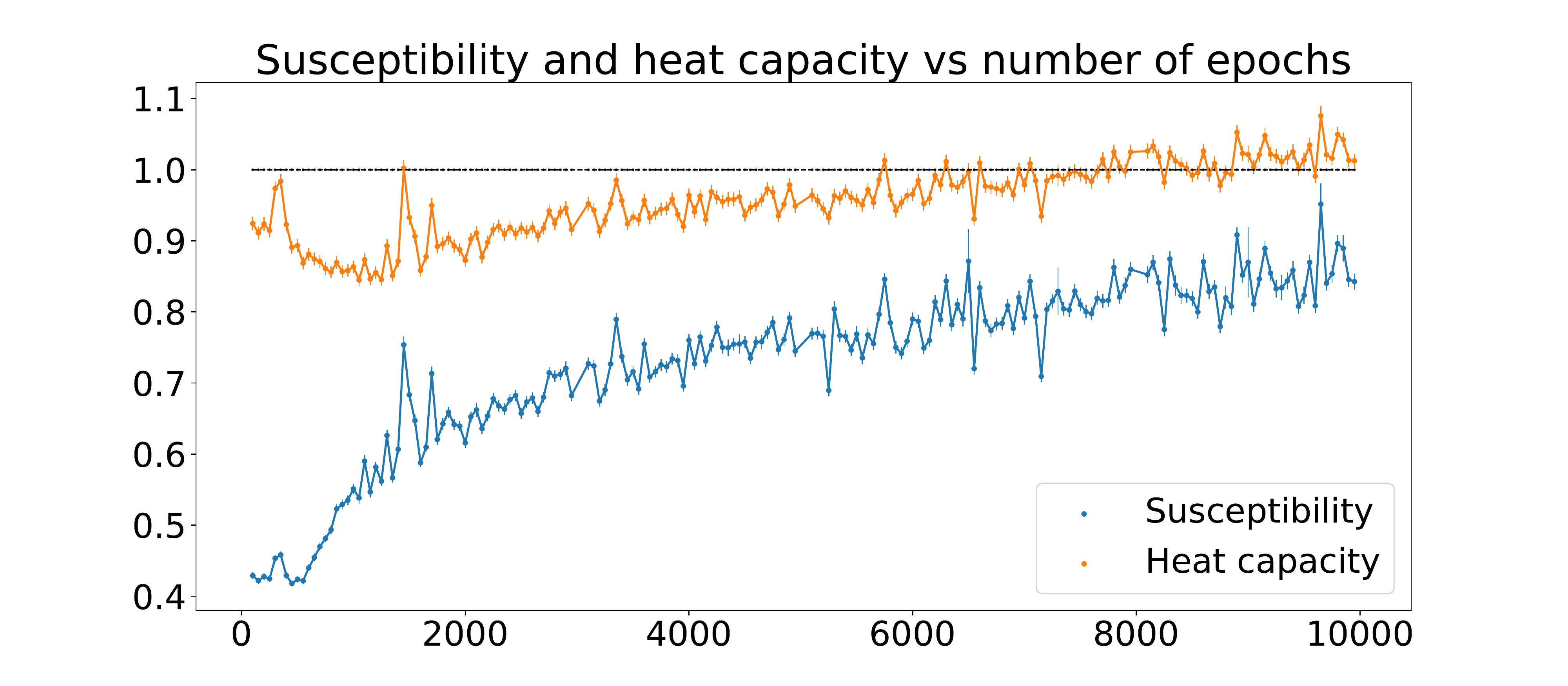}  
	\endminipage\hfill
	\caption{Observables vs epochs for $L^2=16\times16$, $h^2=16\times16$ and batch size 200. The value of each observable, computed from the RBM, is normalised by its expected value, computed from the training set. Magnetisation (blue) and energy (orange) are plotted on the right hand side, susceptibility (blue) and head capacity (orange) are on the left.}
	\label{fig:allobservable16}
\end{figure}

Here we report the results for a training the RBM on $16\times16$ Ising configurations, highlighting the main difficulties arising for larger systems. As done in the case of the 8$\times$8 systems, we first tune the value of $k$ in order to have an increasing log-likelihood. It turns out that the log-likelihood behaviour is sensitive to the value of $k$ and $\alpha$ much more than in the case of smaller 8$\times$8 systems. This is shown in Fig.~\ref{fig:ll_16_diffk}, where the log-likelihood as a function of the training epoch is plotted for different values of $k$ and $\alpha$. On the left hand side we see that using $\alpha=0.1$, the log-likelihood is not an increasing function of the epochs, even when the value of $k$ is increased. However, decreasing the value of $\alpha $ to 0.01 and using $k = 5$ we manage to obtain an increasing behaviour which becomes even more evident when using $k=10$, as shown on the right hand side.
The final setting used to train this machine are reported in Table~\ref{tab:rbm-params-2d}.
The estimators for the training are presented in Fig.~\ref{fig:estimators16} and the observables versus epochs plots in Fig.~\ref{fig:allobservable16}: all the estimators present the correct behaviour. The initial peak visible in the log-likelihood at about 500 epochs, should not be considered as the point where the machine has been best trained, but rather as an initial fluctuation of the log-likelihood that later on dies out. This observation is also confirmed by the loss function and the reconstruction error, which also present a peak around the same number of epochs, implying that the training has not converged yet at this point. The final increase in the log-likelihood at 8000 epochs, is where we started using $k=20$ instead of $k=10$, as reported in Table~\ref{tab:rbm-params-2d}. 

\begin{figure}[!htb]
	\minipage{0.50\textwidth}
	\includegraphics[width=\linewidth]{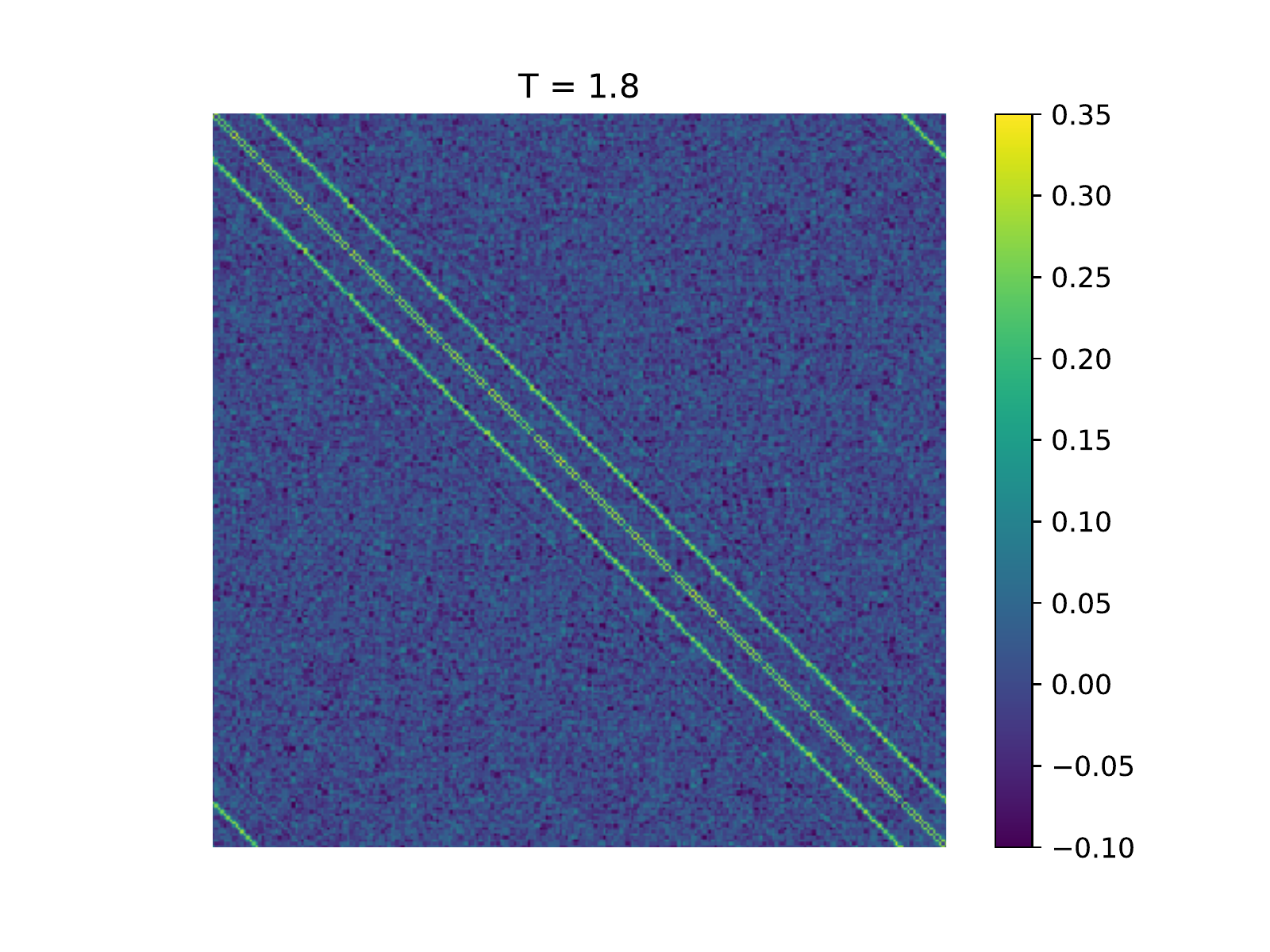}  
	\endminipage\hfill
	\minipage{0.50\textwidth}
	\includegraphics[width=\linewidth]{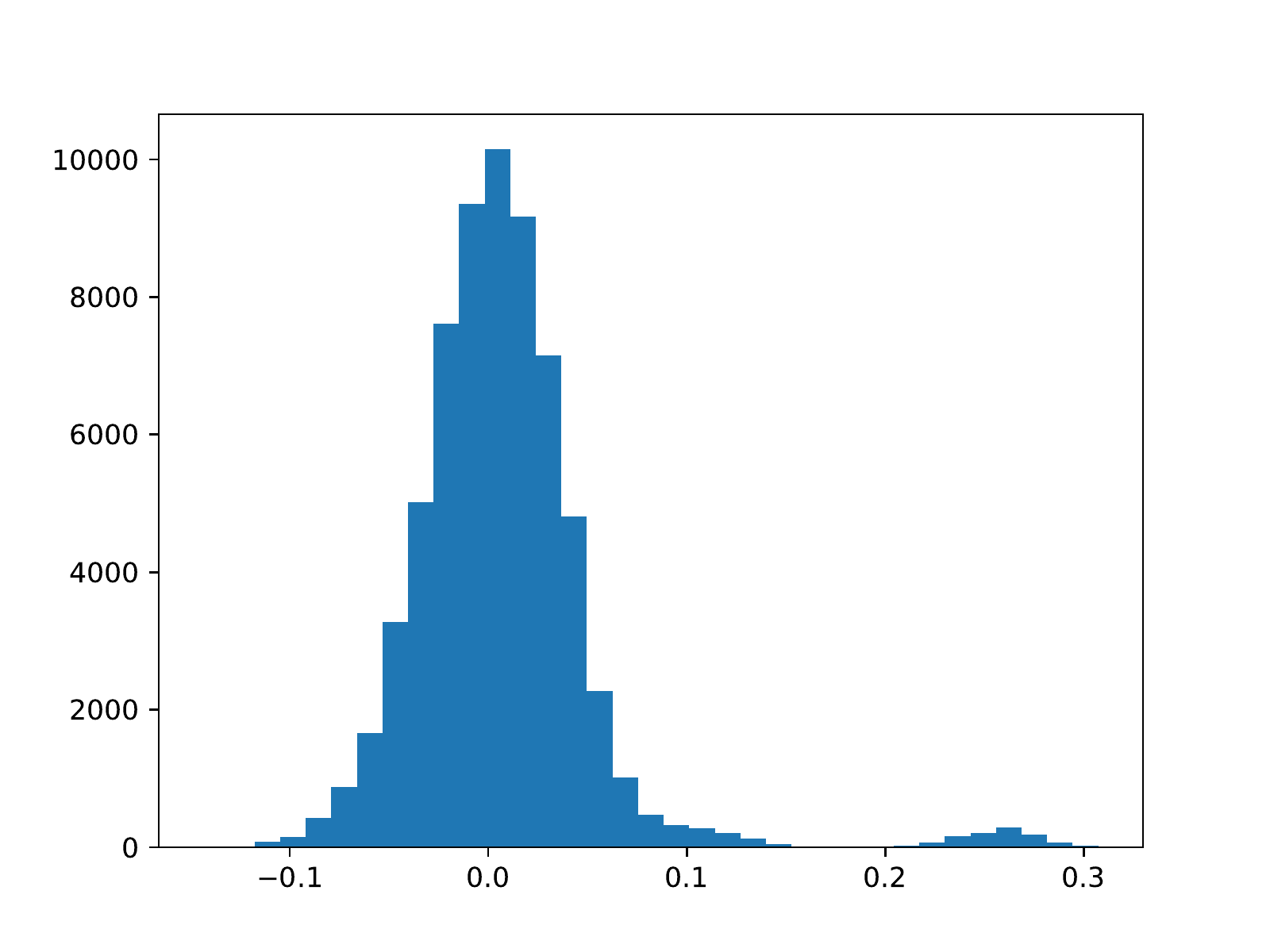}  
	\endminipage\hfill
	\caption{The 2-point interaction matrix, $H_{j_1,j_2}$ (left) and its corresponding histogram (right) for the machine with $L^2=16\times16$, $h^2=16\times16$ and batch size 200. Again, we observed the larger peak centred around zero, corresponding to non nearest neighbour interactions, while there is a second peak representing the coupling with the nearest neighbour spins. }
	\label{fig:matrix_16_batch200}
\end{figure}

From the plot of observables in Fig.~\ref{fig:allobservable16}, it can be seen that magnetization and heat capacity have almost converged to the expected value, and would only require a fine tuning by reducing the value of $\alpha $. The energy and susceptibility are further from the expected values, therefore, longer training is required to obtain more accurate values for these two quantities. Having said that, it appears that all the moments are slowly approaching the expected values, and the fact that the log-likelihood is still increasing with the number of epochs suggests that our machine is indeed learning. We can conclude that for the case of the larger $16\times16$ system the convergence is much slower than the previous smaller $8\times8$ systems.
We can convince ourselves that the machine is actually learning looking at the 2-point interaction matrix for the machine at epoch 10000, plotted in Fig.(\ref{fig:matrix_16_batch200}). Clearly the expected path for the interactions can already be seen in the matrix. The coupling extracted from this matrix is $1/2T \sim 0.258 \pm 0.017 $, as compared to the expected value of $0.277$. This confirms that the machine agrees with the correct theoretical result within 2 sigma. A more accurate result can be obtained if the machine is trained for a longer period of time, as the log-likelihood is still increasing. 

\section{Changing the batch size}\label{app:batch-size}
\begin{figure}[!htb]
\centering
\includegraphics[width=10cm]{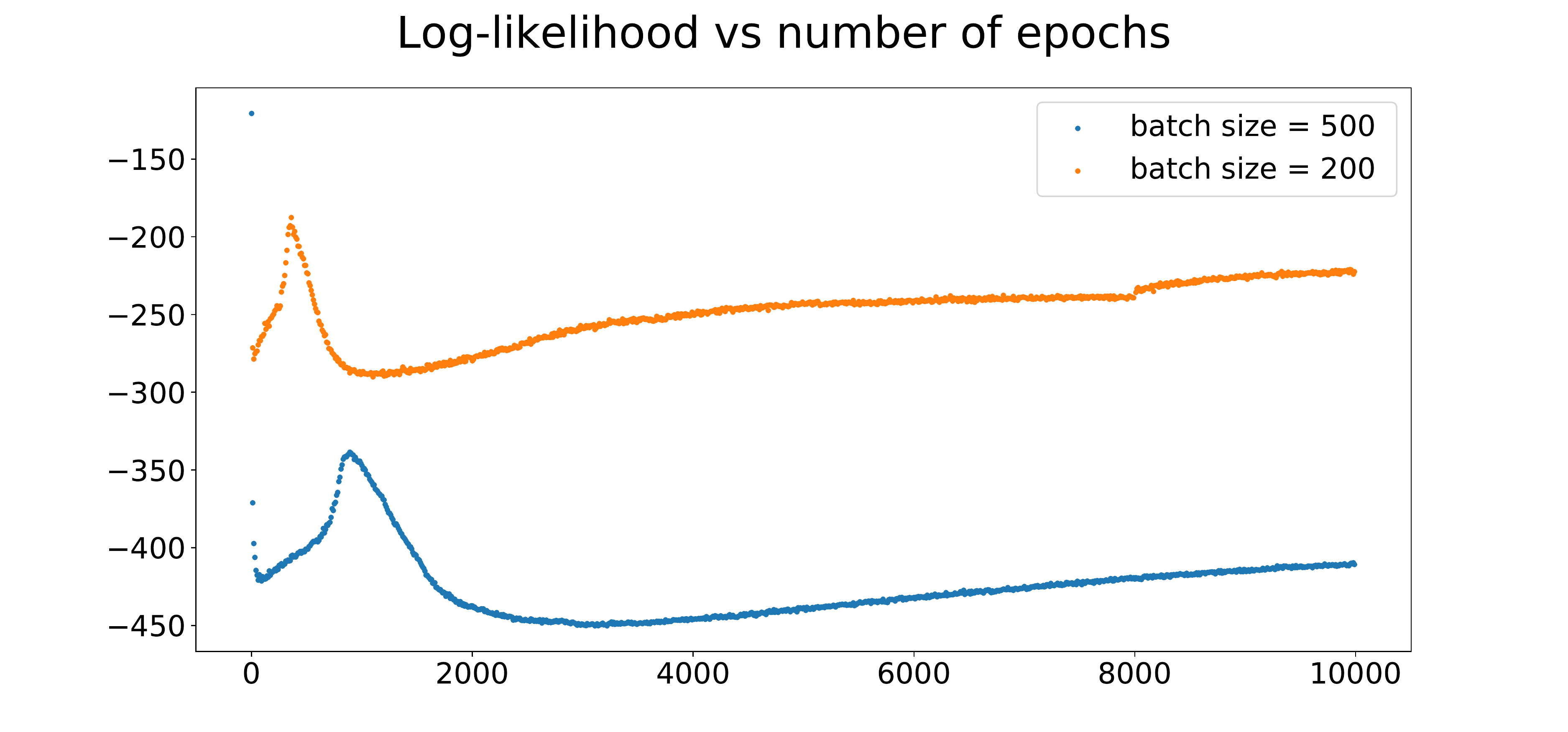}
\caption{The dependence of log likelihood on batch size. The curves correspond to $L^2=16\times16$ lattice with batch size 500 (blue) and batch size 200 (yellow). The choice of a smaller batch size, results in a steeper rise to the log-likelihood.}
\label{fig:batch-size-effect}
\end{figure}

We also examined the effect of changing the batch size for the $L^2=16\times16$ system, keeping the other hyperparameters the same as the successfully trained machine above, \ie, $\alpha=0.01$ and $k=10$. As it can be observed from Fig.~\ref{fig:batch-size-effect}, the log-likelihood has a slower increase with batch size 500, as compared to batch size 200. 

\begin{figure}[!htb]
	\minipage{0.50\textwidth}
	\includegraphics[width=\linewidth]{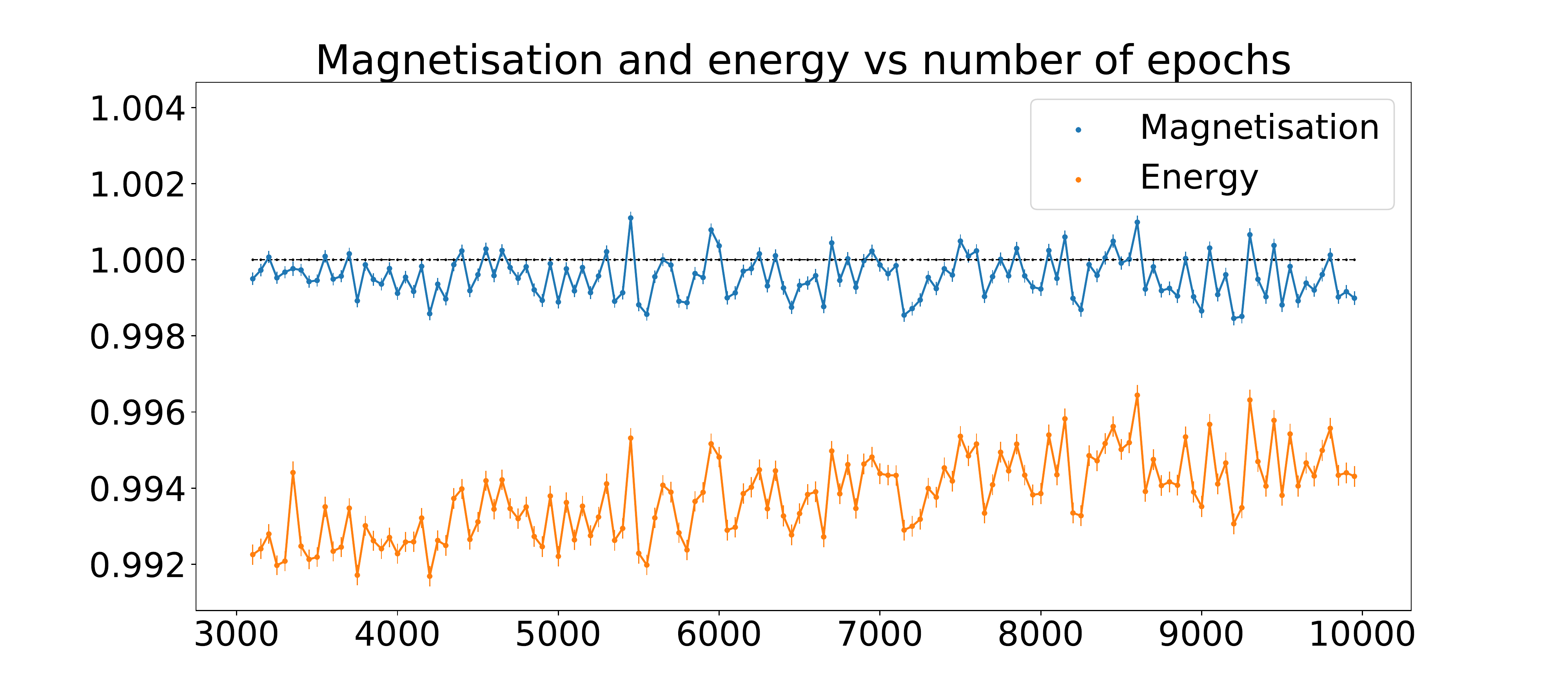}  
	\endminipage\hfill
	\minipage{0.50\textwidth}
	\includegraphics[width=\linewidth]{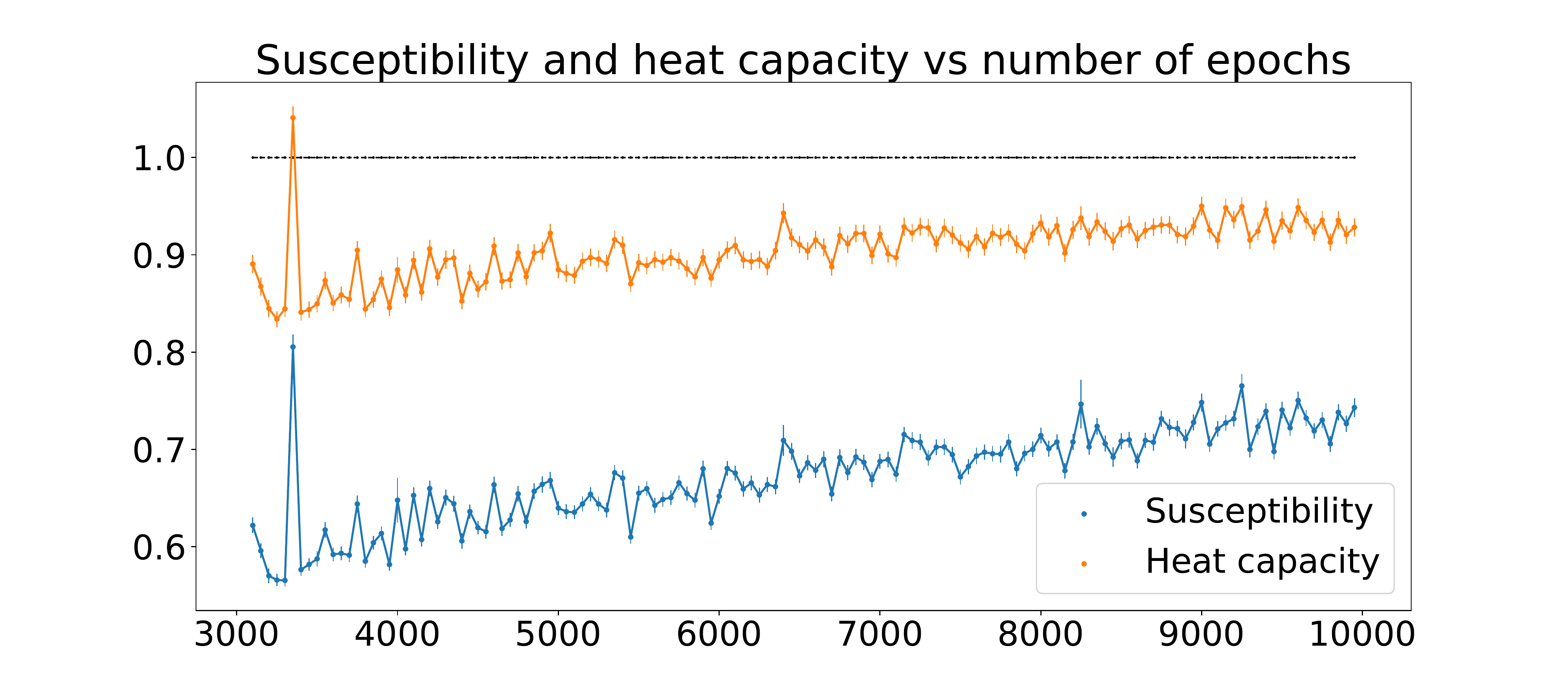}  
	\endminipage\hfill
	\caption{Observables, normalised by their expected values, vs epochs for $L^2=16\times16$, $h^2=16\times16$ and batch size 500. We observe that it takes the observables longer to converge to the correct values, as compared to the case where a smaller batch size is used, \eg, see Fig.~\ref{fig:allobservable16}.}
	\label{fig:batch-500-obs}
\end{figure}

\begin{figure}[!htb]
\minipage{0.50\textwidth}
\includegraphics[width=\linewidth]{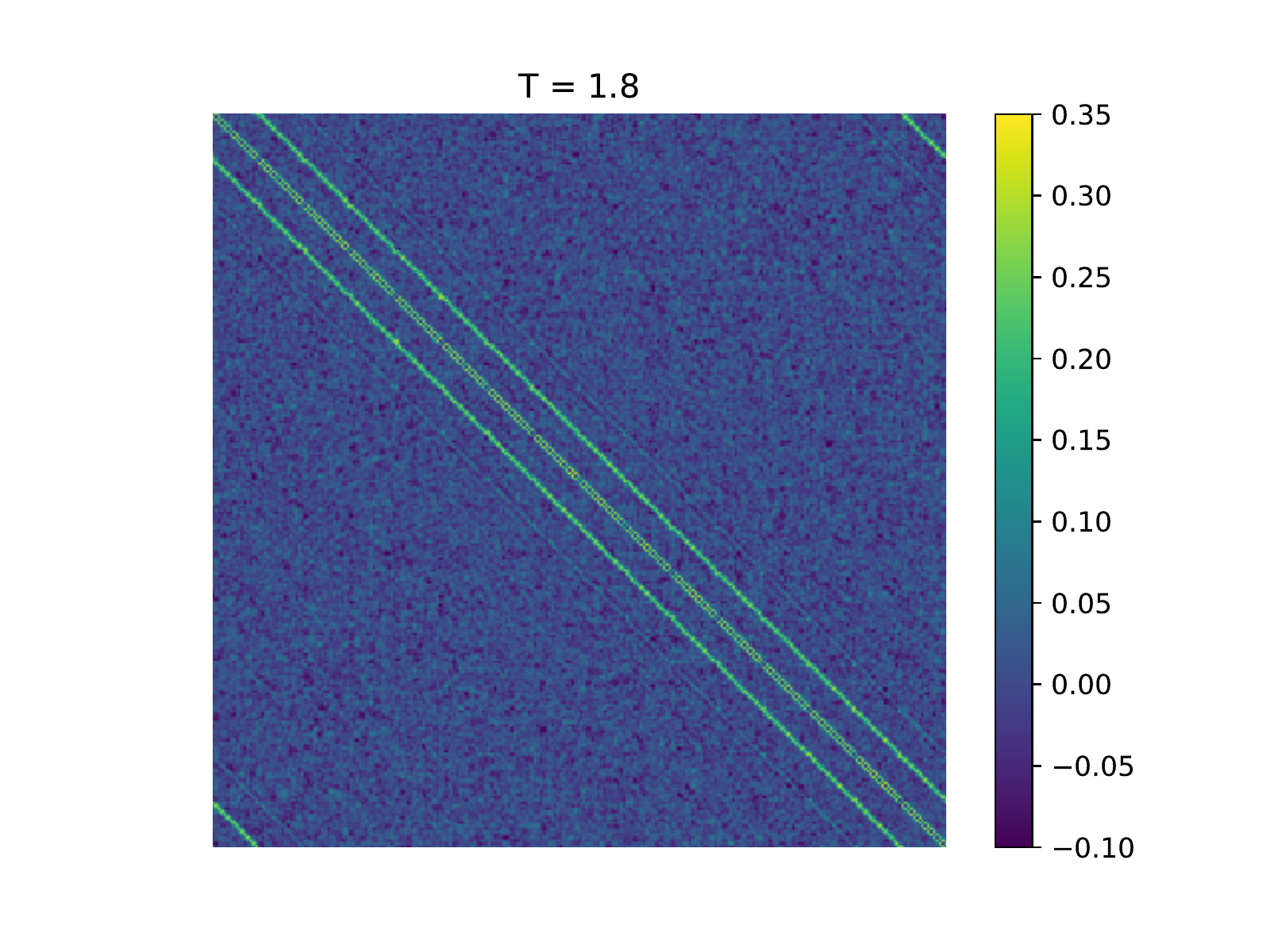}  
\endminipage\hfill
\minipage{0.50\textwidth}
\includegraphics[width=\linewidth]{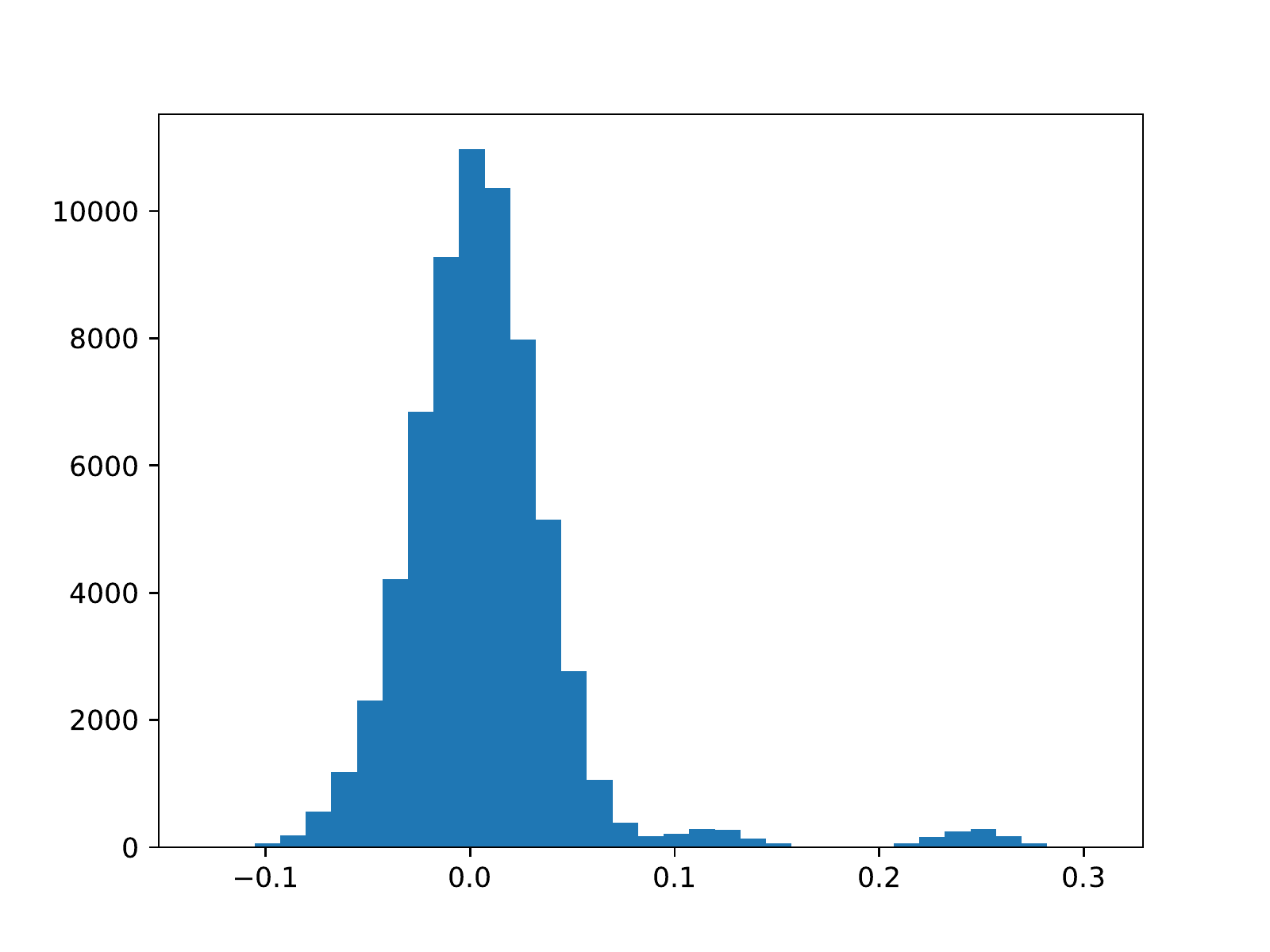}  
\endminipage\hfill
\caption{The 2-point interaction matrix, $H_{j_1,j_2}$ (left) and its corresponding histogram (right) for the machine with $L^2=16\times16$, $h^2=16\times16$ and batch size 500. There is a large peak centred around zero, corresponding to non nearest neighbour interactions, however, a second smaller peak can also be observed next to it. As already discussed, the machine with a larger batch size, \ie 500, has to be trained for longer epochs as compared to the machine with batch size 200, in order to learn that non nearest neighbour interactions are zero. Finally, the distinct peak on the right hand side of the plot represents the expected coupling with the nearest neighbour spins, compare with Fig.\ref{fig:matrix_16_batch200}}
\label{fig:matrix_16_batch500}
\end{figure}

The slower training in this case can also be observed by plotting the trajectories of the observables per epoch, in Fig.~\ref{fig:batch-500-obs}, as compared to Fig.~\ref{fig:allobservable16}. 
This is also consistent with the two-point interaction matrix presented in Fig.~\ref{fig:matrix_16_batch500}, which is more noisy in this case as compared to Fig.~\ref{fig:matrix_16_batch200}.

\section{Changing the number of hidden nodes}
\begin{figure}[!htb]
\centering
\includegraphics[width=0.6\textwidth]{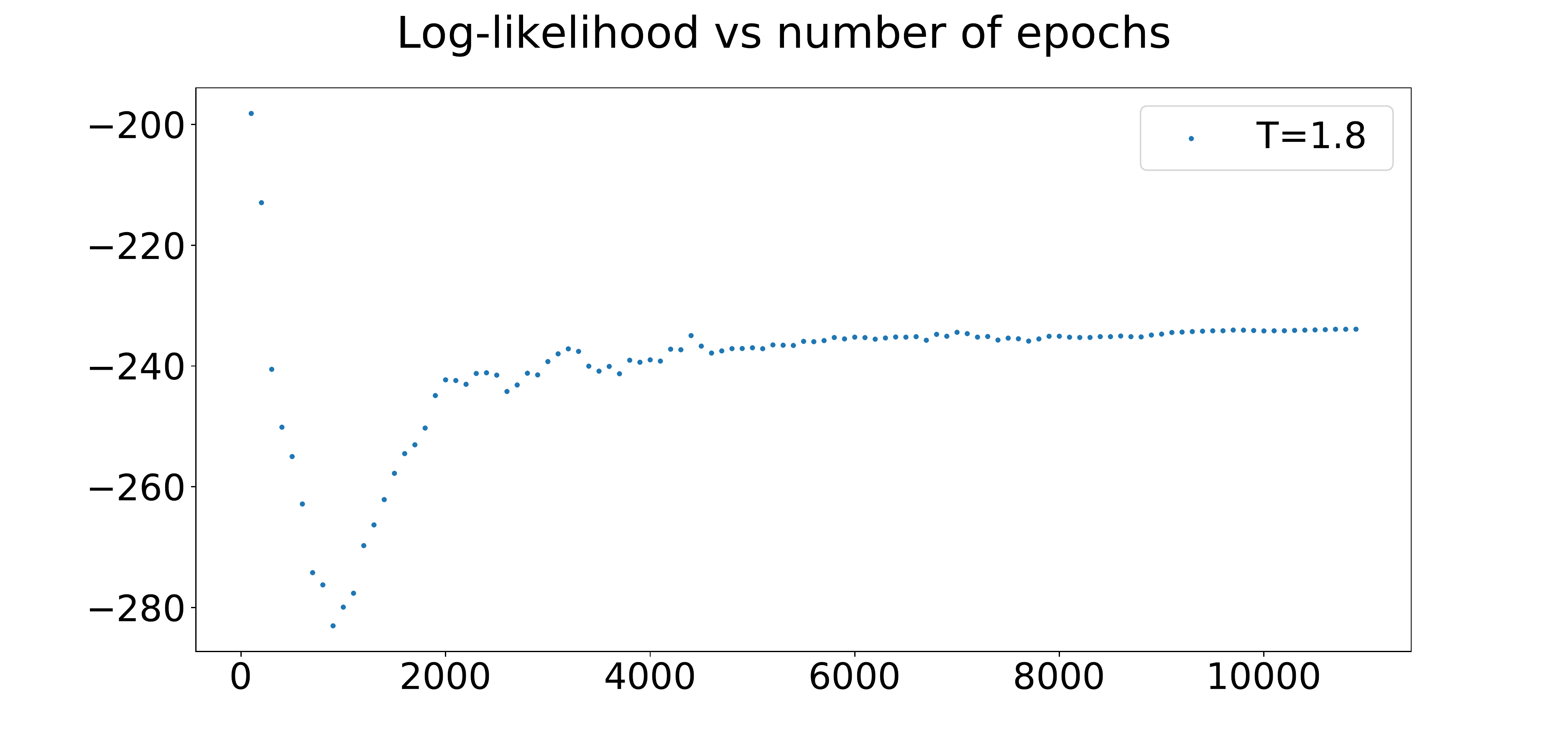}
\caption{Log-likelihood for an RBM with less hidden nodes than visible nodes, $L^2=16\times16$ and $h^2 =  12 \times 12$. The first 4500 epochs were trained using $\alpha=0.01$, $k=10$ and batch size 50.  According the prescription, we then reduced the value of $\alpha$ and increased $k$, \ie, From 4500 to 8000, we set $\alpha=0.001$ and $k=20$. From 8000 to 8700 epochs $\alpha$ and $k$ were kept fixed at their previous value, while the batch size was increase to 200, in order to reduce the fluctuations in the estimate of the log-likelihood. In the last steps we chose $\alpha=0.0001$ and $k=30$, and $\alpha=0.00001$ and $k=40$.}
\label{fig:v16-h12-LL}
\end{figure}

We also attempted to train an RBM, for the $L^2=16\times16$ Ising system, where the number of hidden nodes were less than the visible nodes. More explicitly, we chose $h^2 =  12 \times 12$. The log-likelihood is plotted in Fig.~\ref{fig:v16-h12-LL}. The first 4500 epochs were trained using $\alpha=0.01$ and $k=10$, as these settings were successful in training the $h^2=16\times16$ case in Sec.~\ref{app:training-1616}. However, from the experience gained in Sec.~\ref{app:batch-size}, the batch size was reduced to 50 in order to give an faster increasing slope to the log-likelihood. 

\begin{figure}[!htb]
	\minipage{0.50\textwidth}
	\includegraphics[width=\linewidth]{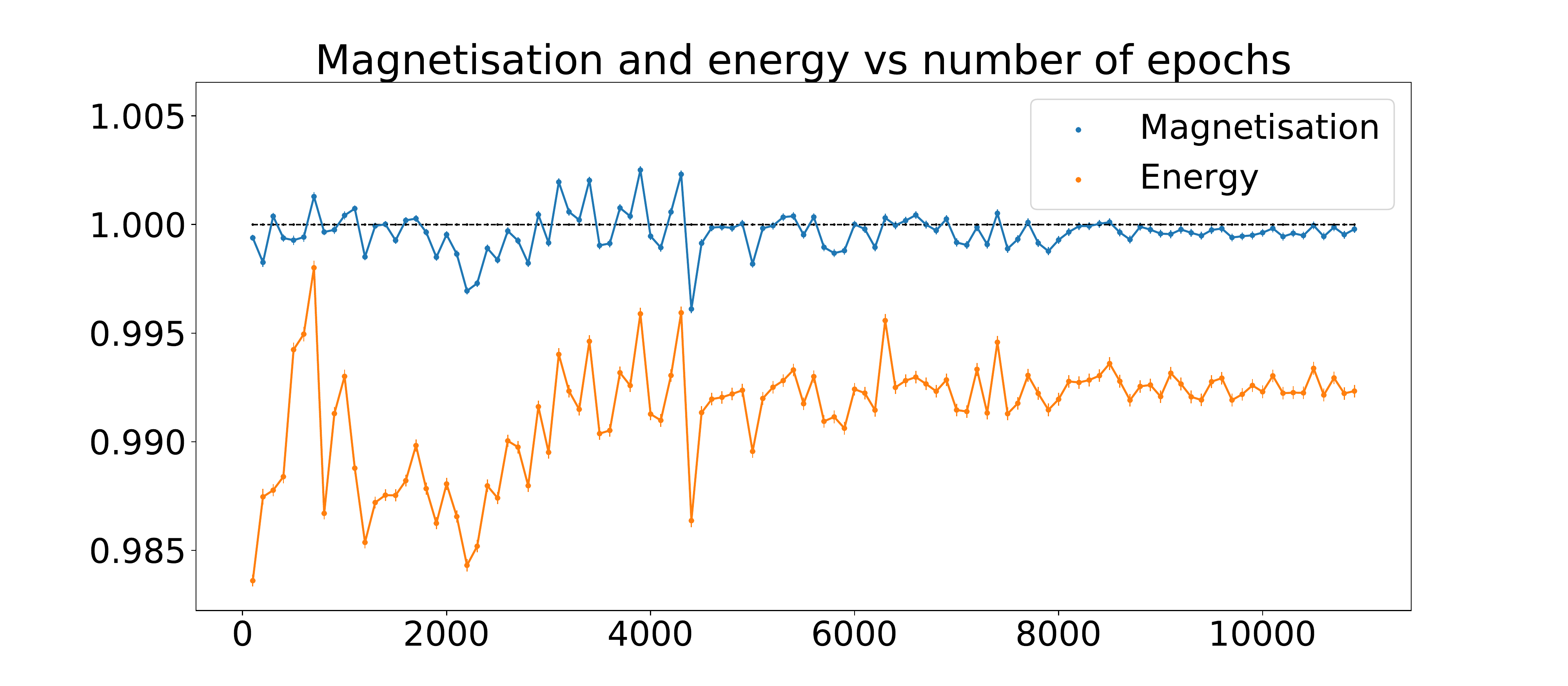}  
	\endminipage\hfill
	\minipage{0.50\textwidth}
	\includegraphics[width=\linewidth]{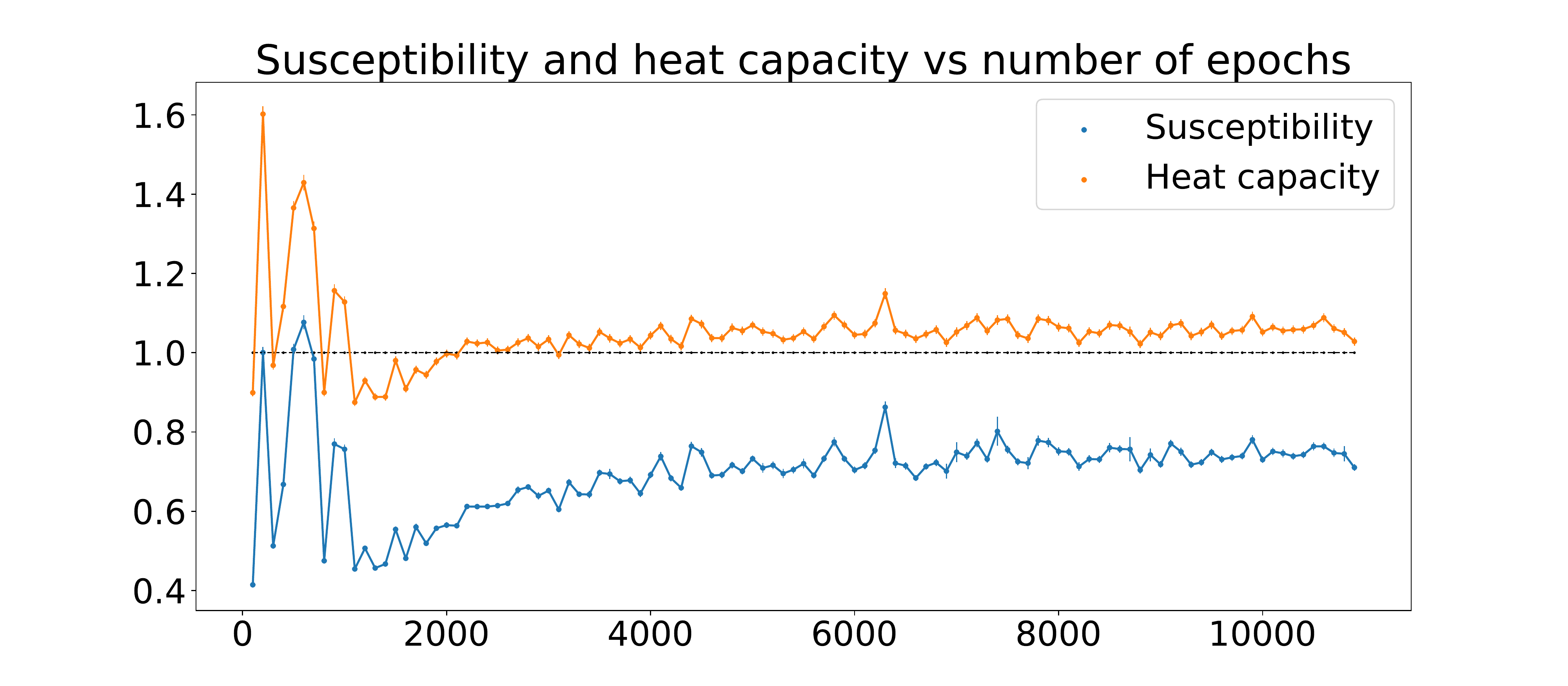}  
	\endminipage\hfill
	\caption{Observables vs epochs for $L^2=16\times16$, $h^2=12\times12$. It can be observed that the machine has to run for more epochs for it to learn the observables and hence the correct structure.}
	\label{fig:h12-k10_a001_k20_a0001}
\end{figure}

\begin{figure}[!htb]
	\minipage{0.50\textwidth}
	\includegraphics[width=\linewidth]{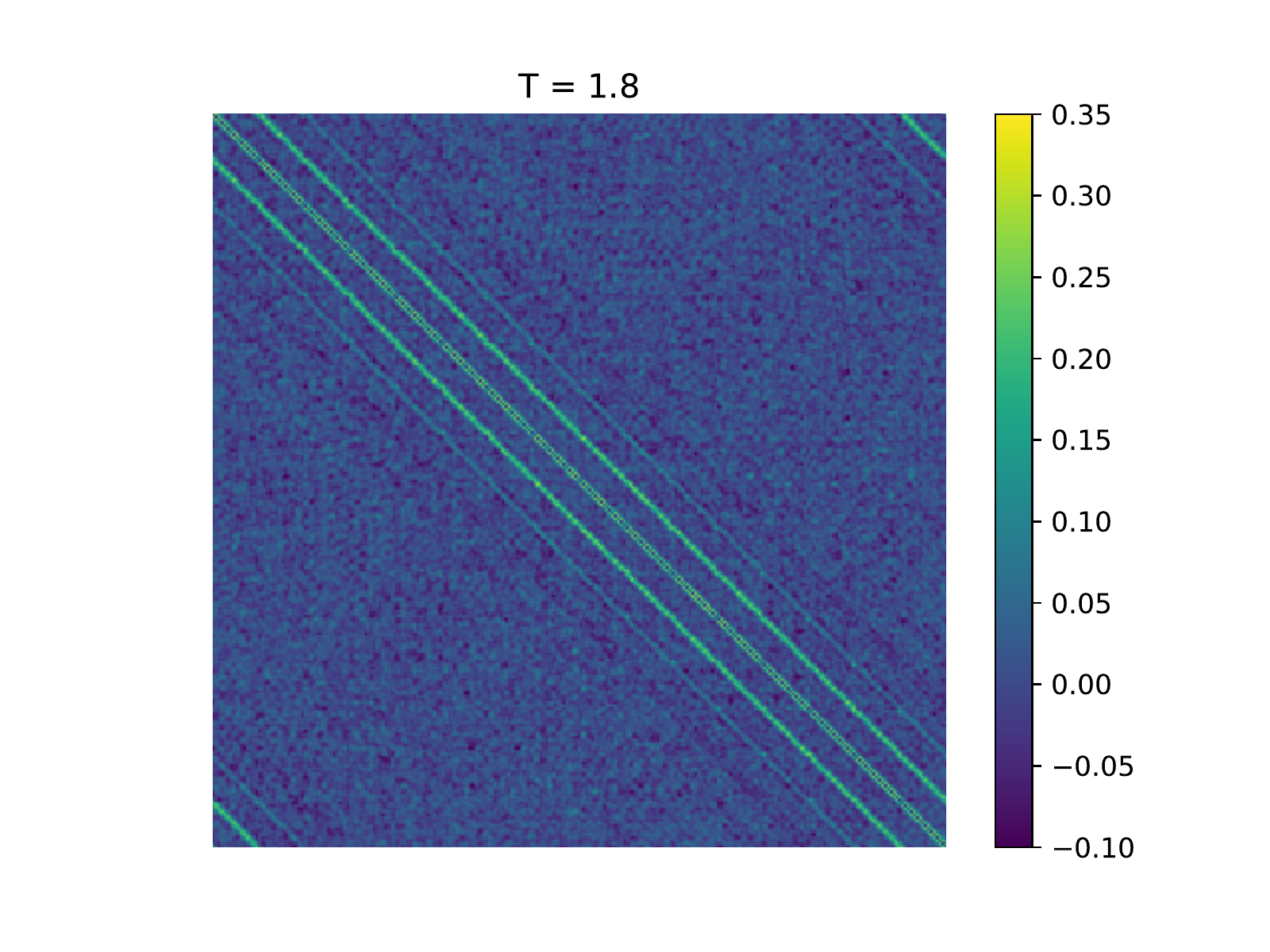}  
	\endminipage\hfill
	\minipage{0.50\textwidth}
	\includegraphics[width=\linewidth]{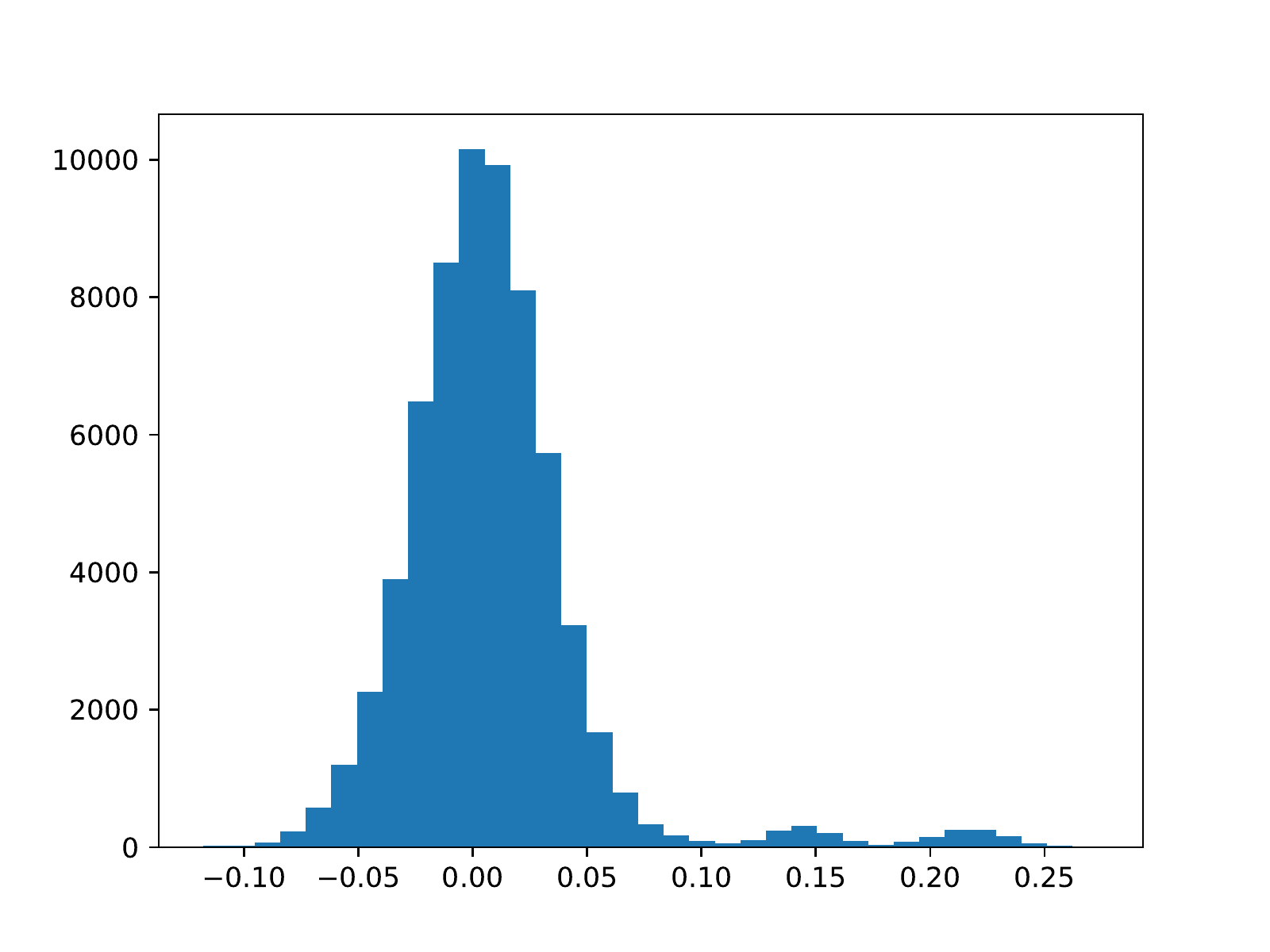}  
	\endminipage\hfill
	\caption{The two-point interaction matrix for the $L^2=16\times16$ system (left) and the corresponding histogram (right) with $h^2 =  12 \times 12$. The first peak is centred around zero, corresponding the the non nearest neighbour interactions. The second peak around 0.15 indicates other non nearest correlations that the machine has to learned to set to zero and are expected to vanish as it trains further. The final peak on the right hand side, corresponds to the correct nearest neighbour coupling. }
	\label{fig:coupling_h12}
\end{figure}

The observables, normalised by their expected values, are plotted in Fig.~\ref{fig:h12-k10_a001_k20_a0001}. We can see that the values are furthest from one as compared to the other cases we have trained so far. This is consistent with the measurement of the coupling matrix at a representative epoch towards the end of this training, as presented in Fig.~\ref{fig:coupling_h12}. The nearest neighbour structure is present, however, further correlations between the spins can also be observed, which are not expected. A third peak in the histogram confirms the latter observation more clearly in Fig.~\ref{fig:coupling_h12}.  \\

According to Fig.~\ref{fig:v16-h12-LL}, the log-likelihood is still increasing, however, the run has become computationally expensive at this stage with $\alpha=0.0001$ and $k=40$. Recall that in Fig.~\ref{fig:rbmweightsprogression}, it is observed that there are correlations which are further than nearest neighbours predicted by the RBM, before the learning is complete. Therefore we could expect that with more training, the RBM may eventually learn the correct nearest neighbour structure, setting the rest of the correlations to zero. Also notice that a $L^2=16\times16$ with hidden nodes of the size $h^2=12\times12$ has $\mathcal{O}(3\times10^5)$ less parameters than the case with  $h^2=16\times16$. Hence, the difficulty in training the machine is not perhaps too surprising. Having said that, further works need to be done to gain a better understanding of the effect of reducing hidden nodes on training an RBM.

\section{3- and 4-point interaction histograms}\label{app:3-4-point-interaction}
\begin{figure}[!htb]
\centering
\includegraphics[width=\textwidth]{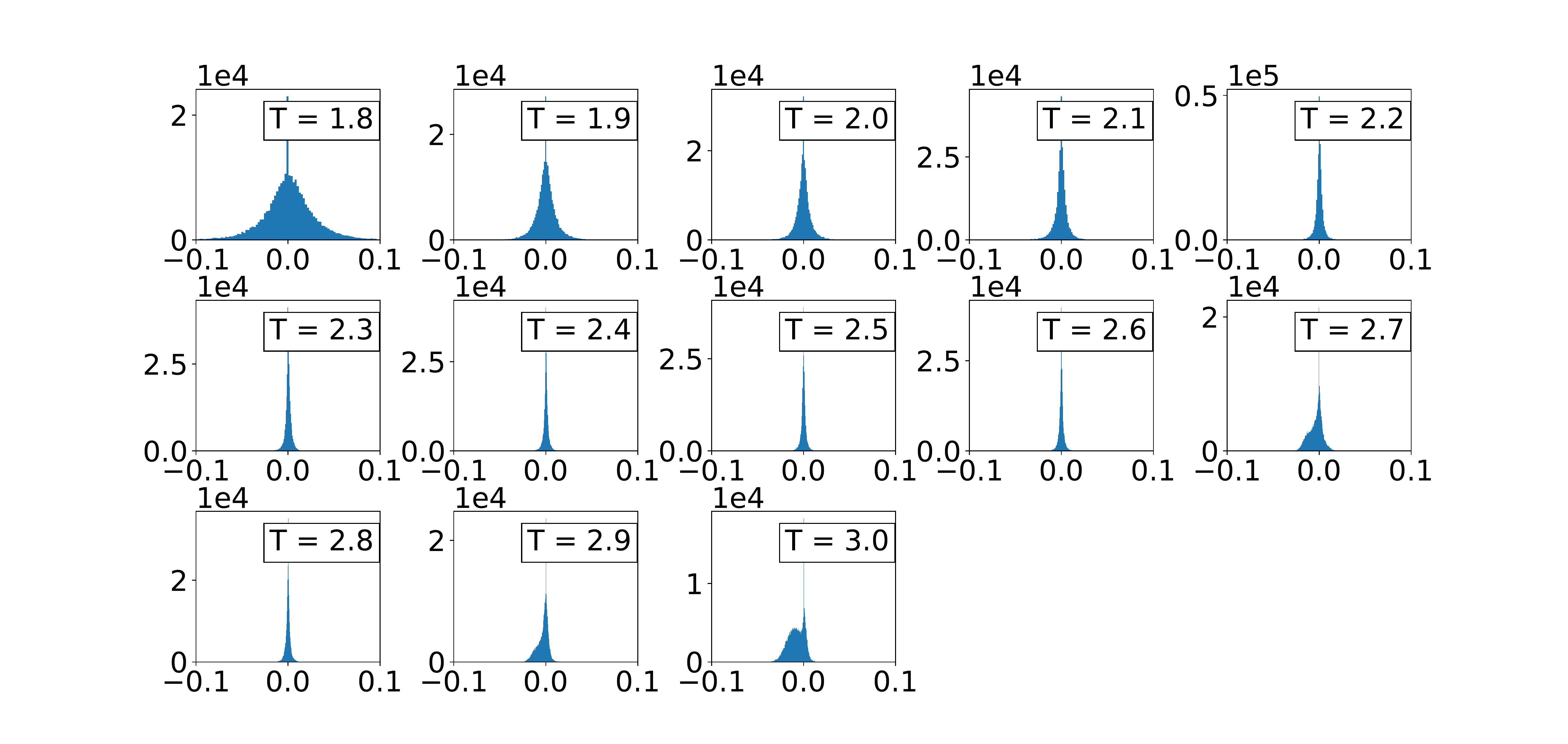}
\caption{The histograms of the entries of the 3-point interaction tensor extracted from RBMs trained at a temperature indicated above each subplot.}
\label{fig:3hist}
\end{figure}

\begin{figure}[!htb]
\centering
\includegraphics[width=\textwidth]{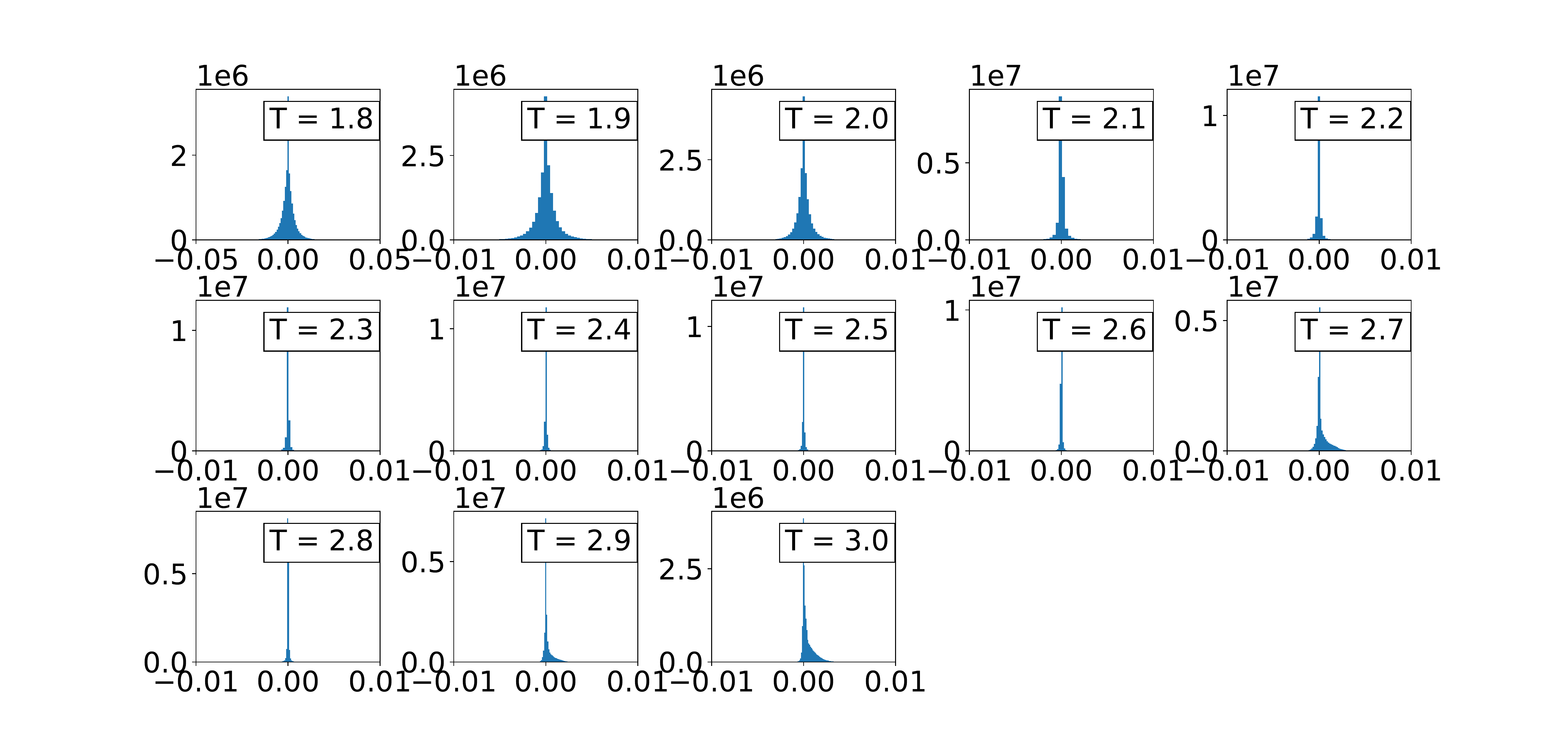}
\caption{The histograms of the entries of the 4-point interaction tensor extracted from RBMs trained at a temperature indicated above each subplot.}
\label{fig:4hist}
\end{figure}

We present the histograms of the 3- and 4-point couplings between the spins, in Fig.~\ref{fig:3hist} and \ref{fig:4hist} respectively. Generally, the single peak expected behaviour is observed. However for some of the higher temperatures, a second mode seems to appear in the histogram of the three-point interaction. More work needs to be done to understand why these machines appear to be learning a higher order interaction than we would not expect.

\section{Metropolis history plots}
\begin{figure}[!htb]
	\minipage{0.3\textwidth}
	\includegraphics[width=\linewidth]{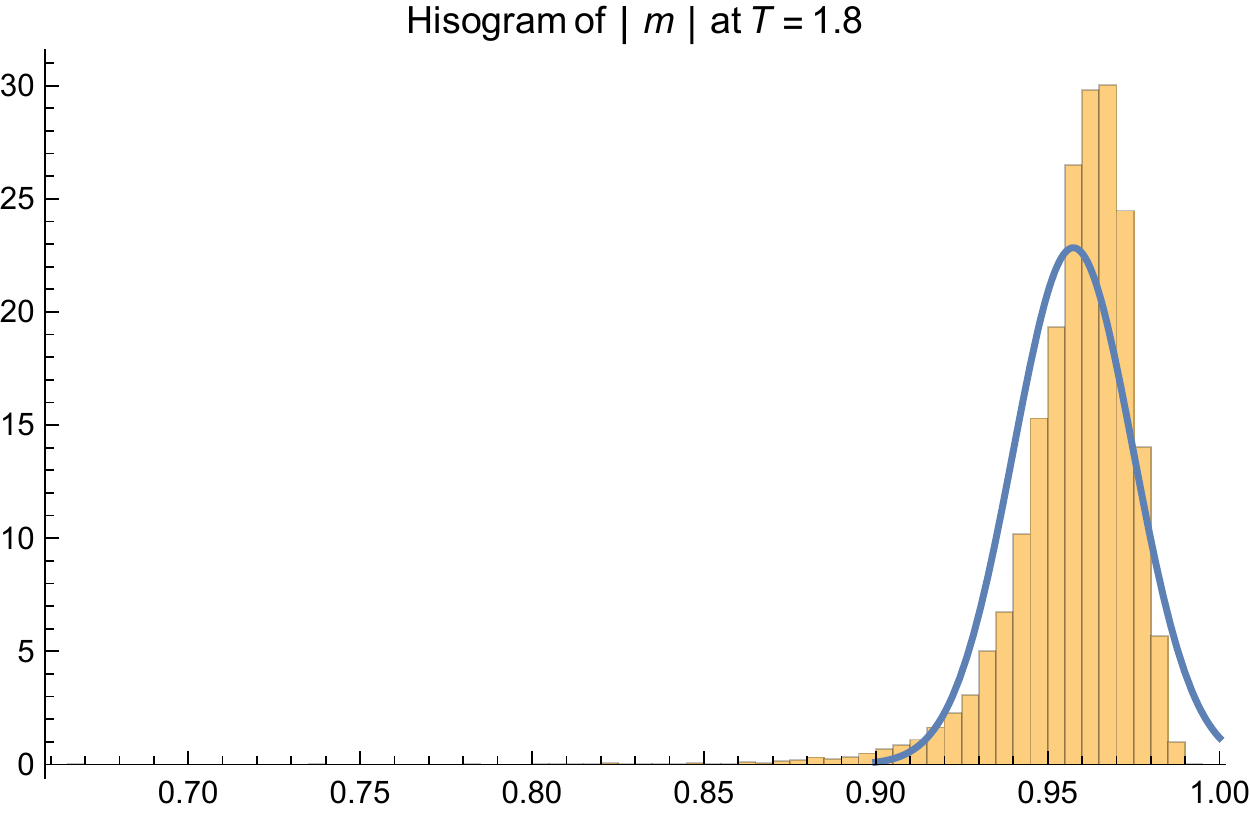}  
	\endminipage\hspace{2cm}
	\minipage{0.3\textwidth}
	\includegraphics[width=\linewidth]{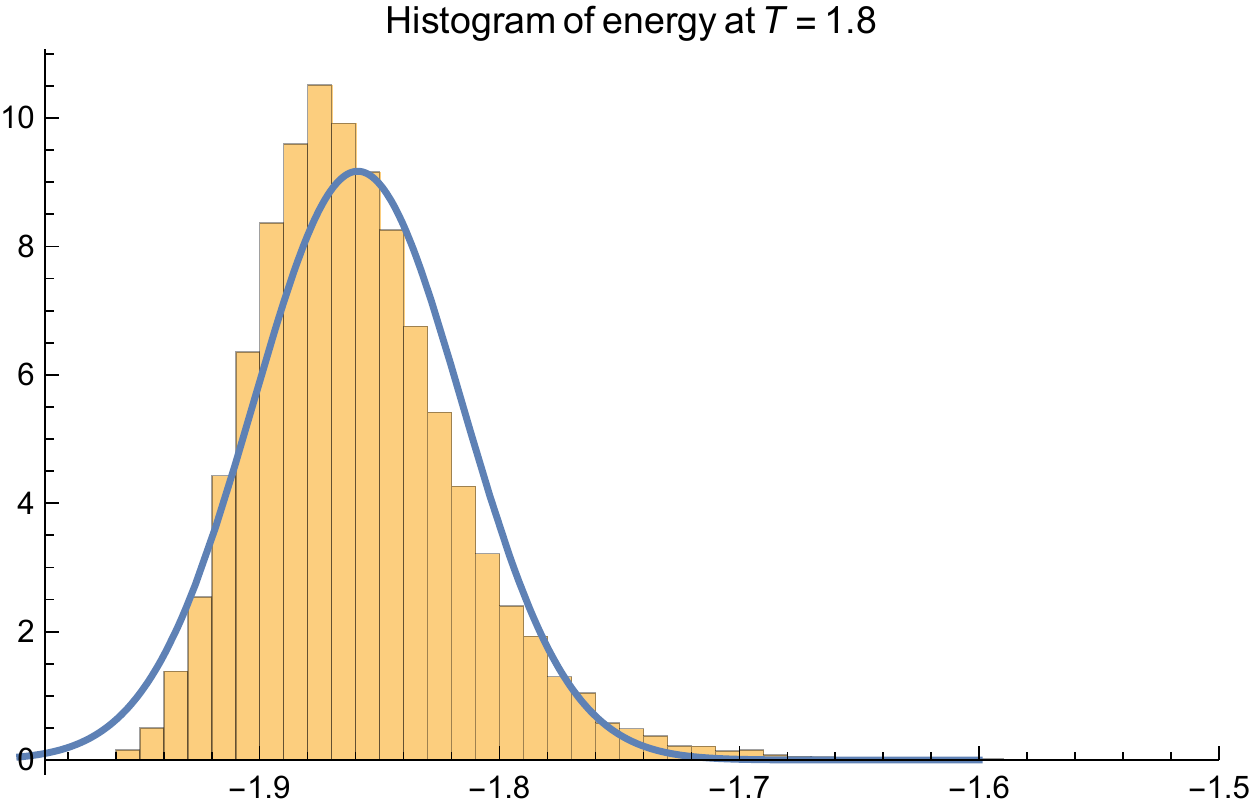}  
	\endminipage\hfill
	\caption{Histogram of $|m|$ (left) and energy (right) for the Metropolis algorithm at $T=1.8$. The blue line represent the normal distribution with values of its mean and standard deviation obtained from the data producing the histogram. }
	\label{fig:hist-metro-18}
\end{figure}

\begin{figure}[!htb]
	\minipage{0.3\textwidth}
	\includegraphics[width=\linewidth]{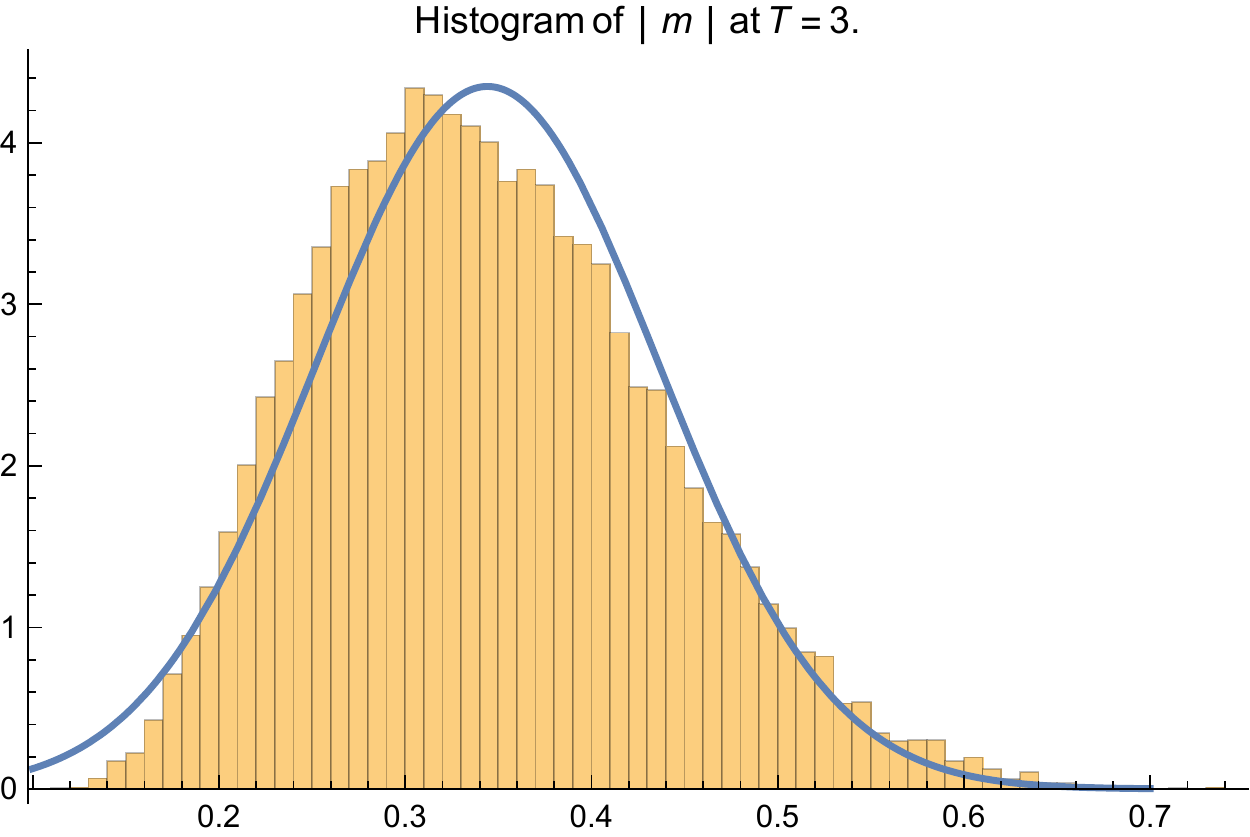}  
	\endminipage\hspace{2cm}
	\minipage{0.3\textwidth}
	\includegraphics[width=\linewidth]{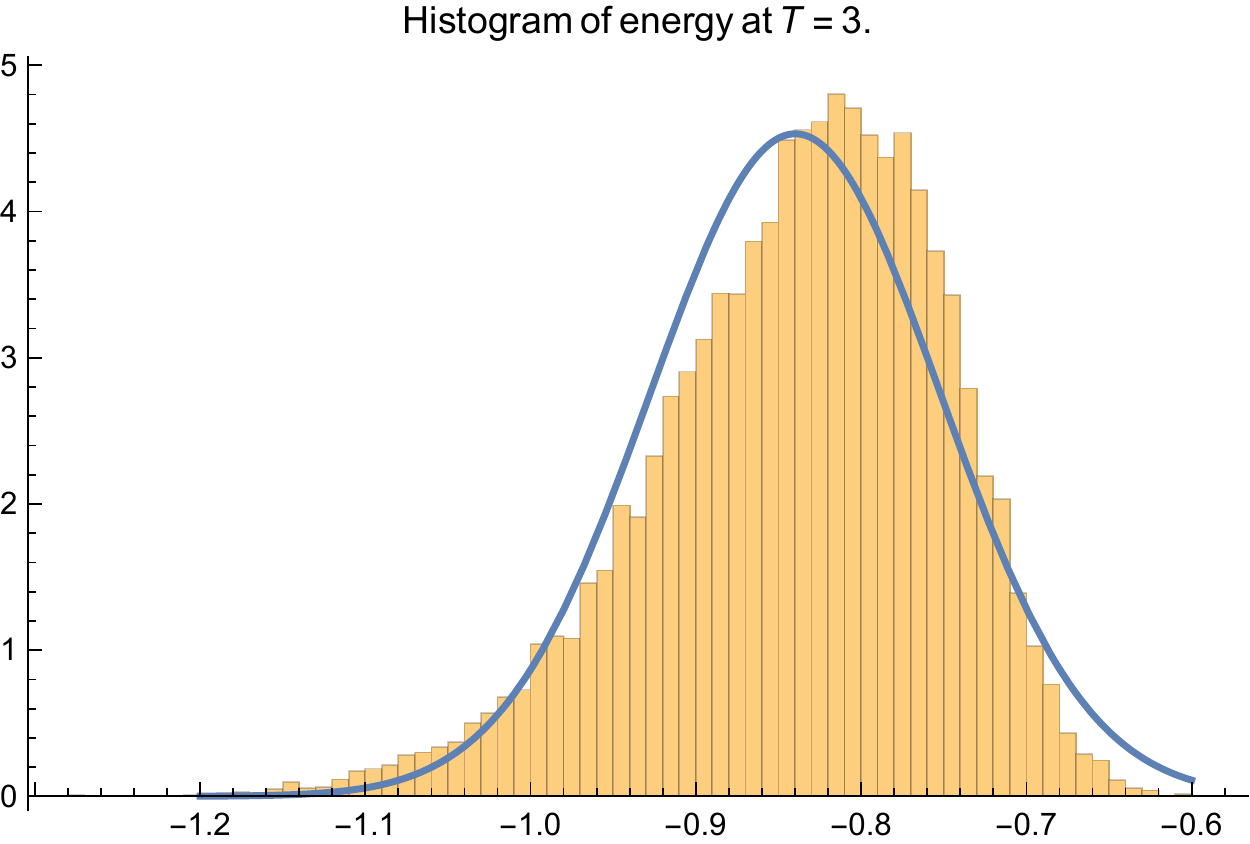}  
	\endminipage\hfill
	\caption{Histogram of $|m|$ (left) and energy (right) for the Metropolis algorithm at $T=3.0$. The blue line represent the normal distribution with values of its mean and standard deviation obtained from the data producing the histogram. }
	\label{fig:hist-metro-30}
\end{figure}

The absolute magnetisation and energy histograms of Ising configurations, over which the final measurements for the observables are made, are presented for two different temperatures in Fig.~\ref{fig:hist-metro-18} and Fig.~\ref{fig:hist-metro-30}. The histograms compare well to their corresponding normal distribution, whose parameters are measured by measuring the mean and the standard deviation from the data producing the histogram. Running the Metropolis chain longer than $1\times10^6$ steps did not change the from of the distributions.

\begin{figure}
\centering
\includegraphics[width=0.6\textwidth]{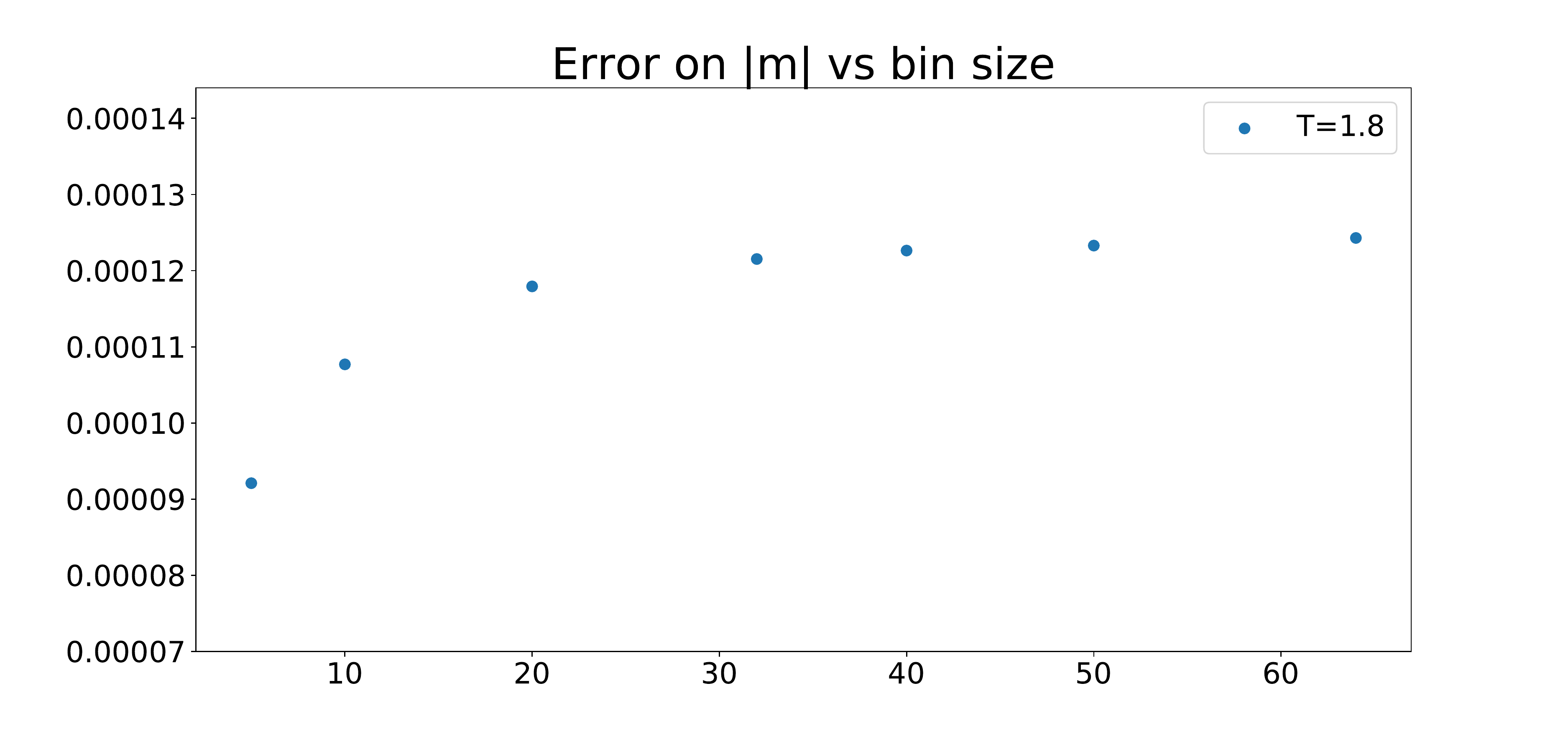}
\caption{Error on $|m|$ plotted for choice of bin size. As the measurements become more independent, the correlation between them decrease and hence the error increases. When the measurements are no longer dependent, the error remains constant.}
\label{fig:metro-bin-size}
\end{figure}

\begin{figure}
\centering
\includegraphics[width=0.6\textwidth]{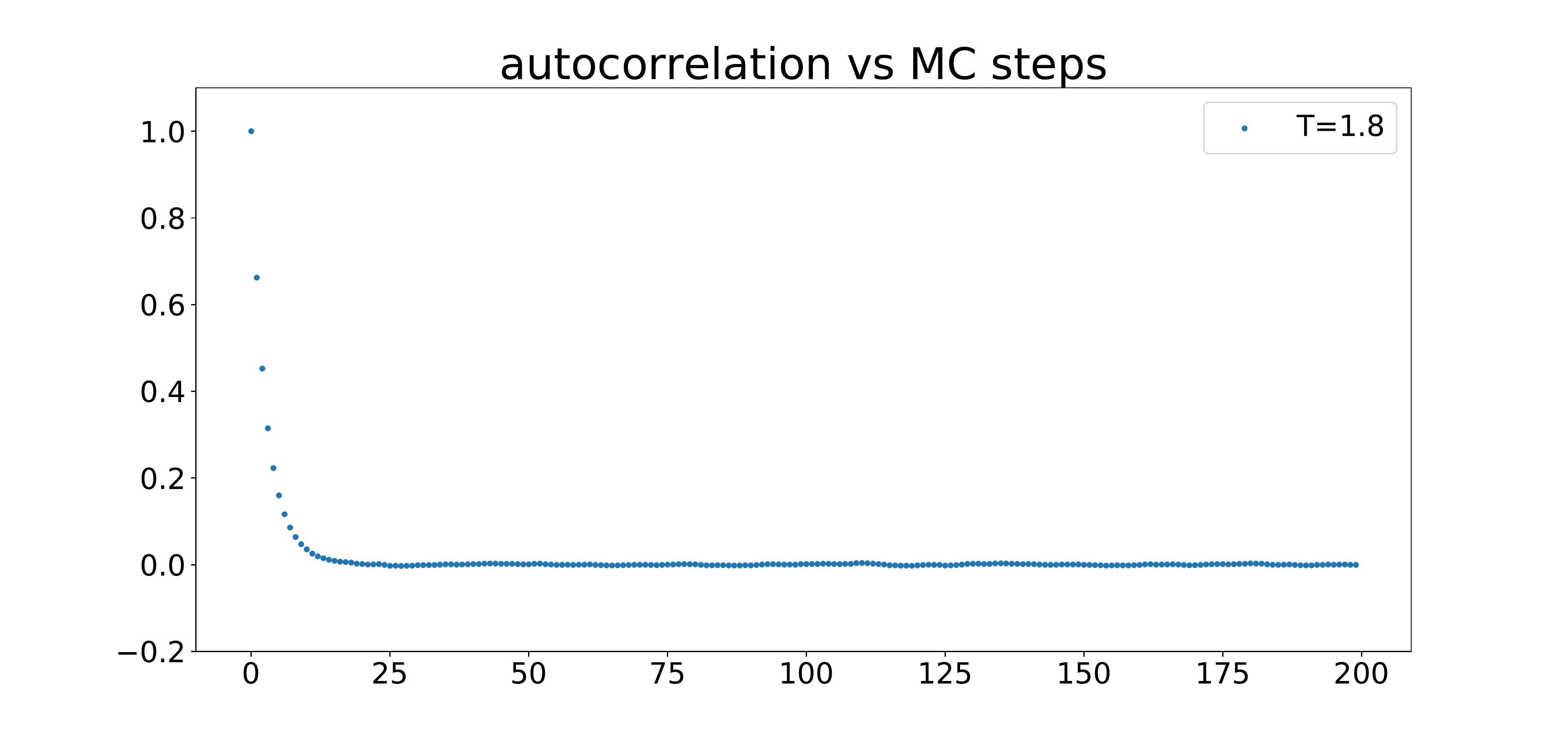}
\caption{Autocorrelation time as a function of MC steps.}
\label{fig:metro-auto-corr}
\end{figure}

Note that in order for the measurements to be independent of each other, we have binned every 50 measurements of $|m|$ and $E$ on each configuration. To decide on the bin size, we plotted the error on the measurement of magnetisation vs choice of bin size. This is shown in Fig.~\ref{fig:metro-bin-size}, while the autocorrelation time, defined in Eq,~\ref{eq:auto-corr-time}, is plotted in Fig.~\ref{fig:metro-auto-corr}

\clearpage

\bibliographystyle{plainnat}

\bibliography{refs}

\end{document}